\documentclass[sigconf,balance=false]{acmart}

\makeatletter
\let\orig@footnotetextcopyrightpermission\footnotetextcopyrightpermission
\renewcommand\footnotetextcopyrightpermission[1]{%
  \orig@footnotetextcopyrightpermission{Manuscript submitted to ACM}}
\makeatother

\usepackage{longtable}
\usepackage{array}
\usepackage{tabularx}
\usepackage{colortbl}
\usepackage{adjustbox}
\usepackage{mdframed}
\usepackage{enumitem}

\usepackage{fvextra}
\usepackage[most]{tcolorbox}

\fvset{fontsize=\footnotesize, breaklines=true, breakanywhere=true,
       breakindent=0pt, breaksymbolleft={}}

\newtcolorbox{promptbox}[1]{%
  enhanced, breakable,
  colback=gray!4, colframe=black!50, boxrule=0.4pt, arc=1.5pt,
  left=6pt, right=6pt, top=4pt, bottom=2pt,
  colbacktitle=gray!15, coltitle=black, fonttitle=\small\bfseries,
  title={#1}}

\definecolor{cbPrim}{HTML}{2F5C8F}   
\definecolor{cbBoth}{HTML}{2E7D5B}   
\definecolor{cbSec}{HTML}{B5651D}    
\definecolor{suKey}{HTML}{2F5C8F}

\definecolor{suHead}{HTML}{DAE5F0}   
\definecolor{suGroup}{HTML}{EEF3F8}  
\definecolor{suZebra}{HTML}{F8FAFC}  
\usepackage{xfp}
\definecolor{suBlue}{HTML}{5B8FC9}   
\definecolor{suUp}{HTML}{1B7F3B}     
\definecolor{suDown}{HTML}{B3261E}   
\definecolor{suZero}{HTML}{8A8A8A}   
\definecolor{traceBackground}{HTML}{F2F2F2}
\definecolor{traceRule}{HTML}{666666}

\newmdenv[
  skipabove=6pt,
  skipbelow=6pt,
  leftmargin=0pt,
  rightmargin=0pt,
  innerleftmargin=9pt,
  innerrightmargin=9pt,
  innertopmargin=6pt,
  innerbottommargin=6pt,
  topline=false,
  rightline=false,
  bottomline=false,
  leftline=true,
  linewidth=2.5pt,
  linecolor=traceRule,
  backgroundcolor=traceBackground
]{reasoningtrace}

\colorlet{suSub}{suHead!35!white}
\newcommand{\sect}[1]{\multicolumn{2}{l}{\cellcolor{suSub}\textbf{#1}}}

\newcommand{\dlt}[1]{%
  \ifdim #1pt>0pt \textcolor{suUp}{#1}%
  \else\ifdim #1pt<0pt \textcolor{suDown}{#1}%
  \else\textcolor{suZero}{#1}\fi\fi}

\AtBeginDocument{%
  }

\setcopyright{acmlicensed}
\copyrightyear{2018}
\acmYear{2018}
\acmDOI{XXXXXXX.XXXXXXX}
\acmConference[Conference acronym 'XX]{Make sure to enter the correct
  conference title from your rights confirmation email}{June 03--05,
  2018}{Woodstock, NY}
\acmISBN{978-1-4503-XXXX-X/2018/06}

\begin{document}

\title[Seeing Is Not Knowing]{Seeing Is Not Knowing: The Epistemic Limits of LLM-Based Mental Health Inference from Youth Social Media Conversations}

\author{Xiaohan Ding}
\email{xding301@gatech.edu}
\affiliation{%
  \department{School of Interactive Computing}
  \institution{Georgia Institute of Technology}
  \city{Atlanta}
  \state{Georgia}
  \country{USA}
}

\author{Mengting Tu}
\email{mtt@gatech.edu}
\affiliation{%
  \department{School of Interactive Computing}   
  \institution{Georgia Institute of Technology}
  \city{Atlanta}
  \state{Georgia}
  \country{USA}
}

\author{Andrew H. Zhao}
\email{azhao63@gatech.edu}
\affiliation{%
  \department{Institute for People and Technology}
  \institution{Georgia Institute of Technology}
  \city{Atlanta}
  \state{Georgia}
  \country{USA}
}

\author{Jamie Zelazny}
\email{jmz22@pitt.edu}
\affiliation{%
  \department{School of Nursing}
  \institution{University of Pittsburgh}
  \city{Pittsburgh}
  \state{Pennsylvania}
  \country{USA}
}

\author{Candice L. Biernesser}
\email{lubbertcl@upmc.edu}
\affiliation{%
  \institution{University of Pittsburgh Medical Center}
  \city{Pittsburgh}
  \state{Pennsylvania}
  \country{USA}
}

\author{Munmun De Choudhury}
\email{munmun.choudhury@cc.gatech.edu}
\affiliation{%
  \department{School of Interactive Computing}
  \institution{Georgia Institute of Technology}
  \city{Atlanta}
  \state{Georgia}
  \country{USA}
}

\renewcommand{\shortauthors}{Ding et al.}


\begin{abstract}
Adolescents increasingly use social media for private conversations, sharing experiences and seeking support. These interactions generate digital traces, prompting efforts to infer mental health and psychosocial states. As large language models become more capable of reasoning about conversations and interpersonal dynamics, a fundamental question remains nonetheless: what can models actually ``\textit{\textbf{know}}'' about an individual’s psychological experience from what they are able to ``\textit{\textbf{see}}''? We evaluated ten reasoning LLMs using donated Instagram Direct Messages from 80 adolescents with a history of depression or suicidal thoughts and behaviors, comparing predictions with self-reports and analyzing 204 reasoning traces. Models performed better when inferring more observable states than participants’ subjective appraisals of social experience, such as perceived social support. Beyond prediction accuracy, our findings highlight the epistemic limits of digital traces for AI design and evaluation: what can be observed, what can reasonably be inferred, and what requires people’s own accounts.
\\
\noindent \textbf{\textit{\textcolor{red}{Content Warning:}} This paper presents findings on sensitive mental health topics such as suicide and self-harm.}
\end{abstract}



\begin{teaserfigure}
\centering
 \includegraphics[width=0.8\textwidth]{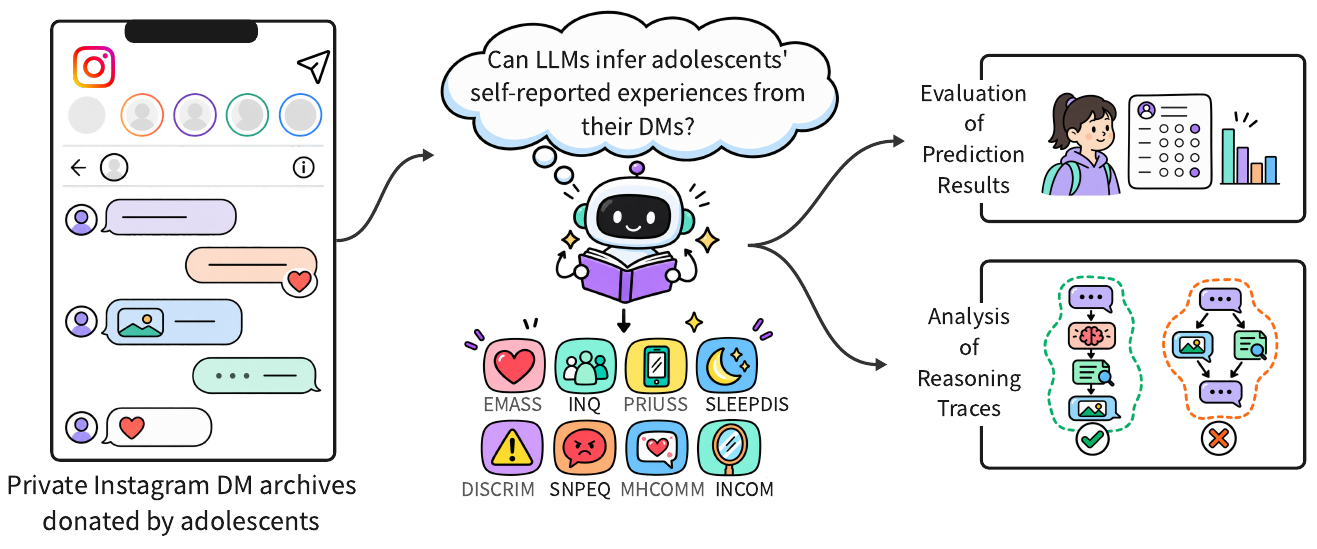}
  \caption{Overview of our study. Adolescents donated their private Instagram DM archives and completed eight self-report measures (left). Ten locally deployed reasoning LLMs predicted participants' questionnaire responses from these DMs (middle). We evaluated prediction performance at the item and construct levels (RQ1, top right) and analyzed model-generated reasoning traces to examine how models selected and interpreted evidence in lower- and higher-error predictions (RQ2, bottom right).}
  \Description{Overview of our study. Adolescents donated their private Instagram DM archives and completed eight self-report measures (left). Ten locally deployed reasoning LLMs predicted participants' questionnaire responses from these DMs (middle). We evaluated prediction performance at the item and construct levels (RQ1, top right) and analyzed model-generated reasoning traces to examine how models selected and interpreted evidence in lower- and higher-error predictions (RQ2, bottom right).}
  \label{fig:teaser}
\end{teaserfigure}


\maketitle

\section{Introduction}
Social media interactions leave traces of relationships as they unfold~\cite{sultan2023leaving, ehrenreich2020adolescents,underwood2015blackberry,edelmann2020computational}. They record moments of support~\cite{haber2007relationship,uchino2009understanding}, conflict~\cite{ehrenreich2014adolescents}, disclosure~\cite{valkenburg2009effects}, and ordinary conversation~\cite{ehrenreich2020adolescents}, along with who participated and when those exchanges occurred. For adolescents, whose social interactions increasingly unfold through private digital communication~\cite{nesi2018transformation}, these records offer a detailed view of observable behavior and relationships~\cite{sultan2023leaving,underwood2012blackberry}. This interpersonal context is particularly relevant to psychological and mental-health inference: developmental and clinical research has long established the importance of peer relationships and interpersonal experiences for adolescent psychological adjustment and well-being~\cite{schwartz2020relations,mitic2021toward}. 

Researchers have long used digital traces to infer personal attributes and psychological states~\cite{kosinski2013private, dechoudhury2013predicting, chancellor2020methods}. 
As large language models (LLMs) become capable of processing long conversational histories~\cite{maharana2024evaluating}, and producing individualized inferences from them~\cite{demszky2023using,xu2024mental}, 
private communication offers an especially rich evidence environment for extending this line of work toward individualized psychological and mental-health inference~\cite{demszky2023using,xu2024mental,ravenda2025llms}. 
Much of this work evaluates whether model outputs correspond to observed labels or responses to psychological and clinical questionnaires~\cite{chancellor2020methods, xu2024mental}. Such correspondence establishes predictive performance, but it does not establish
whether the information available to the model is sufficient to support the inference being made~\cite{ernala2019methodological, jacobs2021measurement, puspo2026mental}. This distinction has become more pronounced with LLMs because their capacity to integrate dispersed conversational evidence and produce coherent individualized judgments can make psychological states appear inferable even when the evidence needed to establish those states is not present in the trace. Essentially, a model can observe what a person says, does, and receives without access to what those interactions mean to that person~\cite{uchino2009understanding, beyens2020effect,orben2019association}. This exposes a fundamental gap in current approaches to digital-trace inference: we know increasingly more about whether models can predict psychological outcomes from behavioral traces, but comparatively less about whether those traces provide adequate evidence for the inference. 
Put differently, \textbf{what can models actually \textit{``know''} about an individual's psychological experience from what they are able to \textit{``see''}}~\cite{ernala2019methodological,chancellor2020methods,sivak2020measuring}? In this sense, \textit{seeing} may not always be \textit{knowing}.


Many of these psychological constructs that serve as ground truth for the models depend on how individuals appraise their own experiences~\cite{cohen1983global, schwarz1999self}. This creates a particular challenge for inference from digital traces because the event or behavior visible to a model may not be equivalent to the experience a questionnaire asks a person to report~\cite{haber2007relationship, nick2018online}. 
A supportive message, for example, shows that support was offered, not necessarily that it was felt; late-night messaging shows nighttime activity, not necessarily sleep interference. Prior HCI research has raised related concerns about proxy-based mental-health inference from social media and the risks of inferring sensitive states from behavioral traces \cite{ernala2019methodological,chancellor2019taxonomy}. Yet the growing use of LLMs for psychological inference introduces a new version of this problem: models can locate, combine, and interpret multiple plausible---even arguably weak---signals into coherent predictions, while the relationship between those signals and the experience being inferred may remain uncertain~\cite{xu2024mental,ji2023survey,jacobs2021measurement}. 



This makes private, longitudinal direct-message (DM) histories on social media a particularly revealing setting in which to examine the epistemic limits of psychological inference. Rather than relying on isolated public posts or aggregate behavioral features, DMs preserve conversational content, interaction partners, directionality, and repeated exchanges across relationships and time. They therefore provide models with comparatively rich evidence about the interpersonal contexts in which adolescents' experiences unfold.

However, the richness of an archive does not mean that its evidence is readily accessible to a model, nor that the evidence it contains is sufficient for the psychological inference being made. Relevant information may be dispersed across an archive or expressed in language unlike formal questionnaire wording. 
LLMs therefore face two distinct challenges: whether they can locate and organize evidence relevant to the target, and whether that evidence can reasonably support the psychological experience they are asked to infer. The first is an evidence-construction problem; the second is an evidence interpretation problem. Separating these problems allows us to examine whether failures arise because models cannot ``see'' relevant evidence, or because what they ``see'' cannot establish what they are being asked to ``know.'' 
We therefore ask:

\begin{itemize}[leftmargin=*]
    \item {\it RQ1: To what extent can LLMs infer individuals’ subjective mental health states from their social-media conversations?}
    \item {\it RQ2: How do LLMs select and interpret conversational evidence when their inferences succeed or fail, and how do these reasoning patterns vary across the mental health states being inferred? }
\end{itemize}

We study these questions using participant-donated Instagram DM archives from 80 adolescents who also completed eight self-report measures spanning negative peer experiences on social media, online mental-health communication, social support, online discrimination, problematic internet use, social-media sleep disturbance, social comparison, and belonging. Their archives contain approximately 2.41 million messages. Across ten locally deployed privacy-preserving reasoning LLMs, we evaluate two approaches for organizing evidence from these archives relevant to questionnaire constructs and assess predictive performance at both the item and construct levels. We further analyze 204 model-generated reasoning traces to examine what evidence models selected, how they interpreted it, and how those interpretations shaped their questionnaire predictions.

Our results reveal a consistent boundary between what conversational traces make visible and what they can support as an inference. Models performed better for constructs more directly reflected in observable interactions, including negative peer experiences, mental-health communication, and discrimination, and worse when inference depended more on participants' subjective perceptions of their experiences. This distinction was especially clear for social support: models frequently treated observable supportive interactions as evidence that participants themselves felt supported, even when their self-reports indicated otherwise. More broadly, numerical closeness to observed questionnaire scores could coexist with weak differentiation among participants, while model-generated reasoning traces showed that models did not reliably recognize when visible interpersonal behavior was insufficient evidence of internal experience. Overall, our work makes three contributions:

\begin{itemize}[leftmargin=*]
\item We characterize an evidence--construct alignment boundary in LLM-based inference from private, longitudinal adolescent communication, showing that prediction depends not only on whether models can locate relevant evidence, but also on whether the evidence available in the trace can reasonably support the experience being inferred. In doing so, we distinguish limitations of evidence access from epistemic limits of the underlying data.

\item We show that numerical closeness can mask a failure to capture individual differences in psychological and psychosocial prediction. Predicted scores could approximate participants' reported levels while understating the differences between them and failing to identify who reported greater or lesser difficulties. This finding challenges the use of score proximity alone as evidence of successful inference about individuals and highlights the need to assess whether predictions capture how people's reported experiences differ.

\item Through analysis of model-generated reasoning traces, we distinguish \emph{evidence selection} from \emph{evidence interpretation} and characterize how models move from observable communication to questionnaire judgments. These patterns identify where models extend visible interpersonal evidence beyond what it can support, informing the design and evaluation of AI systems that account for the epistemic limits of digital traces.
\end{itemize}

\section{Related Work}
\subsection{Psychological Inference from Digital Traces and LLMs}

Over the past decade, social media and other digital traces have been recognized as a non-reactive and longitudinal lens into psychological states and well-being~\cite{dechoudhury2013predicting,chancellor2020methods}. Patterns of platform activity and the language of shared content have been used to infer personality and other sensitive personal attributes~\cite{kosinski2013private}, major depressive disorder~\cite{dechoudhury2013predicting,eichstaedt2018facebook}, and a range of mental health concerns and well-being outcomes~\cite{chancellor2020methods}. Despite differences in platform, modality, and target, these efforts share a common predictive structure: observable digital evidence is mapped to an outcome, and the resulting prediction is evaluated against an external label or self-reported measure.

Large language models extend this line of work by changing how evidence from digital traces can be represented and integrated~\cite{xu2024mental,ravenda2025llms}. Instead of relying  on curated lexica or hand-crafted behavioral features~\cite{schwartz2015data}, LLMs can operate directly on unstructured text~\cite{ravenda2025llms,rathje2024gpt}. Recent work has used them to recover personality dispositions from social media language in zero-shot settings~\cite{peters2024large,rathje2024gpt}, infer personal attributes from seemingly innocuous text~\cite{staab2024beyond}, and predict mental health responses to clinical and psychological questionnaires~\cite{xu2024mental,ravenda2025llms}. These capabilities are particularly relevant to longitudinal communication histories, where information pertinent to an individual's experience may be distributed across conversations, relationships, and time rather than expressed in language that directly mirrors a target construct.

Prior work has identified important limitations in psychological inference from digital traces. Researchers have questioned whether platform-derived labels and behavioral proxies adequately represent the psychological constructs of interest~\cite{ernala2019methodological,jacobs2021measurement, das2022semantic, puspo2026mental}, whether models generalize across populations and settings~\cite{harrigian2020models}, and what ethical risks arise when sensitive states are inferred from behavioral data~\cite{chancellor2019taxonomy,chancellor2019human}. Less attention has been paid to the evidentiary relationship between the information available to a model and the particular psychological claim it is asked to make. 
Predictive models are still commonly evaluated by how closely their outputs correspond to observed labels or self-reported measures~\cite{chancellor2020methods}. Such correspondence is informative about predictive performance, but does not establish whether the evidence available to a model is sufficient to support the inference being made.

Building on these concerns, we distinguish three questions in evaluating psychological inference from digital traces: whether predictions correspond to participants' reported outcomes, whether they preserve meaningful differences between participants, and whether the evidence available to the model can support the psychological inference being made.


\subsection{From Observable Interaction to Subjective Experience}
A substantial body of research has examined how social and behavioral experiences relate to individuals' subjective well-being and psychological assessments. Social support is one prominent example. Prior work has distinguished support that is provided or received from support that individuals perceive as available to them, showing that these forms of support are related but capture different aspects of social experience~\cite{haber2007relationship,uchino2009understanding}. Supportive actions may also go unnoticed by their recipients~\cite{bolger2000invisible}. Related work on social media and well-being has similarly shown substantial individual differences in how comparable patterns of technology use and social interaction are experienced~\cite{beyens2020effect,valkenburg2013differential}. Together, these studies suggest that observable social behavior can be informative about an individual's experience without being equivalent to that experience.

This distinction has become increasingly important as researchers use digital traces to measure psychological and well-being outcomes. Researchers have compared behavioral records from social media, smartphones, and other digital systems with participants' self-reported assessments, finding that the two can capture different dimensions of experience. Passive-sensing research, for example, has described a ``semantic gap'' between behavioral indicators and self-reported well-being~\cite{das2022semantic}. Research on technology use has likewise found that trace-based measures and individuals' perceptions of their technology use are not interchangeable~\cite{verbeij2022experience}. In computational mental health, scholars have raised broader concerns about using observable behavioral proxies to represent psychological constructs whose definitions may not be directly reflected in those behaviors~\cite{ernala2019methodological,jacobs2021measurement}. Together, this work establishes that behavioral and self-reported measures need not be interchangeable, even when they concern the same underlying domain. However, it leaves open how this mismatch changes when models can reason over substantially richer behavioral evidence rather than predefined features or aggregate sensing measures.

Interpersonal relationships provide a particularly important context for understanding adolescent psychological experience. Developmental and clinical research has long emphasized the increasing salience of peer relationships during adolescence and their role in psychological adjustment and mental health~\cite{mitic2021toward,schwartz2020relations}. Support, rejection, conflict, disclosure, and other interpersonal processes can shape, and be shaped by, adolescents’ psychological well-being~\cite{filia2021facilitating}. Thus, interpersonal communication is not simply another source of behavioral data; it provides a naturalistic record of interactions through which psychologically consequential relationships are enacted.

Private and longitudinal communication offers a particularly rich source of behavioral evidence. Adolescent peer interaction increasingly takes place through digital and private communication channels~\cite{nesi2018transformation,ehrenreich2020adolescents}, and prior research has used text messages and participant-donated social-media records to study adolescents' relationships, communication patterns, and negative online experiences~\cite{underwood2012blackberry,underwood2015blackberry,ehrenreich2014adolescents,kim2024respect}. Compared with public posts or aggregate measures of platform activity, these records preserve conversational content~\cite{underwood2012blackberry}, conversation partner~\cite{underwood2012blackberry, ehrenreich2020adolescents}, directionality~\cite{vollet2020exploring}, and exchanges across time~\cite{underwood2012blackberry, vollet2020exploring}. We therefore focus on DMs not because they exhaust the digital evidence potentially available about an individual, but because they provide a naturalistic record of an interpersonal context that psychological research identifies as consequential for adolescent mental health and well-being. 

At the same time, the psychological importance of interpersonal interaction does not mean that interpersonal communication provides direct access to psychological experience. DMs make some aspects of an individual’s social environment visible, while public posts, reactions, viewing behavior, and other platform activity may provide different forms of evidence. Crucially, 
richer conversational context does not necessarily resolve the distinction identified in prior work between observable interaction and subjective experience~\cite{bierstetel2020associations, wolf2021development}. Existing studies of private communication have primarily used these records to characterize behaviors~\cite{kooti2015evolution, brown2022elementary}, relationships~\cite{ireland2011language}, or interactional experiences~\cite{laursen2012sequential}, leaving unresolved whether even contextualized communication can provide an adequate evidentiary basis for inferring individuals’ subjective psychological experiences. 
This question becomes more salient with LLMs~\cite{seo2024chacha}, which can integrate evidence distributed across long communication histories~\cite{maharana2024evaluating}, and translate that evidence directly into psychological judgments~\cite{peters2024large}. It therefore remains unclear whether LLMs' increased capacity to access and integrate conversational evidence narrows the gap between observable interaction and subjective experience or instead enables increasingly plausible inferences that extend beyond what the available evidence can establish. 


Our work combines these lines of research by examining LLM inference from private, longitudinal communication paired with participants' self-reported measures. We ask not only if model predictions correspond to reported outcomes and differentiate among participants, but if the conversational evidence available to the model can reasonably support the experience being inferred.

\begin{table*}[t]
  \caption{Demographic characteristics of the questionnaire sample ($N=237$).}
  \Description{A two-panel table reporting age, sex assigned at birth, and
    gender identity in the left panel, and race and ethnicity in the right
    panel, for the 237 adolescents represented in the questionnaire records.
    Each characteristic appears as a shaded heading row followed by its
    categories and counts. The header row is shaded light blue.}
  \label{tab:demographics}
  \centering
  \small
  \setlength{\tabcolsep}{8pt}
  \renewcommand{\arraystretch}{0.985}
  \setlength{\aboverulesep}{0pt}
  \setlength{\belowrulesep}{0pt}
  \begin{tabular}{lrlr}
    \toprule
    \rowcolor{suHead}
    \textbf{Characteristic} & \textbf{Value} &
    \textbf{Characteristic} & \textbf{Value} \\
    \midrule
    \sect{Age}                  & \sect{Race} \\
    Mean (SD), years            & 15.1 (1.7)   & Black or African American & 124 (52.3\%) \\
    Range, years                & 12--18       & White                     & 80 (33.8\%)  \\
    Missing                     & 6 (2.5\%)    & More than one race        & 17 (7.2\%)   \\
    \sect{Sex assigned at birth} & Asian                    & 2 (0.8\%)    \\
    Female                      & 183 (77.2\%) & Other                     & 7 (3.0\%)    \\
    Male                        & 48 (20.3\%)  & Refused                   & 1 (0.4\%)    \\
    Missing                     & 6 (2.5\%)    & Missing                   & 6 (2.5\%)    \\
    \sect{Gender identity}      & \sect{Ethnicity} \\
    Female                      & 136 (57.4\%) & Non-Hispanic              & 209 (88.2\%) \\
    Male                        & 38 (16.0\%)  & Hispanic                  & 21 (8.9\%)   \\
    Female-to-male transgender  & 23 (9.7\%)   & Unclassified              & 1 (0.4\%)    \\
    Male-to-female transgender  & 7 (3.0\%)    & Missing                   & 6 (2.5\%)    \\
    Non-binary or queer         & 14 (5.9\%)   &                           &              \\
    Other                       & 7 (3.0\%)    &                           &              \\
    Missing                     & 12 (5.1\%)   &                           &              \\
    \bottomrule
  \end{tabular}
\end{table*}
\section{Dataset}
\label{sec:data}
This paper utilizes a dataset of 
participants recruited through a large, federally-funded study of adolescent mental health and suicide risk. This parent study has been conducted across pediatric primary care practices affiliated with two large academic health systems in the midwestern and northeastern United States. The study enrolled adolescents aged 12-18 years across the spectrum of clinical depression and suicide risk, with oversampling of youth at elevated risk for suicidal behavior and a goal of substantial representation of Black youth. Eligibility required adolescents to be able to participate in study procedures and have routine access to a mobile phone. Exclusion criteria included a current manic or psychotic episode, a life-threatening medical condition requiring immediate treatment, an intellectual or developmental disability or autism spectrum disorder that precluded comprehension of study procedures, and lack of routine access to a mobile phone.

Potential participants in the parent study were identified through routine depression and suicide-risk screening in participating practices using the Patient Health Questionnaire-9 Modified for Adolescents (PHQ-9-M). Adolescents were classified as high risk if they endorsed recent suicidal thoughts, past-month suicidal ideation, or a prior suicide attempt; as moderate risk if they had moderate-to-severe depressive symptoms (PHQ-9-M $\geq 11$) without indicators of suicide risk; as subsyndromal if they had a PHQ-9-M score of 5-10 without indicators of suicide risk; and as low risk if they had minimal depressive symptoms (PHQ-9-M $< 5$) without indicators of suicide risk.

The data used in this paper combined two sources: self-report questionnaires, which provided the observed outcomes, and participant-donated Instagram direct-message (DM) archives, which provided the evidence supplied to the models. The questionnaire records contained 237 adolescents, of whom 224 were marked complete. A subset of 103 adolescents donated an Instagram DM archive. Linking the two sources by participant identifier yielded 81 matched participants. One matched participant was absent from the released scored-outcome file and was therefore excluded, leaving 80 participants; per-measure sample sizes ranged from 73 to 80 (Table~\ref{tab:scales}). Table~\ref{tab:demographics} summarizes the demographic characteristics of the 237 adolescents represented in the questionnaire records (demographic breakdowns for the other analytic subsets: the 224 completers, the 103 DM donors, and the 80 matched participants are reported in Appendix~\ref{app:demographics}).

\subsection{Self-report Questionnaires}
\label{sec:outcomes}
The questionnaires measured eight aspects of online experience, mental-health communication, and social behavior. Table~\ref{tab:scales} reports each scale's construct definition and number of scored items. Tables~\ref{tab:items-snpeq-mhcomm}, \ref{tab:items-emass-discrim-priuss-sleepdis}, and~\ref{tab:items-incom-inq} provide the complete item wording.

\begin{table*}[tb]
  \caption{Self-report measures and construct definitions. \emph{Scale} is
    the abbreviation used throughout the paper and \emph{Instrument} the full
    name of the source questionnaire; \emph{Construct} summarizes what the
    measure captures. $k$ is the number of scored items and $n$ the number of
    participants with a valid score on that measure. All measures use a
    past-month recall window.}
  \Description{A five-column table listing eight self-report measures by
    abbreviation, full instrument name, construct definition, number of
    scored items, and number of respondents. The header row is shaded light
    blue and rows alternate white and very light gray.}
  \label{tab:scales}
  \small
  \setlength{\tabcolsep}{5pt}
  \renewcommand{\arraystretch}{1.2}
  \setlength{\aboverulesep}{0pt}
  \setlength{\belowrulesep}{0pt}
  \renewcommand{\tabularxcolumn}[1]{>{\raggedright\arraybackslash}m{#1}}
  \begin{tabularx}{\textwidth}{>{\raggedright\arraybackslash}m{0.075\textwidth}
                               >{\raggedright\arraybackslash}m{0.22\textwidth}
                               X
                               >{\centering\arraybackslash}m{0.03\textwidth}
                               >{\centering\arraybackslash}m{0.03\textwidth}}
    \toprule
    \rowcolor{suHead}
    \textbf{Scale} & \textbf{Instrument} & \textbf{Construct} & $\boldsymbol{k}$ & $\boldsymbol{n}$ \\
    \midrule
    SN-PEQ   & Social Networking--Peer Experiences Questionnaire
      & Negative peer experiences on social media, including blocking,
        hostility, embarrassment, exclusion, rumors, impersonation, and
        adverse romantic interactions. & 12 & 73 \\
    \rowcolor{suZebra}
    MH-Comm  & Online Mental Health Communication
      & Frequency of social-media communication about the participant's or
        friends' mental-health challenges, suicidal thoughts or behavior, and
        self-harm thoughts or behavior. & 4 & 80 \\
    E-MASS   & EMA Social Support
      & Supportive social-media interactions received from friends or
        followers, including help, dependability, listening, and assistance
        with problems. & 5 & 79 \\
    \rowcolor{suZebra}
    Discrim  & ABCD Discrimination Questions (modified)
      & Perceived social-media discrimination based on race or ethnicity,
        country of origin, LGBTQ+ identity, or weight. & 4 & 76 \\
    PRIUSS   & Problematic and Risky Internet Use Screening Scale
      & Social anxiety associated with social-media use, withdrawal when not
        using it, and avoiding other activities to remain online. & 3 & 80 \\
    \rowcolor{suZebra}
    SleepDis & Social Media Sleep Disturbance
      & Social-media use during the hour before sleep and perceived
        interference of social media with sleep. & 2 & 80 \\
    INCOM    & Iowa--Netherlands Comparison Orientation Measure
      & Tendency to compare one's accomplishments, opinions, circumstances,
        and social functioning with those of others. & 11 & 78 \\
    \rowcolor{suZebra}
    INQ      & Interpersonal Needs Questionnaire (modified)
      & Perceived belonging, support, and connection with friends and
        followers on social media. & 4 & 78 \\
    \bottomrule
  \end{tabularx}
\end{table*}
\subsection{Instagram direct messages}
\label{sec:dms}
As part of the social media data donation component, a subset of participants enrolled in the parent study were invited to contribute their social media data for research. Following informed consent and assent, social media data were collected through specialized secure research infrastructure for participant-mediated donation of digital platform data. The infrastructure's name is withheld for anonymous review. This analysis was restricted to participants who consented to social media data donation and had viable Instagram data.

The 80 participants in the analytic sample contributed 100 phase-specific Instagram data-export files containing their DM histories; 20 participants contributed files during two collection phases. Before deduplication, these files contained 2{,}510{,}640 message rows. We treated rows as duplicates when sender identifier, message content, timestamp, and conversation identifier were identical. This procedure removed 101{,}617 rows (4.0\% overall; maximum participant-level proportion, 44.8\%), leaving 2{,}409{,}023 messages, of which 2{,}077{,}677 were text messages. 

\subsubsection{Archive characteristics.}

Across the 80 archives, the corpus contained 7{,}774 conversation threads and 7{,}970 distinct conversation partners. Per participant, the median archive spanned 736 days and contained 55 threads with 48 distinct partners. Across the corpus, participants authored 762{,}110 of the 2{,}409{,}023 messages (31.6\%); the remaining messages were authored by their conversation partners. The corpus contained 331{,}346 non-text messages (13.8\%), including photos, reels, reactions, calls, and attachments. We retained their timestamps and type labels as indicators of interaction.

\subsubsection{Variation across participants.}

The timing of the DM archives relative to questionnaire completion varied across participants. Questionnaires were completed between October 25, 2023, and October 22, 2025, whereas archived messages dated from August 2, 2018, to March 31, 2026. Relative to each participant's questionnaire date, the archives began up to 2{,}269 days earlier and ended up to 779 days later. 

\subsubsection{Ethical and Privacy Safeguards.} 
\label{sec:deid}

The parent study protocol received Institutional Review Board approval for adolescent participants to complete questionnaires and donate social media data. Participants under 18 provided assent alongside parental or guardian
consent, while participants 18 and older provided informed consent. Consent materials described the social-media data being collected, how those data would be stored and protected, and the risks associated with participation.

While deidentification is a typical safeguard adopted in research dealing with private data~\cite{razi2022instagram}, the collection and analysis of private adolescent communication required protections beyond conventional deidentification because donated archives may contain sensitive information about participants as well as messages authored by conversation partners who did not directly participate in the study. Because private conversations necessarily contain contributions from people other than the donating participant, we treated minimizing exposure of both participants' and conversation partners' information as a central requirement of the data pipeline.

Social-media donation occurred through specialized secure research infrastructure designed for participant-mediated donation of digital platform data, following procedures developed in prior work~\cite{razi2022instagram,razi2023sliding,alsoubai2024profiling,ali2023getting}. 
Participants actively downloaded and provided their own platform archives, rather than researchers obtaining data directly from their accounts or platforms. Donated archives were transferred and stored using encrypted, access-controlled infrastructure. Automated de-identification removed or pseudonymized identifying information, including names, locations, account identifiers, URLs, and other identifying text, before research analysis. In the resulting data, the participant was labeled \texttt{SELF}, and conversation partners were assigned numbered pseudonymous labels that remained consistent throughout each archive, preserving conversational and relational structure while reducing
identifiability. Emoji were converted to text descriptions, and non-text messages were identified by type. Importantly, researchers did not manually inspect participants' donated data at any stage of the study, including message content, photographs, or videos; data processing and de-identification were performed computationally within the secure infrastructure. Researcher access followed role-based and least-privilege controls. In essence, we treated deidentification as a risk-reduction measure rather than as guaranteeing anonymity, given the intimate and relational nature of private social media data. Accordingly, the underlying message corpus is not publicly released, and all computational analyses reported here were conducted within the approved secure computing environment.

As noted earlier, all LLM processing reported in this study used locally deployed models within this secure computing environment; private message content was not transmitted to externally hosted model services for prediction or reasoning-trace generation. These safeguards allowed us to preserve the
relational and longitudinal structure necessary for the research questions while limiting additional exposure of participants' and conversation partners' private communication.


\section{RQ1: Predicting Questionnaire Responses from Instagram DMs}

To assess what LLMs can infer from participant-donated social media data, we examine model performance in predicting adolescents' self-reported questionnaire responses from Instagram DMs. 

We compare two approaches to using participants’ DM archives for questionnaire prediction. The \textbf{Note} approach draws on chain-of-thought prompting~\cite{wei2022chain} to guide the model in identifying evidence in DMs and using it to make predictions. This approach thus compares predictions based directly on DM evidence with predictions based on structured understanding (SU) derived from the DMs (\S~\ref{rq1_note}).
The second approach, called the \textbf{RAG} approach, uses retrieval-augmented generation~\cite{lewis2020retrieval} to retrieve evidence from a searchable vector database of the DM archive. Here we compare retrieval using the original questionnaire wording with retrieval using model-generated probes (\S~\ref{rq1_rag}). 
Within each approach, we evaluate whether refining how evidence is organized or retrieved improves prediction performance (\S~\ref{sec:rq1-method}). These analyses address our first research question. 

\begin{figure*}[!t]
    \centering
    \includegraphics[width=2.1\columnwidth]{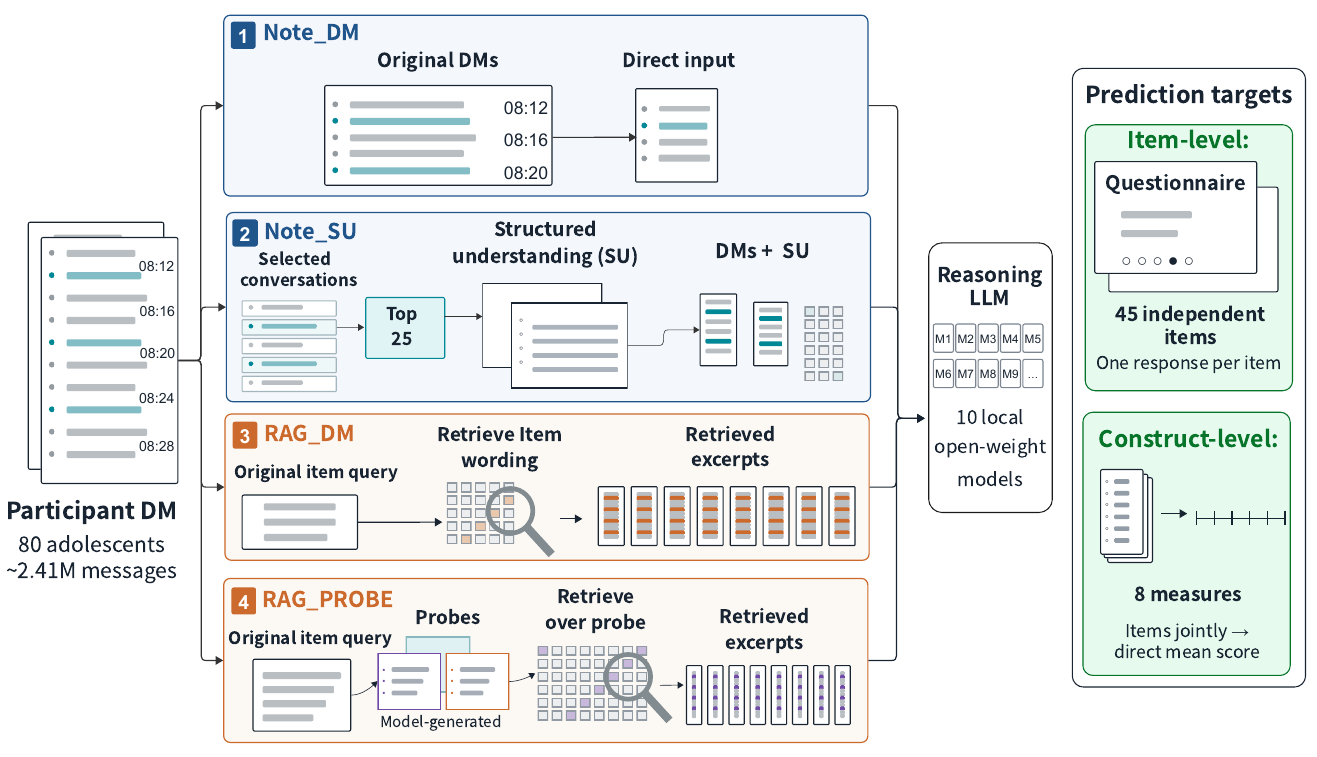}
    \caption{The eight analysis workflows. Four evidence-construction
    conditions were crossed with two prediction targets. Within the Note
    approach, \texttt{Note\_DM} is the baseline. Within the RAG approach,
    \texttt{RAG\_DM} is the baseline.}
  \label{fig:rq1-analysis-workflows}
\end{figure*}

Each of the four evidence-construction conditions was evaluated for both item-level and construct-level prediction, yielding eight analysis workflows (Figure~\ref{fig:rq1-analysis-workflows}). All workflows used the same participant-specific DM data, filtered to the questionnaires' 30-day reference period before any further processing. All participants had DM data within this period.

\subsection{Method}
\label{sec:rq1-method}

\subsubsection{Prediction Targets.}

We evaluate prediction at two levels: (1) \textbf{item-level} prediction of responses to 45 scored questionnaire items and (2) \textbf{construct-level} prediction of eight scale scores, each calculated as the mean of the corresponding measure's scored items. These two prediction targets use different inference settings. For {\it item-level prediction}, each questionnaire item is processed independently in a separate model call, with no context shared across items. The model selects one response option for that item. For {\it construct-level prediction}, all scored items within a questionnaire measure are presented together in a single model call, and the model directly predicts the participant's mean score for the measure. Unlike item-level prediction, this setting allows the model to consider the items jointly when forming its prediction.

\begin{table*}[t]
\caption{The ten open-weight reasoning models, grouped by family and ordered by parameter count within each family. Context windows were set per model to accommodate that model's longest prompt under its own tokenizer.}
  \Description{A four-column table listing ten open-weight reasoning models by
    model name, family, parameter count, and architecture (mixture-of-experts
    or dense). Models are grouped by family with larger models first. The
    header row is shaded light blue and rows alternate white and very light
    gray.}
  \label{tab:models}
  \centering
  \small
  \setlength{\tabcolsep}{8pt}
  \renewcommand{\arraystretch}{1.25}
  \setlength{\aboverulesep}{0pt}
  \setlength{\belowrulesep}{0pt}
  \begin{tabular}{llcc}
    \toprule
    \rowcolor{suHead}
    \textbf{Model} & \textbf{Family} & \textbf{Parameters} & \textbf{Architecture} \\
    \midrule
    \texttt{gpt-oss-120b}                   & gpt-oss     & 120B & MoE   \\
    \rowcolor{suZebra}
    \texttt{gpt-oss-20b}                    & gpt-oss     & 20B  & MoE   \\
    \texttt{GLM-4.5-Air}                    & GLM         & 106B & MoE   \\
    \rowcolor{suZebra}
    \texttt{DeepSeek-R1-Distill-Llama-70B}  & DeepSeek-R1 & 70B  & Dense \\
    \texttt{DeepSeek-R1-Distill-Qwen-32B}   & DeepSeek-R1 & 32B  & Dense \\
    \rowcolor{suZebra}
    \texttt{DeepSeek-R1-Distill-Qwen-14B}   & DeepSeek-R1 & 14B  & Dense \\
    \texttt{DeepSeek-R1-0528-Qwen3-8B}      & DeepSeek-R1 & 8B   & Dense \\
    \rowcolor{suZebra}
    \texttt{Qwen3.6-35B-A3B}                & Qwen        & 35B  & MoE   \\
    \texttt{Qwen3.6-27B}                    & Qwen        & 27B  & Dense \\
    \rowcolor{suZebra}
    \texttt{Magistral-Small-2509}           & Mistral     & 24B  & Dense \\
    \bottomrule
  \end{tabular}
\end{table*}

\subsubsection{Model Selection.}
Models had to support local deployment to keep adolescent DMs within the secure infrastructure (\S~\ref{sec:data}) and expose reasoning traces during inference for qualitative analysis. We then selected ten open-weight reasoning models that varied in family (Table~\ref{tab:models}), scale, and architecture~\cite{openai2025gptoss, glm2025glm45, deepseek2025r1,deepseek2025r1qwen3,qwen2026qwen36_35b,qwen2026qwen36_27b,mistral2025magistral,mistral2025magistralsmall2509}.

\subsubsection{Note Approach}
\label{rq1_note}
Recall that the Note approach compares predictions made directly from DM evidence with predictions made from a structured representation of that evidence.

\textbf{\texttt{Note\_DM}.} In this baseline condition, each participant's raw DM messages from the 30-day observation window are supplied to the model in successive batches. Predictions are then elicited without first generating structured notes or a participant-level summary. This condition provides a reference for evaluating the conversation selection, structured understanding (SU) pipeline used in \texttt{Note\_SU}.

\textbf{\texttt{Note\_SU}.}
This approach first selects conversations that may contain informative evidence. Inspired by~\citet{park2023generative} and~\citet{zhong2024memorybank}, we adopt a ranking scheme to select conversations for model input. We ranked conversations using normalized message volume, recency, and the participant's share of messages, weighted 0.4, 0.3, and 0.3, respectively, and retained the top 25. The model then converts each selected conversation into a structured note containing checkable observations, such as relevant quotations, counts, and dates. These notes are subsequently consolidated into a fixed schema that serves as a shared model input across questionnaire predictions for that participant. Table~\ref{tab:su-schema} describes all SU fields, and Appendix~\ref{app:calculations} provides the formal definitions of conversation ranking and SU consolidation.

\begin{table*}[t]
  \caption{The fixed schema used for each participant's structured
    understanding (SU). \emph{Signal} groups the fields by the kind of
    evidence they capture, \emph{Field} is the schema key, and
    \emph{Evidence recorded} describes what is stored under that key. Every
    field records evidence together with its provenance (who said it, to
    whom, and when) where available.}
  \Description{A three-column table listing the SU schema. The first column
    names one of five signal groups: relationship context, interaction
    dynamics, support and conflict, identity and belonging, and wellbeing
    and coverage. The second column gives the schema field name in blue
    monospace and the third describes the evidence recorded. The header row
    is shaded light blue and signal groups alternate white and very light
    blue-gray.}
  \label{tab:su-schema}
  \small
  \setlength{\tabcolsep}{6pt}
  \renewcommand{\arraystretch}{1.2}
  \setlength{\aboverulesep}{0pt}
  \setlength{\belowrulesep}{0pt}
  \definecolor{suBlock}{HTML}{F1F5F9}
  \definecolor{suKey}{HTML}{2F5C8F}
  \renewcommand{\tabularxcolumn}[1]{>{\raggedright\arraybackslash}m{#1}}
  \begin{tabularx}{\textwidth}{>{\raggedright\arraybackslash}m{0.13\textwidth}
                               >{\raggedright\arraybackslash}m{0.27\textwidth}
                               X}
    \toprule
    \rowcolor{suHead}
    \textbf{Signal} & \textbf{Field} & \textbf{Evidence recorded} \\
    \midrule
    Relationship context
      & \textcolor{suKey}{\texttt{partner\_type\_guess}}
      & Likely relationship to the participant (e.g., close friend, romantic
        partner, family member, classmate, or acquaintance). \\
      & \textcolor{suKey}{\texttt{relationship\_summary}}
      & Brief factual summary of the relationship and the reason supporting
        the partner-type guess. \\
    \rowcolor{suBlock}
    Interaction dynamics
      & \textcolor{suKey}{\texttt{who\_talks\_more}}
      & Who initiates conversations, the balance of participation, and counts
        of extended stretches in which \texttt{SELF} received no reply. \\
    \rowcolor{suBlock}
      & \textcolor{suKey}{\texttt{talk\_about\_being\_on\_the\_phone}}
      & Statements about phone or social-media use, such as being unable to
        stop scrolling or remaining on the phone for long periods. \\
    \rowcolor{suBlock}
      & \textcolor{suKey}{\texttt{night\_time\_and\_tiredness}}
      & Message-based evidence of late-night activity and tiredness. \\
    Support and conflict
      & \textcolor{suKey}{\texttt{help\_and\_comfort}}
      & Comforting, advice, and check-ins, including their direction, specific
        instances, and supporting quotations of at most 15 words. \\
      & \textcolor{suKey}{\texttt{friction\_events}}
      & Arguments, hostile remarks, exclusion, rumors, jealousy, blocking, and
        breakups, with date, initiator, and supporting quotation when
        available. \\
    \rowcolor{suBlock}
    Identity and belonging
      & \textcolor{suKey}{\texttt{remarks\_about\_who\_someone\_is}}
      & Remarks about appearance, origin, accent, identity, or religion,
        together with their target and tone (e.g., hostile, mocking, or
        in-group joking). \\
    \rowcolor{suBlock}
      & \textcolor{suKey}{\texttt{measuring\_self\_against\_others}}
      & Explicit comparisons between the participant and other people. \\
    \rowcolor{suBlock}
      & \textcolor{suKey}{\texttt{feeling\_apart\_or\_together}}
      & Expressions of belonging, connection, separation, or exclusion. \\
    Wellbeing and coverage
      & \textcolor{suKey}{\texttt{hard\_things\_shared}}
      & Disclosures of difficult experiences, such as mood problems,
        self-harm thoughts, or family trouble, including who confided in whom
        and supporting quotations. \\
      & \textcolor{suKey}{\texttt{notable\_events}}
      & Other noteworthy events not captured by the fields above. \\
      & \textcolor{suKey}{\texttt{evidence\_sufficiency}}
      & Explicit assessment of whether the available conversation evidence is
        sufficient to support a conclusion. \\
    \bottomrule
  \end{tabularx}
\end{table*}
\subsubsection{RAG Approach}
\label{rq1_rag}
The RAG approach compares two ways of formulating retrieval queries while holding the indexing and retrieval procedure fixed. Both conditions used the same retrieval infrastructure. We segmented each participant's DM archive within conversation sessions into approximately 1,300-character chunks and embedded them using \texttt{Qwen3-Embedding-4B}~\cite{zhang2025qwen3}. Retrieval combined dense similarity with BM25 lexical matching using reciprocal-rank fusion with a rank constant of $60$~\cite{cormack2009reciprocal}. We retained at most three chunks from any single conversation session and up to 24 chunks for each questionnaire measure.

\textbf{\texttt{RAG\_DM}}. In this approach, the original wording of each scored questionnaire item served as the retrieval query. For item-level prediction, the model received evidence retrieved for that item. For construct-level prediction, evidence retrieved across all scored items in the measure was combined before prediction.

\textbf{\texttt{RAG\_Probe}.} 
Inspired by the retrieval approaches of~\citet{ma2023query} and~\citet{rackauckas2024rag}, we designed a set of conversational retrieval probes intended to capture how the corresponding behaviors or experiences might be expressed in everyday DMs. To generate these probes, the model interpreted the item, inspected candidate messages retrieved from the corpus, and identified common ways in which the relevant behavior or experience could be expressed in conversation. The resulting probes remained general rather than describing specific participant events. For example, an item about difficulty sleeping might produce templates such as \texttt{\{someone\} went to bed but was still awake at \{time\}} or \texttt{\{someone\} can't sleep because they keep \{doing something\}}. These probes were used only as retrieval queries, broadening retrieval beyond the original questionnaire wording and increasing the likelihood of retrieving relevant evidence from the DM archive~\cite{ma2023query, rackauckas2024rag}.

The probes were generated without access to participants' observed questionnaire responses. All other indexing, retrieval, and evidence-selection procedures were identical between \texttt{RAG\_DM} and \texttt{RAG\_Probe}. Appendix~\ref{app:calculations} provides the formal retrieval definitions.

\subsection{Evaluation}
\label{sec:rq1-evaluation}

We assessed prediction performance separately at the item and construct levels. Our primary metrics follow the evaluation framework of~\citet{ravenda2025llms}, with additional construct-level measures used to characterize agreement, association, and score dispersion. We repeated each prediction five times using the same sampling settings and used the median of valid outputs as the final prediction to reduce the influence of variability in model responses and occasional extreme values.

At the item level, we used Average Hit Rate (AHR) and Average Closeness Rate (ACR). AHR measures exact agreement between the predicted and observed response, whereas ACR gives partial credit according to their ordinal distance, normalized by the response range of each item. We evaluated all 45 scored items and report questionnaire-level summaries of item-level AHR and ACR. Appendix~\ref{app:metrics} provides the formal metric definitions.

At the construct level, our primary metrics were Average Difference between Overall Levels (ADODL) and Category Hit Rate (CHR). ADODL measures normalized absolute agreement between predicted and observed questionnaire scores, with a value of one indicating an exact match. CHR measures whether the predicted and observed scores fall within the same one of four equal-width bands of the questionnaire's possible score range. Because these metrics are defined on questionnaire totals, predicted mean item scores were converted to their equivalent totals for these calculations.

To characterize how predictions vary across participants, we additionally report the Concordance Correlation Coefficient (CCC) for absolute agreement, Pearson's $r$ and Spearman's $\rho$ for linear and rank association, and the ratio of predicted to observed score SDs for differences in dispersion. 

\subsection{Results}
\label{sec:rq1-results}

Both evidence-construction refinements (\texttt{Note\_SU} and \texttt{RAG\_Probe}) improved prediction across most measures compared with their respective baselines (\texttt{Note\_DM} and \texttt{RAG\_DM}), with probe-based retrieval yielding the most consistent item-level gains (\S~\ref{sec:rq1-results-1}). Second, prediction performance varied across questionnaire measures at both the item and construct levels (\S~\ref{sec:rq1-results-2}). However, predictions close to observed scale scores did not necessarily preserve differences between participants (\S~\ref{sec:rq1-results-3}). 

\subsubsection{Evidence Construction Improved Prediction, Particularly Through Probe-Based Retrieval}
\label{sec:rq1-results-1}

Both evidence-construction refinements improved item-level prediction, but the gains were broader under probe-based retrieval (Table~\ref{tab:rq1-item-results}). Relative to \texttt{RAG\_DM}, \texttt{RAG\_Probe} increased exact-response accuracy for all eight measures; relative to \texttt{Note\_DM}, \texttt{Note\_SU} increased exact-response accuracy for seven. Across measures, mean AHR increased by \textcolor{suUp}{$\uparrow 3.2$} under \texttt{RAG\_Probe} and \textcolor{suUp}{$\uparrow 3.6$} under \texttt{Note\_SU}. Mean ACR increased by \textcolor{suUp}{$\uparrow 1.2$} and \textcolor{suUp}{$\uparrow 1.4$}, respectively. Thus, conversational probes yielded the more consistent improvement across measures over their corresponding reference condition, while structured notes and SU yielded larger mean gains.

\begin{table*}[t]
\caption{Item-level prediction within each evidence-construction approach. The Note rows compare \texttt{Note\_DM} with \texttt{Note\_SU}, and the RAG rows compare \texttt{RAG\_DM} with \texttt{RAG\_Probe}. Reference and comparison values are means across the ten-model evaluation panel with 95\% participant-bootstrap confidence intervals. Differences are comparison minus reference in percentage points. Cell shading scales with the value within each metric. Green and red differences mark increases and decreases.}
  \Description{A table reporting construct-specific AHR and ACR for the
    reference and comparison conditions in the Note and RAG approaches.
    Comparisons are made within each approach. Mean cells are shaded in a
    light blue tone in proportion to their value; difference cells are
    printed in green for increases and red for decreases.}
  \label{tab:rq1-item-results}
  \centering
  \small
  \setlength{\tabcolsep}{5pt}
  \renewcommand{\arraystretch}{1.2}
  \setlength{\aboverulesep}{0pt}
  \setlength{\belowrulesep}{0pt}
  \definecolor{suWarm}{HTML}{6C93BF}

  \begin{adjustbox}{max width=\textwidth}
  \begin{tabular}{llcccccc}
    \toprule
    \rowcolor{suHead}
    \textbf{Approach} &
    \textbf{Construct} &
    \textbf{AHR reference} &
    \textbf{AHR comparison} &
    $\boldsymbol{\Delta}$\textbf{ AHR} &
    \textbf{ACR reference} &
    \textbf{ACR comparison} &
    $\boldsymbol{\Delta}$\textbf{ ACR} \\
    \midrule

    Note & SN-PEQ
    & \cellcolor{suWarm!18}58.8 {\footnotesize\color{black!60}[54.7, 62.9]}
    & \cellcolor{suWarm!23}72.2 {\footnotesize\color{black!60}[68.0, 76.1]}
    & \textcolor{green!45!black}{+13.4}
    & \cellcolor{suWarm!23}86.1 {\footnotesize\color{black!60}[84.3, 87.9]}
    & \cellcolor{suWarm!27}90.1 {\footnotesize\color{black!60}[88.2, 91.8]}
    & \textcolor{green!45!black}{+4.0} \\

    Note & MH-Comm
    & \cellcolor{suWarm!10}39.5 {\footnotesize\color{black!60}[34.2, 44.8]}
    & \cellcolor{suWarm!11}42.9 {\footnotesize\color{black!60}[37.6, 48.4]}
    & \textcolor{green!45!black}{+3.4}
    & \cellcolor{suWarm!14}78.3 {\footnotesize\color{black!60}[75.5, 80.8]}
    & \cellcolor{suWarm!15}78.8 {\footnotesize\color{black!60}[75.6, 81.7]}
    & \textcolor{green!45!black}{+0.5} \\

    Note & E-MASS
    & \cellcolor{suWarm!5}27.5 {\footnotesize\color{black!60}[22.9, 32.4]}
    & \cellcolor{suWarm!5}26.7 {\footnotesize\color{black!60}[21.9, 31.7]}
    & \textcolor{red!70!black}{-0.8}
    & \cellcolor{suWarm!5}70.2 {\footnotesize\color{black!60}[66.4, 73.6]}
    & \cellcolor{suWarm!4}69.3 {\footnotesize\color{black!60}[65.6, 72.8]}
    & \textcolor{red!70!black}{-0.9} \\

    Note & Discrim
    & \cellcolor{suWarm!23}72.9 {\footnotesize\color{black!60}[68.0, 77.9]}
    & \cellcolor{suWarm!26}80.5 {\footnotesize\color{black!60}[76.8, 84.0]}
    & \textcolor{green!45!black}{+7.6}
    & \cellcolor{suWarm!8}73.2 {\footnotesize\color{black!60}[68.3, 78.1]}
    & \cellcolor{suWarm!16}80.6 {\footnotesize\color{black!60}[76.9, 84.1]}
    & \textcolor{green!45!black}{+7.4} \\

    Note & PRIUSS
    & \cellcolor{suWarm!8}33.9 {\footnotesize\color{black!60}[27.7, 40.5]}
    & \cellcolor{suWarm!8}35.0 {\footnotesize\color{black!60}[27.9, 42.4]}
    & \textcolor{green!45!black}{+1.1}
    & \cellcolor{suWarm!6}71.4 {\footnotesize\color{black!60}[66.9, 75.6]}
    & \cellcolor{suWarm!6}71.4 {\footnotesize\color{black!60}[66.5, 76.1]}
    & \textcolor{black!55}{0.0} \\

    Note & SleepDis
    & \cellcolor{suWarm!3}22.3 {\footnotesize\color{black!60}[17.6, 26.9]}
    & \cellcolor{suWarm!4}24.0 {\footnotesize\color{black!60}[19.0, 29.2]}
    & \textcolor{green!45!black}{+1.7}
    & \cellcolor{suWarm!1}66.5 {\footnotesize\color{black!60}[62.9, 70.0]}
    & \cellcolor{suWarm!0}65.7 {\footnotesize\color{black!60}[61.6, 69.7]}
    & \textcolor{red!70!black}{-0.8} \\

    Note & INCOM
    & \cellcolor{suWarm!2}19.5 {\footnotesize\color{black!60}[17.6, 21.6]}
    & \cellcolor{suWarm!3}20.8 {\footnotesize\color{black!60}[18.8, 22.9]}
    & \textcolor{green!45!black}{+1.3}
    & \cellcolor{suWarm!3}68.0 {\footnotesize\color{black!60}[65.9, 69.9]}
    & \cellcolor{suWarm!3}68.2 {\footnotesize\color{black!60}[66.1, 70.0]}
    & \textcolor{green!45!black}{+0.2} \\

    Note & INQ
    & \cellcolor{suWarm!2}18.0 {\footnotesize\color{black!60}[14.8, 21.3]}
    & \cellcolor{suWarm!2}18.7 {\footnotesize\color{black!60}[15.5, 22.0]}
    & \textcolor{green!45!black}{+0.7}
    & \cellcolor{suWarm!8}73.1 {\footnotesize\color{black!60}[70.2, 75.9]}
    & \cellcolor{suWarm!9}73.6 {\footnotesize\color{black!60}[70.7, 76.4]}
    & \textcolor{green!45!black}{+0.5} \\

    \midrule

    RAG & SN-PEQ
    & \cellcolor{suWarm!19}62.1 {\footnotesize\color{black!60}[58.1, 65.9]}
    & \cellcolor{suWarm!24}74.9 {\footnotesize\color{black!60}[70.7, 78.6]}
    & \textcolor{green!45!black}{+12.8}
    & \cellcolor{suWarm!24}87.4 {\footnotesize\color{black!60}[85.7, 89.0]}
    & \cellcolor{suWarm!28}91.0 {\footnotesize\color{black!60}[89.2, 92.7]}
    & \textcolor{green!45!black}{+3.6} \\

    RAG & MH-Comm
    & \cellcolor{suWarm!8}34.4 {\footnotesize\color{black!60}[29.1, 39.4]}
    & \cellcolor{suWarm!10}38.0 {\footnotesize\color{black!60}[33.5, 42.6]}
    & \textcolor{green!45!black}{+3.6}
    & \cellcolor{suWarm!10}74.3 {\footnotesize\color{black!60}[71.4, 77.0]}
    & \cellcolor{suWarm!13}77.2 {\footnotesize\color{black!60}[74.7, 79.6]}
    & \textcolor{green!45!black}{+2.9} \\

    RAG & E-MASS
    & \cellcolor{suWarm!5}27.0 {\footnotesize\color{black!60}[22.6, 31.5]}
    & \cellcolor{suWarm!6}28.2 {\footnotesize\color{black!60}[24.1, 32.4]}
    & \textcolor{green!45!black}{+1.2}
    & \cellcolor{suWarm!4}69.7 {\footnotesize\color{black!60}[66.3, 72.9]}
    & \cellcolor{suWarm!4}69.6 {\footnotesize\color{black!60}[66.1, 72.9]}
    & \textcolor{red!70!black}{-0.1} \\

    RAG & Discrim
    & \cellcolor{suWarm!27}82.7 {\footnotesize\color{black!60}[78.2, 86.9]}
    & \cellcolor{suWarm!28}84.6 {\footnotesize\color{black!60}[80.5, 88.3]}
    & \textcolor{green!45!black}{+1.9}
    & \cellcolor{suWarm!19}82.8 {\footnotesize\color{black!60}[78.3, 87.0]}
    & \cellcolor{suWarm!21}84.6 {\footnotesize\color{black!60}[80.6, 88.4]}
    & \textcolor{green!45!black}{+1.8} \\

    RAG & PRIUSS
    & \cellcolor{suWarm!7}31.9 {\footnotesize\color{black!60}[26.6, 37.6]}
    & \cellcolor{suWarm!8}33.6 {\footnotesize\color{black!60}[27.6, 40.2]}
    & \textcolor{green!45!black}{+1.7}
    & \cellcolor{suWarm!6}71.2 {\footnotesize\color{black!60}[67.4, 75.0]}
    & \cellcolor{suWarm!7}71.6 {\footnotesize\color{black!60}[67.3, 75.8]}
    & \textcolor{green!45!black}{+0.4} \\

    RAG & SleepDis
    & \cellcolor{suWarm!4}23.3 {\footnotesize\color{black!60}[18.4, 28.5]}
    & \cellcolor{suWarm!4}24.4 {\footnotesize\color{black!60}[19.8, 29.1]}
    & \textcolor{green!45!black}{+1.1}
    & \cellcolor{suWarm!1}66.9 {\footnotesize\color{black!60}[63.5, 70.1]}
    & \cellcolor{suWarm!1}66.5 {\footnotesize\color{black!60}[63.1, 70.2]}
    & \textcolor{red!70!black}{-0.4} \\

    RAG & INCOM
    & \cellcolor{suWarm!3}21.1 {\footnotesize\color{black!60}[19.0, 23.2]}
    & \cellcolor{suWarm!3}22.0 {\footnotesize\color{black!60}[19.9, 24.0]}
    & \textcolor{green!45!black}{+0.9}
    & \cellcolor{suWarm!5}70.2 {\footnotesize\color{black!60}[68.3, 71.9]}
    & \cellcolor{suWarm!5}70.1 {\footnotesize\color{black!60}[68.3, 71.8]}
    & \textcolor{red!70!black}{-0.1} \\

    RAG & INQ
    & \cellcolor{suWarm!0}13.9 {\footnotesize\color{black!60}[11.2, 16.5]}
    & \cellcolor{suWarm!1}15.9 {\footnotesize\color{black!60}[13.0, 18.8]}
    & \textcolor{green!45!black}{+2.0}
    & \cellcolor{suWarm!5}70.0 {\footnotesize\color{black!60}[66.9, 72.8]}
    & \cellcolor{suWarm!6}71.3 {\footnotesize\color{black!60}[68.4, 73.9]}
    & \textcolor{green!45!black}{+1.3} \\

    \bottomrule
  \end{tabular}
  \end{adjustbox}
\end{table*}

The largest item-level gains also favored measures with identifiable events in the message history. Under \texttt{RAG\_Probe}, AHR increased most for SN-PEQ (\textcolor{suUp}{$\uparrow 12.8$}) and MH-Comm (\textcolor{suUp}{$\uparrow 3.6$}); their corresponding ACR changes were \textcolor{suUp}{$\uparrow 3.6$} and \textcolor{suUp}{$\uparrow 2.9$}. Under \texttt{Note\_SU}, the largest AHR increases were observed for SN-PEQ (\textcolor{suUp}{$\uparrow 13.4$}) and Discrim (\textcolor{suUp}{$\uparrow 7.6$}), with corresponding ACR increases of \textcolor{suUp}{$\uparrow 4.0$} and \textcolor{suUp}{$\uparrow 7.4$}. ACR changes were generally smaller in percentage-point magnitude than AHR changes.

\subsubsection{Prediction Performance Varied Across Questionnaire Measures}
\label{sec:rq1-results-2}
Across the four evidence-construction conditions, exact-response accuracy was highest for Discrim (72.9\%--84.6\%) and SN-PEQ (58.8\%--74.9\%), whereas AHR did not exceed 22.0\% for INQ and INCOM. ACR was generally higher than AHR, showing that many incorrect predictions were nevertheless close to the observed response on the ordinal scale. AHR therefore characterizes exact-response recovery within each measure's response scale, while evidence--construct alignment is examined separately in RQ2.

The same measure dependence appeared at the construct level (Table~\ref{tab:rq1-construct-accuracy}). SN-PEQ had the highest score closeness and same-band agreement across conditions, with ADODL ranging from 86.6\% to 91.6\% and CHR from 73.2\% to 90.4\%. E-MASS had the lowest ADODL and CHR in three of the four conditions, with ADODL ranging from 69.1\% to 72.7\% and CHR from 15.8\% to 29.2\%. The gap between metrics was also consequential: INQ attained ADODL values of 74.5\%--76.1\%, yet its CHR remained between 24.0\% and 31.9\%. A scale-score prediction can therefore be numerically close without placing the participant in the same score band.

\begin{table*}[t]
  \caption{Construct-level prediction accuracy within each evidence-construction approach.
    The Note rows compare \texttt{Note\_DM} with \texttt{Note\_SU}, and the
    RAG rows compare \texttt{RAG\_DM} with \texttt{RAG\_Probe}. Reference
    and comparison values are means across the ten-model evaluation panel with
    95\% participant-bootstrap confidence intervals. Differences are comparison
    minus reference in percentage points. Green and red differences mark increases and decreases.}
  \Description{A table reporting construct-specific ADODL and CHR for the
    reference and comparison conditions in the Note and RAG approaches.
    Comparisons are made within each approach. Mean cells are shaded light
    blue in proportion to their value; difference cells are printed in green
    for increases and red for decreases.}
  \label{tab:rq1-construct-accuracy}
  \centering
  \small
  \setlength{\tabcolsep}{5pt}
  \renewcommand{\arraystretch}{1.2}
  \setlength{\aboverulesep}{0pt}
  \setlength{\belowrulesep}{0pt}
  \definecolor{suBlue}{HTML}{6C93BF}

  \begin{adjustbox}{max width=\textwidth}
  \begin{tabular}{llcccccc}
    \toprule
    \rowcolor{suHead}
    \textbf{Approach} &
    \textbf{Construct} &
    \textbf{ADODL reference} &
    \textbf{ADODL comparison} &
    $\boldsymbol{\Delta}$\textbf{ ADODL} &
    \textbf{CHR reference} &
    \textbf{CHR comparison} &
    $\boldsymbol{\Delta}$\textbf{ CHR} \\
    \midrule

    Note & SN-PEQ
    & \cellcolor{suBlue!22}86.6 {\footnotesize\color{black!60}[84.6, 88.6]}
    & \cellcolor{suBlue!27}90.6 {\footnotesize\color{black!60}[89.3, 91.8]}
    & \textcolor{green!45!black}{+4.0}
    & \cellcolor{suBlue!22}73.2 {\footnotesize\color{black!60}[64.5, 81.2]}
    & \cellcolor{suBlue!26}85.4 {\footnotesize\color{black!60}[79.6, 90.4]}
    & \textcolor{green!45!black}{+12.2} \\

    Note & MH-Comm
    & \cellcolor{suBlue!5}72.7 {\footnotesize\color{black!60}[68.3, 76.9]}
    & \cellcolor{suBlue!11}77.9 {\footnotesize\color{black!60}[74.9, 80.7]}
    & \textcolor{green!45!black}{+5.2}
    & \cellcolor{suBlue!5}30.4 {\footnotesize\color{black!60}[20.9, 40.5]}
    & \cellcolor{suBlue!8}36.1 {\footnotesize\color{black!60}[28.6, 43.6]}
    & \textcolor{green!45!black}{+5.7} \\

    Note & E-MASS
    & \cellcolor{suBlue!1}69.1 {\footnotesize\color{black!60}[65.3, 72.8]}
    & \cellcolor{suBlue!0}69.9 {\footnotesize\color{black!60}[65.9, 73.8]}
    & \textcolor{green!45!black}{+0.8}
    & \cellcolor{suBlue!0}15.8 {\footnotesize\color{black!60}[8.9, 23.3]}
    & \cellcolor{suBlue!1}19.3 {\footnotesize\color{black!60}[14.2, 24.7]}
    & \textcolor{green!45!black}{+3.5} \\

    Note & Discrim
    & \cellcolor{suBlue!11}77.5 {\footnotesize\color{black!60}[73.0, 81.8]}
    & \cellcolor{suBlue!17}82.2 {\footnotesize\color{black!60}[78.9, 85.3]}
    & \textcolor{green!45!black}{+4.7}
    & \cellcolor{suBlue!9}41.0 {\footnotesize\color{black!60}[31.2, 50.7]}
    & \cellcolor{suBlue!13}50.0 {\footnotesize\color{black!60}[42.4, 57.6]}
    & \textcolor{green!45!black}{+9.0} \\

    Note & PRIUSS
    & \cellcolor{suBlue!7}74.3 {\footnotesize\color{black!60}[70.4, 78.2]}
    & \cellcolor{suBlue!7}74.6 {\footnotesize\color{black!60}[70.6, 78.5]}
    & \textcolor{green!45!black}{+0.3}
    & \cellcolor{suBlue!5}28.4 {\footnotesize\color{black!60}[20.3, 37.2]}
    & \cellcolor{suBlue!6}33.1 {\footnotesize\color{black!60}[25.1, 41.4]}
    & \textcolor{green!45!black}{+4.7} \\

    Note & SleepDis
    & \cellcolor{suBlue!8}75.4 {\footnotesize\color{black!60}[71.4, 79.2]}
    & \cellcolor{suBlue!8}75.1 {\footnotesize\color{black!60}[71.6, 78.5]}
    & \textcolor{red!70!black}{-0.3}
    & \cellcolor{suBlue!8}36.5 {\footnotesize\color{black!60}[27.7, 45.3]}
    & \cellcolor{suBlue!6}31.9 {\footnotesize\color{black!60}[24.7, 39.5]}
    & \textcolor{red!70!black}{-4.6} \\

    Note & INCOM
    & \cellcolor{suBlue!21}85.4 {\footnotesize\color{black!60}[83.1, 87.4]}
    & \cellcolor{suBlue!17}82.8 {\footnotesize\color{black!60}[80.5, 84.9]}
    & \textcolor{red!70!black}{-2.6}
    & \cellcolor{suBlue!15}56.2 {\footnotesize\color{black!60}[47.9, 64.4]}
    & \cellcolor{suBlue!11}44.2 {\footnotesize\color{black!60}[37.8, 50.8]}
    & \textcolor{red!70!black}{-12.0} \\

    Note & INQ
    & \cellcolor{suBlue!7}74.5 {\footnotesize\color{black!60}[71.0, 77.8]}
    & \cellcolor{suBlue!9}75.6 {\footnotesize\color{black!60}[72.1, 78.8]}
    & \textcolor{green!45!black}{+1.1}
    & \cellcolor{suBlue!3}24.0 {\footnotesize\color{black!60}[15.8, 32.2]}
    & \cellcolor{suBlue!6}31.9 {\footnotesize\color{black!60}[24.0, 39.7]}
    & \textcolor{green!45!black}{+7.9} \\

    \midrule

    RAG & SN-PEQ
    & \cellcolor{suBlue!25}88.8 {\footnotesize\color{black!60}[87.2, 90.4]}
    & \cellcolor{suBlue!28}91.6 {\footnotesize\color{black!60}[90.2, 92.8]}
    & \textcolor{green!45!black}{+2.8}
    & \cellcolor{suBlue!26}83.8 {\footnotesize\color{black!60}[76.5, 90.4]}
    & \cellcolor{suBlue!28}90.4 {\footnotesize\color{black!60}[85.0, 95.0]}
    & \textcolor{green!45!black}{+6.6} \\

    RAG & MH-Comm
    & \cellcolor{suBlue!16}81.9 {\footnotesize\color{black!60}[78.7, 84.7]}
    & \cellcolor{suBlue!16}82.0 {\footnotesize\color{black!60}[78.7, 85.0]}
    & \textcolor{green!45!black}{+0.1}
    & \cellcolor{suBlue!11}44.5 {\footnotesize\color{black!60}[34.2, 54.8]}
    & \cellcolor{suBlue!13}50.7 {\footnotesize\color{black!60}[42.5, 58.8]}
    & \textcolor{green!45!black}{+6.2} \\

    RAG & E-MASS
    & \cellcolor{suBlue!5}72.7 {\footnotesize\color{black!60}[68.4, 76.8]}
    & \cellcolor{suBlue!1}69.4 {\footnotesize\color{black!60}[65.2, 73.4]}
    & \textcolor{red!70!black}{-3.3}
    & \cellcolor{suBlue!5}29.2 {\footnotesize\color{black!60}[20.1, 38.9]}
    & \cellcolor{suBlue!4}26.1 {\footnotesize\color{black!60}[19.6, 32.9]}
    & \textcolor{red!70!black}{-3.1} \\

    RAG & Discrim
    & \cellcolor{suBlue!20}85.1 {\footnotesize\color{black!60}[81.4, 88.6]}
    & \cellcolor{suBlue!23}87.2 {\footnotesize\color{black!60}[83.8, 90.3]}
    & \textcolor{green!45!black}{+2.1}
    & \cellcolor{suBlue!15}57.0 {\footnotesize\color{black!60}[46.5, 67.6]}
    & \cellcolor{suBlue!17}61.0 {\footnotesize\color{black!60}[52.1, 69.6]}
    & \textcolor{green!45!black}{+4.0} \\

    RAG & PRIUSS
    & \cellcolor{suBlue!6}73.3 {\footnotesize\color{black!60}[68.5, 77.8]}
    & \cellcolor{suBlue!7}74.0 {\footnotesize\color{black!60}[69.6, 78.2]}
    & \textcolor{green!45!black}{+0.7}
    & \cellcolor{suBlue!8}36.3 {\footnotesize\color{black!60}[26.0, 47.3]}
    & \cellcolor{suBlue!8}35.8 {\footnotesize\color{black!60}[26.0, 45.9]}
    & \textcolor{red!70!black}{-0.5} \\

    RAG & SleepDis
    & \cellcolor{suBlue!3}71.1 {\footnotesize\color{black!60}[67.0, 75.1]}
    & \cellcolor{suBlue!7}73.9 {\footnotesize\color{black!60}[70.2, 77.4]}
    & \textcolor{green!45!black}{+2.8}
    & \cellcolor{suBlue!4}27.4 {\footnotesize\color{black!60}[18.5, 36.3]}
    & \cellcolor{suBlue!5}29.5 {\footnotesize\color{black!60}[22.1, 37.3]}
    & \textcolor{green!45!black}{+2.1} \\

    RAG & INCOM
    & \cellcolor{suBlue!14}79.9 {\footnotesize\color{black!60}[77.5, 82.2]}
    & \cellcolor{suBlue!18}83.3 {\footnotesize\color{black!60}[80.6, 85.8]}
    & \textcolor{green!45!black}{+3.4}
    & \cellcolor{suBlue!9}41.0 {\footnotesize\color{black!60}[35.1, 46.8]}
    & \cellcolor{suBlue!13}51.4 {\footnotesize\color{black!60}[43.1, 59.7]}
    & \textcolor{green!45!black}{+10.4} \\

    RAG & INQ
    & \cellcolor{suBlue!9}75.8 {\footnotesize\color{black!60}[72.1, 79.2]}
    & \cellcolor{suBlue!9}76.1 {\footnotesize\color{black!60}[72.2, 79.7]}
    & \textcolor{green!45!black}{+0.3}
    & \cellcolor{suBlue!5}27.9 {\footnotesize\color{black!60}[21.2, 34.9]}
    & \cellcolor{suBlue!6}30.6 {\footnotesize\color{black!60}[21.5, 39.6]}
    & \textcolor{green!45!black}{+2.7} \\

    \bottomrule
  \end{tabular}
  \end{adjustbox}
\end{table*}

Both refinements also improved construct-level prediction across most measures, with distinct gains across ADODL and CHR. Under \texttt{Note\_SU}, ADODL and CHR each increased for six of the eight measures, averaging \textcolor{suUp}{$\uparrow 1.7$} and \textcolor{suUp}{$\uparrow 3.3$}, respectively. Under \texttt{RAG\_Probe}, ADODL increased for seven measures and CHR for six, averaging \textcolor{suUp}{$\uparrow 1.1$} and \textcolor{suUp}{$\uparrow 3.6$}. Among SN-PEQ, MH-Comm, and Discrim, \texttt{RAG\_Probe} produced higher comparison-condition ADODL and CHR values than \texttt{Note\_SU}.

The improvements were not uniform. E-MASS was the clearest counterexample under probe-based retrieval: relative to \texttt{RAG\_DM}, \texttt{RAG\_Probe} decreased ADODL by \textcolor{suDown}{$\downarrow 3.3$} and CHR by \textcolor{suDown}{$\downarrow 3.1$}. This shows that making retrieved evidence more targeted generally improved prediction, but did not ensure better performance for every measure. This exception becomes important when considering whether apparently relevant DM evidence actually represents the self-reported experience being predicted.

\subsubsection{Score Proximity Did Not Ensure Participant Differentiation}
\label{sec:rq1-results-3}
Score proximity did not reliably translate into the preservation of between-participant differences. Under both comparison conditions, positive concordance, linear association, and rank-order association were concentrated in three constructs: SN-PEQ, MH-Comm, and Discrim (Table~\ref{tab:rq1-construct-diagnostics}). \texttt{RAG\_Probe} showed the clearest participant differentiation for MH-Comm ($\mathrm{CCC}=.35$, $r=.38$, $\rho=.35$) and Discrim ($\mathrm{CCC}=.27$, $r=.31$, $\rho=.31$), and SN-PEQ ($\mathrm{CCC}=.20$, $r=.27$, $\rho=.25$). \texttt{Note\_SU} reproduced the same construct-specific pattern, yielding CCC estimates of .28 for MH-Comm and .17 for both SN-PEQ and Discrim. Outside these three constructs, participant-level correspondence was negligible or negative: across E-MASS, PRIUSS, SleepDis, INCOM, and INQ, CCC ranged from $-.12$ to .05, Pearson's $r$ from $-.23$ to .06, and Spearman's $\rho$ from $-.21$ to .07. Participant differentiation was therefore not uniformly absent, but concentrated in a small subset of constructs.

\begin{table*}[t]
\caption{Construct-level concordance, participant differentiation, ranking, and score variability. Each metric cell shows the reference estimate $\rightarrow$ the comparison estimate. The Note rows compare \texttt{Note\_DM} with \texttt{Note\_SU}, and the RAG rows compare \texttt{RAG\_DM} with \texttt{RAG\_Probe}.}
  \Description{A table reporting construct-specific CCC, Pearson correlation,
    Spearman correlation, and SD Ratio for Note and RAG.
    Arrows show change from the reference condition to the comparison condition
    within each approach. Comparison values for the concordance
    and correlation metrics are green for increases and red for decreases.}
  \label{tab:rq1-construct-diagnostics}
  \centering
  \small
  \setlength{\tabcolsep}{5pt}
  \renewcommand{\arraystretch}{1.1}
  \setlength{\aboverulesep}{0pt}
  \setlength{\belowrulesep}{0pt}

  \newcommand{\up}[2]{{\color{black!55}$#1$}\,{\color{black!40}$\rightarrow$}\,\textcolor{green!45!black}{$#2$}}
  \newcommand{\dn}[2]{{\color{black!55}$#1$}\,{\color{black!40}$\rightarrow$}\,\textcolor{red!70!black}{$#2$}}
  \newcommand{\eq}[2]{{\color{black!55}$#1$}\,{\color{black!40}$\rightarrow$}\,\textcolor{black!55}{$#2$}}
  \newcommand{\nt}[2]{{\color{black!55}$#1$}\,{\color{black!40}$\rightarrow$}\,$#2$}

  \begin{adjustbox}{max width=\textwidth}
  \begin{tabular}{llcccc}
    \toprule
    \rowcolor{suHead}
    \textbf{Approach} &
    \textbf{Construct} &
    \textbf{CCC} &
    \textbf{Pearson $r$} &
    \textbf{Spearman $\rho$} &
    \textbf{SD Ratio} \\
    \midrule

    Note & SN-PEQ   & \up{.138}{.17}    & \up{.165}{.23}   & \up{.140}{.21}   & \nt{1.146}{1.02} \\
    \rowcolor{suZebra}
    Note & MH-Comm  & \up{.133}{.28}    & \up{.135}{.30}   & \up{.093}{.28}   & \nt{.822}{.82} \\
    Note & E-MASS   & \up{-.054}{-.04}  & \dn{-.060}{-.10} & \dn{-.071}{-.12} & \nt{.898}{.62} \\
    \rowcolor{suZebra}
    Note & Discrim  & \up{.107}{.17}    & \up{.108}{.20}   & \up{.141}{.19}   & \nt{1.027}{.78} \\
    Note & PRIUSS   & \dn{-.011}{-.06}  & \dn{-.041}{-.10} & \dn{.020}{-.10}  & \nt{.308}{.43} \\
    \rowcolor{suZebra}
    Note & SleepDis & \dn{.058}{.05}    & \dn{.114}{.06}   & \dn{.120}{.02}   & \nt{.714}{.63} \\
    Note & INCOM    & \dn{.038}{-.07}   & \dn{.078}{-.13}  & \dn{.135}{-.12}  & \nt{.704}{.38} \\
    \rowcolor{suZebra}
    Note & INQ      & \dn{-.038}{-.06}  & \dn{-.048}{-.12} & \dn{-.051}{-.08} & \nt{.536}{.37} \\

    \midrule

    RAG & SN-PEQ   & \up{.064}{.20}    & \up{.089}{.27}   & \up{.086}{.25}   & \nt{1.154}{1.10} \\
    \rowcolor{suZebra}
    RAG & MH-Comm  & \up{.122}{.35}    & \up{.175}{.38}   & \up{.158}{.35}   & \nt{1.229}{.92} \\
    RAG & E-MASS   & \up{-.139}{-.08}  & \dn{-.159}{-.22} & \up{-.174}{-.16} & \nt{.937}{.58} \\
    \rowcolor{suZebra}
    RAG & Discrim  & \up{.112}{.27}   & \up{.120}{.31}  & \up{.007}{.31}  & \nt{1.287}{.95} \\
    RAG & PRIUSS   & \dn{-.078}{-.12}  & \dn{-.126}{-.23} & \dn{-.141}{-.21} & \nt{.620}{.55} \\
    \rowcolor{suZebra}
    RAG & SleepDis & \up{.008}{.04}    & \up{.040}{.05}   & \up{-.015}{.01}  & \nt{.863}{.74} \\
    RAG & INCOM    & \dn{.086}{-.06}   & \dn{.112}{-.14}  & \dn{.096}{-.12}  & \nt{.687}{.45} \\
    \rowcolor{suZebra}
    RAG & INQ      & \up{-.001}{.04}   & \up{-.001}{.05}  & \up{-.014}{.07}  & \nt{.568}{.50} \\

    \bottomrule
  \end{tabular}
  \end{adjustbox}
\end{table*}

Reduced predicted-score variability accompanied the weak correspondence observed elsewhere. Fourteen of the sixteen comparison-condition SD ratios were below one, indicating that the models generally produced narrower score distributions than those observed in the participant data. Compression was most pronounced for INQ and INCOM: their SD ratios were .37 and .38 under \texttt{Note\_SU}, and .50 and .45 under \texttt{RAG\_Probe}. INCOM provides a particularly clear example of why score proximity alone was insufficient. Under \texttt{RAG\_Probe}, it attained an ADODL of 83.3\%, yet its predicted scores retained less than half of the observed variability and were negatively associated with participants' observed scores ($r=-.14$). Thus, a model could generate scores that were close in magnitude while still compressing differences among participants and failing to preserve their relative standing.

The converse pattern appeared where \texttt{RAG\_Probe} differentiated participants most successfully. Its predicted-score variability closely approached the observed variability for Discrim (SD ratio $=.95$) and MH-Comm (SD ratio $=.92$), which were also its two strongest constructs on both Pearson and Spearman association. Taken together, these results separate two forms of performance that aggregate proximity measures can conflate: approximating observed score levels and preserving meaningful differences between participants.

Summarily, the RQ1 results show that improved evidence construction helped models recover questionnaire responses, with probe-based retrieval yielding the most consistent item-level gains. The strongest evidence of successful inference was concentrated in SN-PEQ, MH-Comm, and Discrim, whose target events or communication were directly observable in DMs. For measures involving more subjective appraisals, high scale-score proximity often coexisted with weak participant differentiation and compressed predictions. RQ2 therefore examines what models treated as evidence and how they mapped observable interactions to participants' self-reported experiences.

\section{RQ2: Evidence Use in Model-Generated Reasoning Traces}
\label{sec:rq2}

Prediction accuracy alone, however, does not reveal what evidence models relied on or how they interpreted that evidence. Messages directly document communication about mental health or negative peer experiences when these events occur in the conversation, providing construct-aligned evidence for these targets. In contrast, observable behaviors such as late-night messaging, frequent online activity, or supportive language serve as indirect proxies for participants’ reported sleep-related distress, loss of control, or perceived social support. Prediction performance therefore depends not only on what information is available in the messages, but also on how models interpret its relevance to participants’ self-reported experiences. In RQ2, we therefore examined the evidence cited in model-generated reasoning traces and how models mapped that evidence onto each target construct.

\begin{figure*}[!t]
    \centering
    \includegraphics[width=2.0\columnwidth]{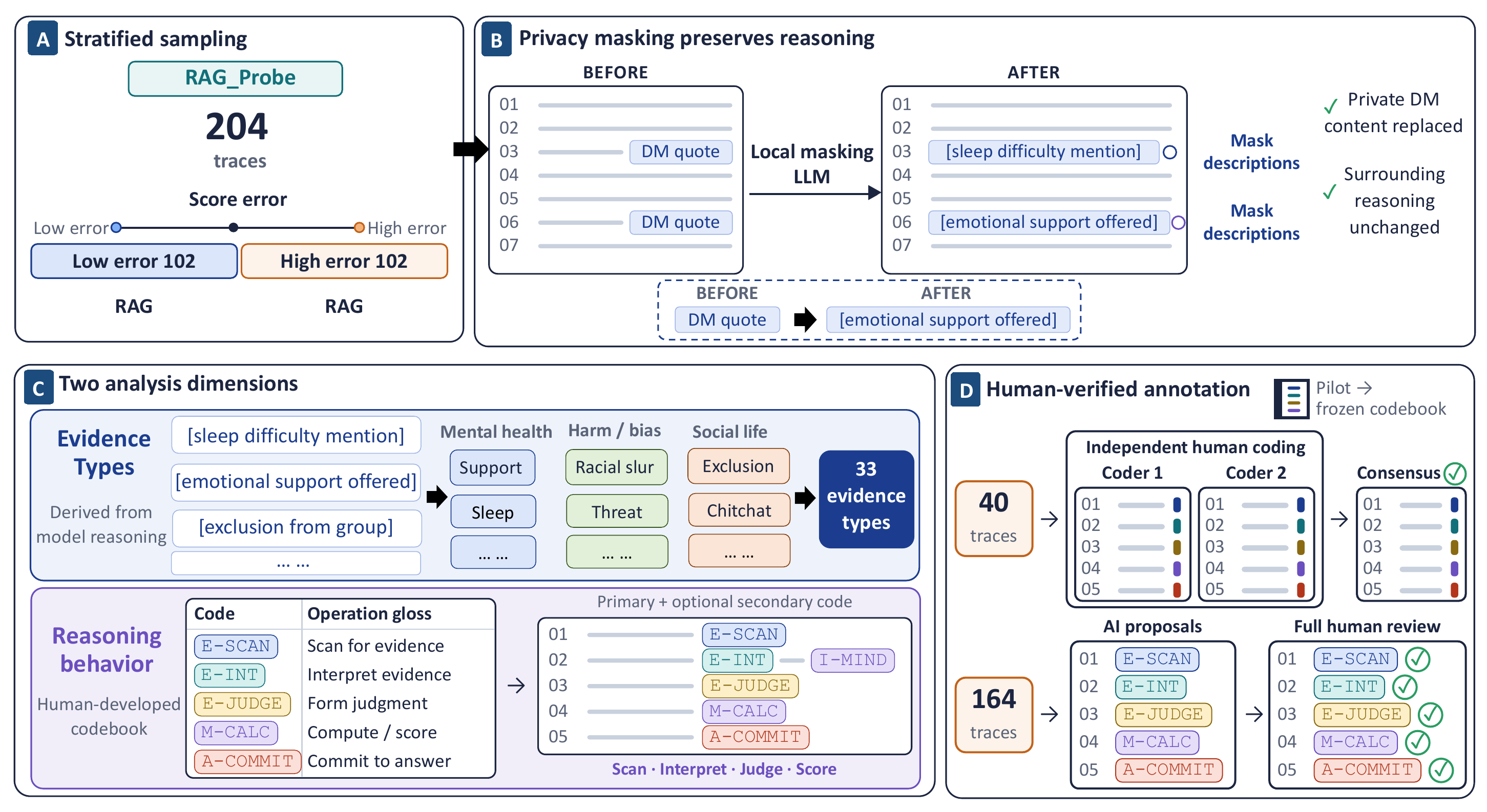}
    \caption{
    Overview of the RQ2 analysis workflow. 
    (A) We constructed a stratified sample of 204 model-generated reasoning traces, balanced across lower- and higher-error predictions. 
    (B) To protect private message content while retaining the context needed for analysis, DM excerpts in the traces were replaced with short semantic descriptions while the surrounding model reasoning was preserved. 
    (C) We analyzed each trace along two dimensions: the conversational evidence invoked by the model, organized into 33 evidence types, and the reasoning behaviors used to interpret that evidence and produce a prediction. 
    (D) We developed and validated the annotation procedure through independent human coding and consensus on a 40-trace set, then used AI-generated coding proposals followed by full human review for the remaining 164 traces.
    }
  \label{fig:rq2-analysis-workflows}
\end{figure*}

To answer this question, we conducted a systematic content analysis of model-generated reasoning traces. We analyzed both the DM evidence cited by the models and the interpretive patterns through which they connected that evidence to their predictions (Figure~\ref{fig:rq2-analysis-workflows} summarizes the analysis workflow).

\subsection{Methods}
\label{sec:rq2-method}

\subsubsection{Corpus and Trace Sampling}

We analyzed construct-level model-generated reasoning traces from \texttt{RAG\_Probe} -- the best performing workflow in RQ1. We focused on this condition because retrieval searched across each participant's complete eligible 30-day DM history, allowing us to examine how models interpreted the retrieved evidence. The sampling frame was restricted to participant-construct predictions for which an observed construct score was available, because classification as lower or higher error required comparison with the participant's observed response. We focused on construct-level traces because the construct was the theoretical and comparative unit of RQ2. Each construct-level call presented all scored items from a measure together, allowing us to examine how models integrated evidence across the construct's facets and translated those judgments into a scale-level prediction.

We used a conventional $\chi^2$ power calculation as a planning benchmark for the size of the qualitative corpus~\cite{cohen2013statistical}. With eight construct categories ($df=7$), $\alpha=.05$, a medium effect size (Cohen's $w=.30$), and .90 power, the calculation yielded a target of 204 traces. This calculation was used to set the corpus size rather than as the basis for a null-hypothesis significance test in the qualitative analysis. We ranked construct-level predictions by error and used stratified sampling to select 102 lower-error and 102 higher-error traces. The sample covered all ten models and eight constructs.

\subsubsection{Trace Preparation and Annotation Framework}

We processed each trace at two levels. First, a locally deployed masking LLM replaced private-DM quotations and model-generated paraphrases with bracketed semantic descriptions while preserving the surrounding reasoning text. We reviewed these model-generated descriptions and consolidated synonymous descriptions into 33 evidence types; for example, \texttt{[Self-harm message]} and \texttt{[Self-harm language]} were standardized as \texttt{[Self-harm mention]}. 
Table~\ref{tab:rq2-evidence-types} lists the 33 evidence types we used. 

\begin{table*}[t]
  \caption{The 33 evidence types consolidated from LLM-generated
    privacy-masking descriptions, eleven in each of three domains. These
    descriptions replaced private DM content that the prediction models
    quoted or paraphrased in their model-generated reasoning traces; they
    were not researcher-authored paraphrases of the original messages.}
  \Description{A three-column table listing eleven evidence types in each of
    three domains: Mental Health and Affect; Harm, Bias, and Harassment; and
    Everyday Social Life. Each row holds one evidence type per domain. The
    header row is shaded light blue and rows alternate white and very light
    gray.}
  \label{tab:rq2-evidence-types}
  \small
  \setlength{\tabcolsep}{8pt}
  \renewcommand{\arraystretch}{1.2}
  \setlength{\aboverulesep}{0pt}
  \setlength{\belowrulesep}{0pt}
  \renewcommand{\tabularxcolumn}[1]{>{\raggedright\arraybackslash}m{#1}}
  \begin{tabularx}{\textwidth}{XXX}
    \toprule
    \rowcolor{suHead}
    \textbf{Mental Health \& Affect} & \textbf{Harm, Bias \& Harassment} & \textbf{Everyday Social Life} \\
    \midrule
    Suicidal ideation               & Racial slur                        & Everyday chitchat            \\
    \rowcolor{suZebra}
    Self-harm mention               & Race-related remark                & Casual greeting              \\
    Expression of sadness           & Sexual-orientation-related remark  & Expression of affection      \\
    \rowcolor{suZebra}
    Anxiety or stress mention       & Gender-related remark              & Plan or invitation           \\
    Sleep difficulty mention        & Appearance-related insult          & Argument or conflict         \\
    \rowcolor{suZebra}
    Mental health disclosure        & Insult or name-calling             & Rumor or gossip              \\
    Loneliness or belonging concern & Threat or intimidation             & Social exclusion mention     \\
    \rowcolor{suZebra}
    Emotional support offered       & Sexual harassment message          & Social comparison remark     \\
    Emotional support requested     & Sexual content                     & Family conflict mention      \\
    \rowcolor{suZebra}
    Substance use mention           & Unwanted contact                   & School or academic mention   \\
    Other personal disclosure       & Violence or abuse disclosure       & Social media usage mention   \\
    \bottomrule
  \end{tabularx}
\end{table*}

Second, we segmented each model-generated reasoning trace into individual sentences and treated each sentence as the unit of reasoning-trace annotation. Each sentence received one primary code for its main function and an optional secondary code when it explicitly performed an additional coded operation. The codebook follows the progression from task framing through evidence handling to answer production (Table~\ref{tab:rq2-reasoning-codebook}).

We used explicit boundary rules to distinguish finding evidence, interpreting it, and deciding whether it supported a response. \texttt{E-SCAN}, \texttt{E-INT}, and \texttt{E-JUDGE} were mutually exclusive as primary codes and required a concrete connection to supplied evidence. For example, reporting timestamps after midnight was coded as \texttt{E-SCAN}; inferring that late-night messaging was common was coded as \texttt{E-INT} with \texttt{I-BEHAV}; inferring that the participant felt sleep-related distress was coded as \texttt{E-INT} with \texttt{I-MIND}. Codes were assigned independently to each sentence, and final answers were separated from the judgments that preceded them.

\begin{table*}[t]
  \caption{Codebook for sentence-level reasoning-trace annotation. Code color
    indicates the level at which
    a code may be assigned: \textcolor{cbPrim}{\texttt{blue}} codes are
    primary only, \textcolor{cbBoth}{\texttt{green}} codes may be primary or
    secondary, and \textcolor{cbSec}{\texttt{orange}} codes are secondary
    only. Secondary codes were assigned only when the additional operation was explicit in the same
    sentence.}
  \Description{A two-column table listing each reasoning-trace annotation code and
    its operational definition. Codes are printed in monospace and colored
    by level: blue for primary-only codes, green for codes that may be
    primary or secondary, and orange for secondary-only codes. The header
    row is shaded light blue and rows alternate white and very light gray.}
  \label{tab:rq2-reasoning-codebook}
  \small
  \setlength{\tabcolsep}{6pt}
  \renewcommand{\arraystretch}{1.15}
  \setlength{\aboverulesep}{0pt}
  \setlength{\belowrulesep}{0pt}
  \definecolor{cbPrim}{HTML}{2F5C8F}   
  \definecolor{cbBoth}{HTML}{2E7D5B}   
  \definecolor{cbSec}{HTML}{B5651D}    
  \renewcommand{\tabularxcolumn}[1]{>{\raggedright\arraybackslash}m{#1}}
  \begin{tabularx}{\textwidth}{>{\raggedright\arraybackslash}m{0.11\textwidth} X}
    \toprule
    \rowcolor{suHead}
    \textbf{Code} & \textbf{Operational definition} \\
    \midrule
    \textcolor{cbPrim}{\texttt{T-SET}}    & Opens the trace by re-explaining the task or listing the questionnaire items. \\
    \rowcolor{suZebra}
    \textcolor{cbPrim}{\texttt{T-SET-M}}  & Opens a later reasoning step without adding substantive evidence, interpretation, or judgment. \\
    \textcolor{cbBoth}{\texttt{T-RUBRIC}} & Interprets task instructions, scoring rules, construct boundaries, evidence requirements, time windows, or output requirements. \\
    \rowcolor{suZebra}
    \textcolor{cbBoth}{\texttt{E-SCAN}}   & Locates, reads, quotes, or lists concrete supplied evidence without interpreting it. \\
    \textcolor{cbBoth}{\texttt{E-INT}}    & Explains, summarizes, classifies, compares, contextualizes, or infers meaning from concrete supplied evidence. \\
    \rowcolor{suZebra}
    \textcolor{cbBoth}{\texttt{E-JUDGE}}  & Decides whether evidence counts, is direct or indirect, or supports a particular questionnaire item or response. \\
    \textcolor{cbBoth}{\texttt{M-CALC}}   & Assigns or transforms scores, probabilities, frequencies, or other numerical quantities. \\
    \rowcolor{suZebra}
    \textcolor{cbBoth}{\texttt{U-HEDGE}}  & Expresses unresolved doubt, active uncertainty, or competing alternatives. \\
    \textcolor{cbBoth}{\texttt{C-CHECK}}  & Returns to, verifies, corrects, retracts, or reopens an earlier reasoning path. \\
    \rowcolor{suZebra}
    \textcolor{cbPrim}{\texttt{A-CONSOL}} & Consolidates or restates a current result before the final answer. \\
    \textcolor{cbPrim}{\texttt{A-COMMIT}} & Commits to the final fixed answer, including final structured output. \\
    \rowcolor{suZebra}
    \textcolor{cbSec}{\texttt{I-BEHAV}}   & Infers participant behavior beyond what the supplied evidence directly establishes. \\
    \textcolor{cbSec}{\texttt{I-MIND}}    & Infers the participant's internal feeling, appraisal, or likely self-report beyond what the supplied evidence directly establishes. \\
    \rowcolor{suZebra}
    \textcolor{cbSec}{\texttt{I-TASK}}    & Uses external or general knowledge about the task or construct to perform the primary operation. \\
    \textcolor{cbSec}{\texttt{C-CONF}}    & Reassures itself that a conclusion is correct without performing a substantive verification. \\
    \bottomrule
  \end{tabularx}
\end{table*}

\subsubsection{Reasoning-Trace Annotation Procedure and Reliability}

Two annotators independently reviewed pilot traces and met four times to develop the reasoning-trace annotation codebook. They compared proposed categories, merged overlapping codes, clarified inclusion and exclusion criteria, and added boundary examples before finalizing the codebook for reliability assessment. The annotators then independently coded a 40-trace reliability subset comprising 6,658 sentence-level units. Before adjudication, Cohen's $\kappa$ was $.865$ for primary codes, $.753$ for secondary codes, and $.826$ for the combined coding. The annotators subsequently reviewed every disagreement and assigned a consensus code.

Following established human--AI annotation workflows~\cite{mirzakhmedova2024large, ding2025multi, chen2024large}, we used GPT-5.6 Sol to generate initial sentence-level annotations for the remaining 164 traces~\cite{openai2026chatgpt}. Based on the configuration used in previous studies~\cite{ding2025multi, leung2026annotategpt}, the model received a three-shot prompt constructed from consensus-coded examples and assigned a primary code and, when applicable, a secondary code to each sentence. A human annotator then checked every model-generated annotation against the finalized codebook and corrected it when necessary. Agreement between the initial model-generated annotations and the final human-verified annotations was Cohen's $\kappa=.91$, which we report as workflow concordance.

\subsubsection{Analysis Strategy}

We used qualitative comparative content analysis supported by descriptive counts. Evidence-type instances characterized the conversational content cited by models, individual sentences served as the units for reasoning-trace annotation, and complete traces represented prediction pathways. We calculated the frequencies of the 33 evidence types overall and separately by construct; these frequencies describe evidence cited by the models, not the prevalence of participants' experiences. We summarized reasoning behaviors using sentence-level code distributions, reported separately for lower-error traces, higher-error traces, and the full sampled corpus.

We compared low-error and high-error traces and conducted close readings of contrasting traces to examine how models connected evidence to constructs, inferred experiences beyond the evidence, treated missing evidence, and converted judgments into scores.

\subsection{Results}
 
Models were most successful for MH-Comm, SN-PEQ, and Discrim, whose target behaviors or events were represented in the DMs. The most informative failures occurred when models found plausible cues but treated observable communication as evidence of a participant's subjective experience. E-MASS, PRIUSS, and SleepDis illustrate this gap particularly clearly. Appendix~\ref{app:rq2-sample-composition} reports reasoning-code distributions.

\subsubsection{What Models Treated as Evidence Across Constructs}

The models drew on a recurring vocabulary of 33 evidence types rather than an unrestricted collection of idiosyncratic message details. Across the 11,590 evidence-type occurrences, the twelve most frequent types accounted for 74.5\% of all evidence instances. Everyday chitchat was the most frequent type at 15.02\%, followed by expressions of affection at 11.50\% and emotional support offered at 9.35\%. The remaining frequently used evidence included appearance-related insults, sleep difficulty, suicidal ideation, arguments, identity-related remarks, social media use, and social exclusion. Thus, the evidence used for questionnaire prediction was grounded not only in overt mental-health language, but also in the ordinary social texture of participants' conversations.

\begin{figure*}[t]
  \centering
  \includegraphics[width=\textwidth]{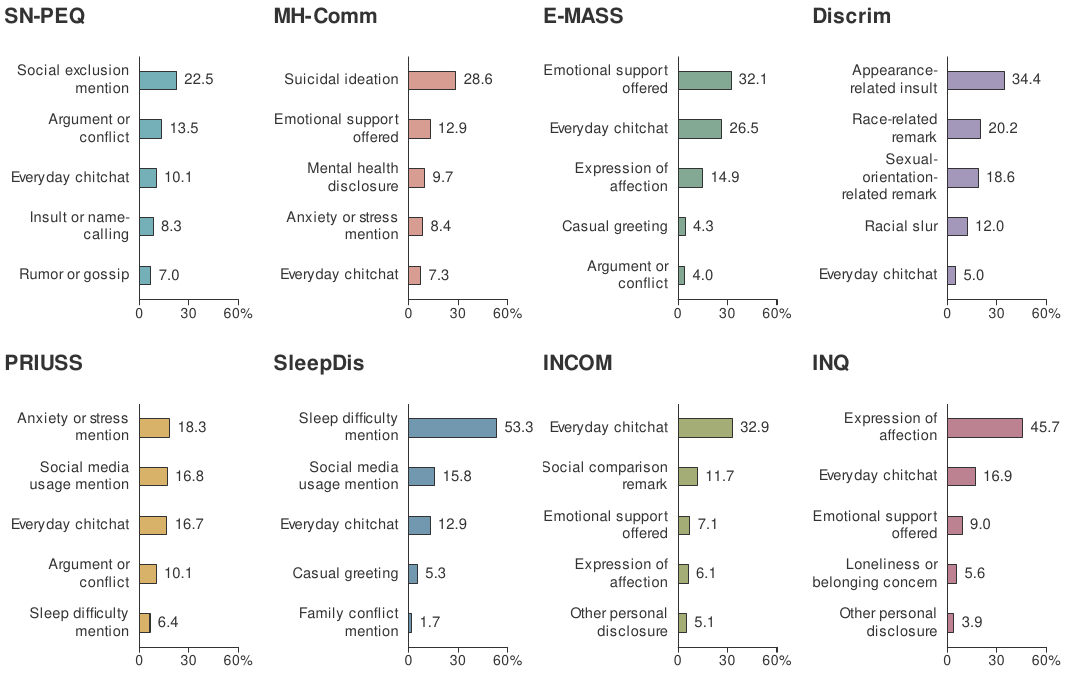}
    \caption{The five most frequent evidence types within each construct. Percentages are calculated among all evidence-type occurrences within the corresponding construct. Bars are ordered by frequency and use a common linear percentage scale; labels at the bar ends report the corresponding percentages. These distributions describe the evidence types that appeared in the reasoning traces.}
    \Description{Eight horizontal bar charts arranged in two rows and four columns, one for each construct: SN-PEQ, MH-Comm, E-MASS, Discrim, PRIUSS, SleepDis, INCOM, and INQ. Each chart shows the five most frequent evidence types in descending order, with evidence types on the vertical axis and within-construct percentages on a shared horizontal scale from 0 to 60 percent. Each bar is labelled with its percentage.}
  \label{fig:rq2-leading-evidence}
\end{figure*}

Evidence use was strongly construct-specific (Figure~\ref{fig:rq2-leading-evidence}). Sleep difficulty mentions constituted 53.3\% of the evidence used for SleepDis, while expressions of affection constituted 45.7\% of the evidence used for INQ. Emotional support offered was the leading evidence type for E-MASS at 32.1\%, and everyday chitchat led INCOM at 32.9\%. The more predictable constructs were associated with evidence types that more directly resembled their questionnaire content. The five most frequent evidence types accounted for 90.4\% of Discrim evidence, 67.0\% of MH-Comm evidence, and 61.4\% of SN-PEQ evidence. For these constructs, the dominant evidence consisted of discriminatory remarks, mental-health disclosures, exclusion, conflict, insults, and rumors.

These concentrations had two different meanings. For MH-Comm, SN-PEQ, and Discrim, the most frequent evidence types often identified events that the questionnaire itself asked about. For E-MASS, INQ, INCOM, PRIUSS, and SleepDis, the most frequent evidence types were frequently related to the construct without establishing the participant's self-reported experience. Evidence-type frequency therefore described what models repeatedly noticed, not whether the noticed content was a valid measure of the construct.

\subsubsection{When Visible Evidence Aligned with the Target Construct}

MH-Comm provided the clearest alignment between an observable behavior and the target construct. The questionnaire asks how often participants communicated about their own or their friends' mental-health difficulties, suicidal thoughts, and self-harm. These acts of communication can occur directly in DMs. Among the 1,706 MH-Comm evidence-type occurrences, suicidal ideation accounted for 28.6\%, emotional support offered for 12.9\%, mental-health disclosure for 9.7\%, and anxiety or stress mentions for 8.4\%. In a low-error trace, the model separated direct communication about the participant's own experience from the absence of comparable evidence about friends.

\begin{reasoningtrace}{``The participant directly mentions suicidal thoughts. This is direct evidence for communication about their own mental-health challenges. I do not see specific communication about a friend's mental-health difficulties.''} \textnormal{(gpt-oss-120b, MH-Comm, low-error trace, paraphrased)}\end{reasoningtrace}

High-error traces revealed that direct evidence still required correct attribution and frequency mapping. One high-error trace retrieved several suicidal-ideation references and messages in which the participant responded to friends' disclosures, then pooled these different directions into a high communication score. It predicted 3.5 for a participant whose observed construct score was 0. The model had found highly relevant content, but it did not preserve whose difficulty was being discussed or convert message instances into the questionnaire's frequency scale appropriately. 

SN-PEQ showed a similar advantage because blocking, exclusion, conflict, insults, rumors, and embarrassing posts appeared as identifiable events in the conversations. Social exclusion was its leading evidence type at 22.5\%; other evidence types included argument or conflict at 13.5\%, insult or name-calling at 8.3\%, and rumor or gossip at 7.0\%. Accurate traces did more than locate negative language. They checked whether an event matched a particular questionnaire item and whether it involved the participant.

\begin{reasoningtrace}{``The blocking exchange corresponds to the blocking item, and the public post corresponds to the embarrassing-post item. The identity-based insult was directed at someone else, so it should not count for this participant.''} \textnormal{(DeepSeek-R1-Distill-Qwen-32B, SN-PEQ, low-error trace, paraphrased)}\end{reasoningtrace}

High-error SN-PEQ traces often found the same broad categories but blurred the boundary between witnessing a negative interaction and experiencing it. One trace treated discussion of another person's blocking as indirect evidence for the participant and struggled to decide whether an insult in a group conversation satisfied an item about a privately received mean message. The relevant event was visible, but its direction and setting remained consequential.

Discrim showed the strongest concentration of construct-specific evidence. Appearance-related insults, race-related remarks, sexual-orientation-related remarks, and racial slurs together accounted for 85.3\% of its evidence-type occurrences. Low-error traces explicitly separated the presence of an identity-related remark from evidence that the participant was its target.

\begin{reasoningtrace}{``The race-related and weight remarks concern the participant directly, so they count for those items. The sexuality-related remark concerns another person, and the participant does not describe being targeted, so it should not count.''} \textnormal{(Magistral-Small-2509, Discrim, low-error trace, paraphrased)}\end{reasoningtrace}

In contrast, high-error traces sometimes treated any identity-related language in the retrieved messages as evidence of discrimination against the participant. Retrieval surfaced construct-specific evidence for Discrim, while accurate inference additionally required grounding the identity and target of each remark. Across these constructs, the common advantage was not simply greater evidence volume. The target events could be directly observed, and accurate predictions emerged when the model preserved the actor, target, and questionnaire item while mapping those events to a score.

\subsubsection{When Observable Evidence Was Mistaken for Subjective Experience}

E-MASS made the epistemic boundary between visible interaction and subjective experience especially clear. Emotional support offered was its leading evidence type at 32.1\%, making support one of the most visible signals in the entire annotated corpus. 
In one high-error trace, the model identified supportive replies, listening, and practical help, classified these messages as direct evidence, and assumed that direct evidence implied a high self-report score. It predicted 3.8 for a participant whose observed score was 0.

\begin{reasoningtrace}{``Friends responded supportively, listened to the participant's concerns, and offered practical help. These are direct instances of support. If an item has direct evidence, the participant should answer higher.''} \textnormal{(DeepSeek-R1-Distill-Llama-70B, E-MASS, high-error trace, paraphrased)}\end{reasoningtrace}

The opposite error also occurred. A high-error RAG trace retrieved several supportive exchanges but rejected many of them because the participant offered support, the friend requested support, or the interaction appeared casual. It then predicted 0 for a participant whose observed score was 3.2. Together, these cases show that message content could establish that supportive communication occurred without determining how supported the participant felt. More supportive communication could also reflect a greater need for help rather than the resolution of that need. This recurring inference helps explain the negative E-MASS concordance observed in the RQ1 results.

SleepDis showed that evidence concentration and construct validity are distinct. Sleep difficulty mentions accounted for 53.3\% of its evidence-type occurrences, the highest concentration observed for any construct. Yet the models also relied on hourly message histograms, late-night message shares, and last-message times. These features captured messaging behavior, whereas the second SleepDis item measured whether social media interfered with sleep. In one high-error trace, the model recognized this mismatch but still translated the behavioral proxy into a low score.

\begin{reasoningtrace}{``The median last-message time is 5 PM and the 90th percentile is 8 PM, so the participant probably does not use social media in the hour before sleep. There is no explicit statement connecting social media to sleep interference, so the second answer should be zero.''} \textnormal{(Qwen3.6-35B-A3B, SleepDis, high-error trace, paraphrased)}\end{reasoningtrace}


PRIUSS demonstrated the same epistemic boundary for problematic internet use. Message volume, active hours, phone-related comments, anxiety, and staying awake were observable. PRIUSS instead measured perceived social anxiety caused by social media, withdrawal when not using it, and avoidance of other activities to remain online. One high-error trace repeatedly searched for an explicit link, found none, and converted that absence into a score of 0 for a participant whose observed score was 3.33.

\begin{reasoningtrace}{``There is no direct evidence of anxiety caused by social media, withdrawal when not using it, or avoiding activities to remain online. The tiredness is not tied to social media. With no evidence, all three item scores should be zero.''} \textnormal{(DeepSeek-R1-Distill-Qwen-14B, PRIUSS, high-error trace, paraphrased)}\end{reasoningtrace}

Other PRIUSS traces moved in the opposite direction by treating frequent messaging or late activity as evidence of dependence. Both paths substituted observable use for the participant's appraisal of control. Observable use intensity therefore did not indicate whether participants experienced their internet use as compulsive or disruptive.

INQ and INCOM extended this subjectivity gap to ordinary social interaction. Expressions of affection constituted 45.7\% of INQ evidence, while everyday chitchat constituted 32.9\% of INCOM evidence. In a high-error INQ trace, affectionate emojis, reciprocal conversation, and shared activities were interpreted as direct evidence of belonging and closeness. The model assigned the maximum score of 7 to a participant whose observed score was 2.

\begin{reasoningtrace}{``The participant uses affectionate language and has many reciprocal conversations, which shows that they are socially engaged and close to their friends. All four belonging items therefore have direct support.''} \textnormal{(gpt-oss-20b, INQ, high-error trace, paraphrased)}\end{reasoningtrace}

For INCOM, the reverse pattern occurred when the absence of explicit comparison language was interpreted as evidence that the participant did not compare themselves with others. These cases establish the same boundary from two directions. Frequent or affectionate interaction does not rule out loneliness, and the absence of self-comparison language does not establish that comparison is absent. Across E-MASS, SleepDis, PRIUSS, INQ, and INCOM, the critical error occurred when observable communication was converted into an unobserved appraisal, feeling, or causal experience.

\subsubsection{From Retrieved Evidence to Questionnaire Judgments}
Across constructs, model-generated reasoning traces under \texttt{RAG\_Probe} showed recurring reasoning behaviors: interpreting questionnaire requirements (\textcolor{cbBoth}{\texttt{T-RUBRIC}}), locating evidence (\textcolor{cbBoth}{\texttt{E-SCAN}}), interpreting its meaning (\textcolor{cbBoth}{\texttt{E-INT}}), judging whether it supported an item (\textcolor{cbBoth}{\texttt{E-JUDGE}}), and assigning scores (\textcolor{cbBoth}{\texttt{M-CALC}}) before committing to an answer (\textcolor{cbPrim}{\texttt{A-COMMIT}}). These behaviors did not follow a fixed sequence. Models sometimes assigned tentative scores before completing their evidence assessment, expressed uncertainty (\textcolor{cbBoth}{\texttt{U-HEDGE}}), or returned to earlier interpretations (\textcolor{cbBoth}{\texttt{C-CHECK}}). Across constructs, models needed to establish different things: whose experience an event represented, how an interaction related to a self-report, or whether a behavioral cue established the connection required by an item (Table~\ref{tab:rq2-inferential-questions}).

\begin{table*}[t]
  \caption{Three reasoning questions identified through qualitative analysis of model-generated reasoning traces under \texttt{RAG\_Probe}. The table groups constructs by the questions models addressed and the reasoning behaviors used to answer them. Individual traces could address more than one question.}
  \Description{A three-column table organizing eight constructs by the questions models addressed and their reasoning behaviors: attributing events and matching items; relating interactions to self-reports; and establishing the connection between behavior and the experience specified by an item. The header row is shaded light blue and rows alternate white and very light gray.}
  \label{tab:rq2-inferential-questions}
  \centering
  \small
  \setlength{\tabcolsep}{6pt}
  \renewcommand{\arraystretch}{1.2}
  \setlength{\aboverulesep}{0pt}
  \setlength{\belowrulesep}{0pt}
  \renewcommand{\tabularxcolumn}[1]{>{\raggedright\arraybackslash}m{#1}}
  \begin{tabularx}{\textwidth}{>{\raggedright\arraybackslash}m{0.27\textwidth}
                               >{\raggedright\arraybackslash}m{0.16\textwidth}
                               X}
    \toprule
    \rowcolor{suHead}
    \textbf{Reasoning question} & \textbf{Constructs} & \textbf{Reasoning behaviors} \\
    \midrule
    Whose experience does the event represent, and does it match the item? &
    Discrim, SN-PEQ, MH-Comm &
    Identifying who was involved, distinguishing communication about SELF from communication about others, matching events to items, and estimating frequency. \\
    \rowcolor{suZebra}
    How does the interaction relate to the participant's self-report? &
    E-MASS, INQ, INCOM &
    Distinguishing support given from support received, interpreting relationships, and assessing whether specific comparisons support a broader comparison tendency. \\
    Does the behavior establish the connection required by the item? &
    SleepDis, PRIUSS &
    Checking whether messaging times indicate use before sleep and whether observed activity supports the specific social-media-related experience named in the item. \\
    \bottomrule
  \end{tabularx}
\end{table*}

For Discrim, SN-PEQ, and MH-Comm, models checked who was involved in an event or conversation and whether it matched the questionnaire item. In the low-error examples, models did not stop at locating identity-related language, a blocking exchange, or a mental-health disclosure through \textcolor{cbBoth}{\texttt{E-SCAN}}. They interpreted who was involved through \textcolor{cbBoth}{\texttt{E-INT}} and assessed whether that involvement matched the item through \textcolor{cbBoth}{\texttt{E-JUDGE}}. The Discrim example excluded a remark concerning another person; the SN-PEQ example distinguished events matching different questionnaire items; and the MH-Comm example separated communication about the participant's own difficulties from communication about friends. In the contrasting high-error examples, models located relevant content but attributed others' experiences to the participant or pooled exchanges that needed to be assessed separately. Accurate judgments depended on preserving who was affected and whose experience was discussed when deciding whether the evidence supported an item.

For E-MASS, INQ, and INCOM, models interpreted specific interactions to estimate how participants would describe their social experiences. What the evidence needed to establish varied across items and measures. E-MASS traces needed to distinguish support given, received, and exchanged reciprocally, and then match the interaction to the particular support item. A message offering help could show that support was offered without establishing every other aspect of support. In INQ traces, models went beyond describing interactions: affectionate messages and reciprocal exchanges were interpreted as evidence of the participant's own belonging or closeness. The secondary code \textcolor{cbSec}{\texttt{I-MIND}} marks sentences in which models infer an internal feeling or likely self-report beyond what the supplied interaction establishes. For INCOM, models needed to distinguish specific comparisons from a general tendency: an explicit comparison could inform a particular item, but a local exchange did not establish a general comparison tendency, and silence about comparison did not establish disagreement. Across these examples, models used their interpretations of communication to predict what the participant would report. An \textcolor{cbSec}{\texttt{I-MIND}} annotation identifies this reasoning behavior; whether the inference was supported depended on the evidence and the requirements of the item.

SleepDis and PRIUSS required models to establish a connection between observable behavior and the experience specified by the questionnaire. For SleepDis, \textcolor{cbBoth}{\texttt{E-SCAN}} could locate message timestamps, but interpreting those timestamps as use before sleep required information about bedtime. Inferring perceived sleep interference required a further connection between social-media use and the participant's experience. PRIUSS similarly required more than identifying activity, anxiety, or tiredness: the evidence needed to address the particular social-media-related effect named in the item. These traces illustrate the distinction between \textcolor{cbSec}{\texttt{I-BEHAV}}, which marks inferences about behavior beyond what the supplied evidence establishes, and \textcolor{cbSec}{\texttt{I-MIND}}, which marks inferences about internal experience or likely self-report. In the high-error examples, models sometimes assigned scores before establishing the required connection, as reflected in \textcolor{cbBoth}{\texttt{E-JUDGE}} and \textcolor{cbBoth}{\texttt{M-CALC}}, either treating a behavioral cue as sufficient support or converting the absence of an explicit connection into a low response.

Taken together, the traces show that similar reasoning behaviors addressed different questions across constructs. Models needed to identify who experienced an event and whether it matched an item, distinguish social interactions from participants' reports of those interactions, or check whether observed behavior supported the experience specified by an item. When interpreting evidence (\textcolor{cbBoth}{\texttt{E-INT}}) and judging whether it supported an item (\textcolor{cbBoth}{\texttt{E-JUDGE}}), models could either match the evidence to the question or rely on a plausible interpretation that went beyond it. Expressing uncertainty or checking earlier judgments did not necessarily prevent models from using unsupported interpretations to assign scores. The reasoning pathways for each construct are provided in Appendix Table~\ref{tab:rq2-construct-pathways}.
\section{Discussion}

\subsection{Beyond Prediction Accuracy: Evaluating the Evidentiary Basis of Mental-Health Inference}
Evaluations of mental-health inference from digital traces commonly ask whether a model's predictions correspond to observed labels or self-reported measures~\cite{chancellor2020methods,ernala2019methodological}. Our findings show why this correspondence is necessary but incomplete. RQ1 demonstrated that improving access to relevant evidence can improve prediction, particularly when conversational probes help retrieve events expressed differently from questionnaire wording. At the same time, predictions that were numerically close to observed scale scores often showed weak or negative associations with those scores across participants and substantially less variation, a pattern that aligns with the broader concerns raised by \citet{yarkoni2017choosing} about the distinction between predictive performance and explanatory validity. A model may therefore approximate the location of reported scores without recovering which participants reported higher or lower levels of the construct.

These findings distinguish three questions that are often collapsed into a single evaluation of accuracy. \emph{Outcome correspondence} asks whether a prediction matches or approaches a participant's reported response. \emph{Participant differentiation} asks whether predictions preserve meaningful differences and rankings across participants rather than concentrating around a narrow range~\cite{bisbee2024synthetic, park2024diminished}. \emph{Evidentiary support} asks whether the information available to the model could reasonably establish the construct being inferred~\cite{wallach2025evaluating}. The first two questions concern properties of model outputs and can be evaluated quantitatively. The third concerns the relationship between the target construct and the evidence from which the prediction was produced. RQ2 showed that this relationship can remain weak even when a trace contains abundant, topically relevant material: supportive exchanges, nighttime activity, and frequent messaging were visible, but the participant's felt support, sleep interference, or perceived loss of control often were not.

Building on longstanding concerns about construct validity and measurement~\cite{borsboom2004concept,flake2020measurement}, we use \emph{evidence--construct alignment} to describe the relationship between what a target construct requires for inference and what a particular trace makes available. Alignment is not a fixed property of a questionnaire construct. It depends on the target, the data source, and the evidence present in a given case. Communication about mental health, blocking, or a discriminatory remark may be represented directly in DMs when the relevant actor, target, direction, and timeframe are preserved. Subjective appraisal, understood as a person's evaluation of a situation~\cite{lazarus1984stress}, may become visible in a trace when the participant explicitly expresses it. In its absence, however, a related behavior remains evidence of that behavior rather than direct evidence of the appraisal. A supportive reply can establish that support was offered while remaining insufficient to establish that it was experienced as available or adequate~\cite{olteanu2019social, wethington1986perceived}. Improving retrieval or model reasoning may make such evidence easier to locate and synthesize, but it cannot by itself supply an appraisal that the trace does not contain.

Accuracy should therefore be interpreted in relation to the evidentiary demands of the target rather than treated as sufficient evidence that a model recovered an individual's psychological state. Before evaluating an inference task, researchers should specify whether the target can be directly expressed or enacted in the data, whether the trace preserves the relevant actor, target, direction, and timeframe~\cite{vanhee2018automatic}, whether the construct requires subjective appraisal or causal attribution~\cite{olteanu2019social}, and whether an absence of trace evidence can support a negative judgment. These questions do not replace conventional performance metrics; they clarify what those metrics demonstrate.

\subsection{Designing AI That Distinguishes Observation from Inference}

The distinction between observable interaction and subjective experience also changes what an inference system should communicate. Evidence may directly establish one claim while serving only as a proxy for another. 
In the model-generated reasoning traces, these distinctions were often collapsed as models converted an observed event into a stronger psychological claim. Systems should therefore preserve what their evidence directly establishes instead of treating topical relevance as permission to infer the target construct.

This requirement is not solved by retrieval alone. The RQ1 results show that probe-based retrieval can improve access to messages expressed in language that differs from questionnaire wording. Yet RQ2 shows that construct-related evidence can remain insufficient for the inference requested of the model. Retrieval relevance and inferential sufficiency are therefore distinct design objectives: retrieval asks whether a passage concerns the target topic, whereas a sufficiency assessment asks whether that passage supports the specific psychological claim. This distinction parallels information science's long-standing separation of topical relevance from utility or situational relevance~\cite{saracevic2007relevance}, and recent work on retrieval-augmented generation has similarly framed the problem in terms of whether retrieved context is sufficient to support the requested inference~\cite{joren2025sufficient}. For constructs such as perceived support, problematic use, or sleep interference, a retriever may successfully surface supportive exchanges, frequent activity, or nighttime messages while leaving the participant's appraisal or the required causal relationship unresolved. More relevant evidence is not necessarily more sufficient evidence~\citep{joren2025sufficient,shi2023distracted}.

AI systems can preserve this distinction by representing the epistemic status of their outputs, for example, by presenting claims as grounded in identifiable evidence rather than as free-standing assertions~\cite{rashkin2023measuring}. Rather than collapsing all evidence into a questionnaire score, a system could separate what was \emph{observed}, what was \emph{inferred}, and what remained \emph{unestablished}. For example, it could report that friends offered help in several conversations, identify perceived support as one possible interpretation, and state that the archive does not establish whether the participant experienced those interactions as supportive or adequate. Such calibrated claims retain the evidentiary value of the trace without presenting an interpretation as an observation, consistent with evidence that language expressions of uncertainty measurably shape user reliance and trust~\cite{kim2024imnotsure,zhou2024relying}. They also make the model's inferential step available for scrutiny instead of concealing it within a definitive psychological judgment.

Our findings further suggest that insufficient evidence should be a legitimate system outcome. Several reasoning traces acknowledged missing causal or experiential evidence but nevertheless presumed that its absence meant a low response value in the corresponding questionnaire: effectively treating \textit{absence of evidence} as \textit{evidence of absence}. Prompting a model to factor in uncertainty is therefore not equivalent to ensuring that uncertainty constrains its predictions~\cite{zhou2024relying,xiong2024can}. Future systems could introduce an explicit construct-sufficiency check between evidence retrieval and prediction, requiring the model to identify whether the available evidence establishes the relevant actor, target, timeframe, appraisal, and causal relation. When those requirements are not met, the system should be able to abstain, mark the inference as unresolved~\cite{wen2025know}, or request information that bears directly on the participant's experience. These mechanisms remain to be evaluated.

\subsection{When Inference Requires People's Own Accounts}
For constructs defined partly by subjective appraisal, self-report is more than a label that a model should learn to reproduce. It provides access to how a person understood, felt, or evaluated an experience~\cite{gairola2026social}, information that may not be expressed in a behavioral trace~\cite{park2024personally}. This does not make self-report an infallible account of an underlying psychological state. Questionnaire responses remain shaped by recall, interpretation, and the conditions under which they are elicited~\cite{tourangeau2000psychology, schwarz2001asking}. Their importance here is more specific: when a construct asks whether support felt available~\cite{barrera1986distinctions}, whether social media caused distress~\cite{keles2020systematic}, or whether use felt difficult to control~\cite{van2016social}, the participant's appraisal is part of what the construct measures. 

Our findings echo prior work showing that `third-party' interpretations of online interactions can differ from participants' own appraisals~\cite{park2024personally}. This distinction has implications for how trace-based inference relates to conventional assessment. Although self-reports can serve as reference outcomes for training and evaluating predictive models~\cite{xu2024mental, shi2025mentalqlm}, the ability to predict these responses does not by itself establish that behavioral traces can substitute for participants' own accounts~\cite{yang2024mentallama, chancellor2020methods}. In our study, DMs provided stronger evidence when questionnaire targets were themselves communicative events, but offered less direct evidence for experiences that depended on an unexpressed appraisal. For those targets, trace-based inference may complement a person's account by identifying events, interaction patterns, or questions worth examining. Better models and richer archives may improve prediction, but such improvements alone would not establish that a model has captured the meaning those events had for the participant.

A complementary approach would change the role of AI from answering on a person's behalf to supporting reflection or inquiry. Our findings suggest a direction that instead of concluding that a participant feels supported because friends offered help, a system could identify the supportive exchanges and ask how those interactions were experienced. Instead of inferring sleep disturbance from nighttime messages, it could surface the observed activity and ask whether social media affected the person's sleep. Such interactions preserve the distinction between documenting an event and interpreting its personal meaning. They also allow conversational traces to contribute useful context without forcing them to stand in for information they do not contain~\cite{kwasnicka2015data, cole2023using}.

People should therefore retain interpretive authority over experiences that cannot be established from behavioral traces alone~\cite{park2024personally}. This is especially important when systems make sensitive inferences about adolescents~\cite{chancellor2019taxonomy}, whose private conversations include not only their own disclosures but also the words and experiences of peers~\cite{underwood2012blackberry, ehrenreich2020adolescents}. Inviting a person's account should not be framed merely as collecting an additional feature when model evidence is sparse. It recognizes that some psychological claims concern meanings available most directly to the person who experienced them. A system calibrated to this boundary should distinguish what it observed, state the limits of what it can infer, and defer to the participant when the target depends on how an experience was perceived or appraised.

\subsection{Ethical Considerations of Psychological Inference from Private Communication}

Our findings also raise ethical considerations that extend beyond protecting the privacy and security of the underlying data. 
Prior work with private adolescent social-media data has highlighted the distinctive privacy and ethical challenges created by
such relational data~\cite{razi2022instagram}. In this paper, we mitigated these risks through the use of participant-mediated donation, automated
deidentification, restricted researcher access, and local model deployment, as described in Section~\ref{sec:deid}. These protections
reduce exposure of private communication, but they do not by themselves resolve what inferences should be drawn from it. 

Our results raise a related ethical concern about the inferences drawn from these data. Behavioral traces may allow models to derive sensitive information that participants did not explicitly disclose, while models may also produce plausible judgments about subjective experiences that the available evidence does not adequately support. These possibilities create different risks: a system may infer something that is supported by the
data but that an individual did not intend to reveal, or it may attribute a psychological state to an individual that the data do not warrant. 
The latter is particularly important in light of our findings that LLMs did not reliably recognize when observable interpersonal behavior was
insufficient to establish subjective experience. Thus, responsible use of LLMs with private behavioral traces requires attention both to
privacy aspects around what information models are permitted to access and to epistemic ones around what claims they are permitted to make from that information.

\subsection{Limitations and Future Work}
Our findings characterize the language model's epistemic boundary of inference from private Instagram conversations, an evidence environment that preserves relational exchanges but represents one part of adolescents' broader digital lives. Different social-media platforms support different forms of disclosure, interaction, and self-presentation: private messages foreground interpersonal exchange, whereas public posts, comments, group interactions, and platform-level activity may make other aspects of experience visible. Future work can examine how evidence--construct alignment changes across these environments and whether evidence that supports an inference on one platform retains the same meaning when it appears on another. Cross-platform analysis would thereby extend the conditions under which the boundary can be examined without assuming that signals are interchangeable across platforms.

The evidentiary scope of the present analysis was also shaped by the modalities available to the models. Message text and interaction metadata were represented directly, whereas the semantic content of shared images, video, audio, reels, and other media was not. Multimodal models could determine whether these sources reveal events, relationship contexts, or explicit self-expressions that text alone does not capture. The relevant test, however, is not simply whether adding modalities improves predictive accuracy, but whether the added information provides stronger evidence for the construct being inferred. Multimodal evidence may therefore expand what a model can see without eliminating the distinction between observation and subjective experience.

Future research should bring these directions together through multimodal, cross-platform, and longitudinal evidence paired with participants' contemporaneous accounts. Such work can test, construct by construct, whether an additional source supplies direct evidence or strengthens a previously indirect inference. It should evaluate these changes through outcome correspondence, participant differentiation, and evidentiary support, while allowing AI systems to leave an inference unresolved when the combined evidence remains insufficient.

\section{Conclusion}

Using Instagram DM archives donated by 80 adolescents, eight self-report measures, and ten locally deployed reasoning LLMs, RQ1 evaluated predictions produced through the Note and RAG evidence-construction approaches. Probe-based retrieval produced broader improvements over its within-approach baseline, and models were more successful when questionnaire targets---such as mental-health communication, negative peer experiences, and discrimination---could be expressed or enacted directly in the messages. Yet numerical proximity to participants' scale scores did not consistently preserve differences among participants. RQ2 connected these outcomes to the evidentiary paths recorded in model-generated reasoning traces. Models could use direct conversational evidence effectively when they preserved the actor, target, direction, and timeframe, but they frequently treated supportive language, messaging activity, and nighttime communication as evidence of felt support, problematic use, and sleep-related distress.

The central boundary revealed by these findings is epistemic rather than merely technical. Better retrieval can make relevant evidence easier to find, and stronger models may interpret that evidence more consistently, but neither improvement can establish an experience that the available trace does not contain. Evaluating mental-health inference therefore requires more than prediction accuracy: it requires examining whether predictions differentiate among people and whether the available evidence can reasonably support the psychological claim being made. Systems should distinguish observation from inference, represent when a construct remains unestablished, and preserve people's interpretive authority when the target depends on their own perception or appraisal. In this sense, social-media traces can complement people's accounts by documenting communicative events and interaction patterns, but they should not be treated as substitutes for those accounts.

\bibliographystyle{ACM-Reference-Format}
\bibliography{sample-base}

@book{cohen2013statistical,
  title={Statistical power analysis for the behavioral sciences},
  author={Cohen, Jacob},
  year={2013},
  publisher={Routledge}
}

@inproceedings{gairola2026social,
  title={" Social Media Killed Our Generation": Teenagers' Felt Experiences of Harm on Social Media},
  author={Gairola, Ritika and Gray, Colin M and Dong, Jingxin and Jeong, Kyung Jin and Otenen, Ege and Sarria, Juan J},
  booktitle={Proceedings of the 2026 CHI Conference on Human Factors in Computing Systems},
  pages={1--15},
  year={2026}
}

@article{razi2023sliding,
  title={Sliding into my DMs: Detecting uncomfortable or unsafe sexual risk experiences within Instagram direct messages grounded in the perspective of youth},
  author={Razi, Afsaneh and Alsoubai, Ashwaq and Kim, Seunghyun and Ali, Shiza and Stringhini, Gianluca and De Choudhury, Munmun and Wisniewski, Pamela J},
  journal={Proceedings of the ACM on Human-Computer Interaction},
  volume={7},
  number={CSCW1},
  pages={1--29},
  year={2023},
  publisher={ACM New York, NY, USA}
}

@article{alsoubai2024profiling,
  title={Profiling the offline and online risk experiences of youth to develop targeted interventions for online safety},
  author={Alsoubai, Ashwaq and Razi, Afsaneh and Agha, Zainab and Ali, Shiza and Stringhini, Gianluca and De Choudhury, Munmun and Wisniewski, Pamela J},
  journal={Proceedings of the ACM on human-computer interaction},
  volume={8},
  number={CSCW1},
  pages={1--37},
  year={2024},
  publisher={ACM New York, NY, USA}
}

@article{ali2023getting,
  title={Getting meta: A multimodal approach for detecting unsafe conversations within instagram direct messages of youth},
  author={Ali, Shiza and Razi, Afsaneh and Kim, Seunghyun and Alsoubai, Ashwaq and Ling, Chen and De Choudhury, Munmun and Wisniewski, Pamela J and Stringhini, Gianluca},
  journal={Proceedings of the ACM on human-computer interaction},
  volume={7},
  number={CSCW1},
  pages={1--30},
  year={2023},
  publisher={ACM New York, NY, USA}
}

@inproceedings{puspo2026mental,
  title={Mental health disorder detection beyond social media: A systematic review of available datasets},
  author={Puspo, Sadiya Sayara Chowdhury and Bucur, Ana-Maria and Chancellor, Stevie and Uzuner, Ozlem and Zampieri, Marcos},
  booktitle={Proceedings of the Fifteenth Language Resources and Evaluation Conference (LREC 2026)},
  volume={11},
  number={16},
  pages={6235--6250},
  year={2026}
}

@article{mitic2021toward,
  title={Toward an integrated model of supportive peer relationships in early adolescence: A systematic review and exploratory meta-analysis},
  author={Mitic, Marija and Woodcock, Kate A and Amering, Michaela and Krammer, Ina and Stiehl, Katharina AM and Zehetmayer, Sonja and Schrank, Beate},
  journal={Frontiers in psychology},
  volume={12},
  pages={589403},
  year={2021},
  publisher={Frontiers Media SA}
}

@article{schwartz2020relations,
  title={Relations of friendship experiences with depressive symptoms and loneliness in childhood and adolescence: A meta-analytic review.},
  author={Schwartz-Mette, Rebecca A and Shankman, Jessica and Dueweke, Aubrey R and Borowski, Sarah and Rose, Amanda J},
  journal={Psychological bulletin},
  volume={146},
  number={8},
  pages={664},
  year={2020},
  publisher={American Psychological Association}
}

@article{filia2021facilitating,
  title={Facilitating improvements in young people’s social relationships to prevent or treat depression: A review of empirically supported interventions},
  author={Filia, Kate and Eastwood, Oliver and Herniman, Sarah and Badcock, Paul},
  journal={Translational Psychiatry},
  volume={11},
  number={1},
  pages={305},
  year={2021},
  publisher={Nature Publishing Group UK London}
}

@inproceedings{razi2022instagram,
  title={Instagram data donation: a case study on collecting ecologically valid social media data for the purpose of adolescent online risk detection},
  author={Razi, Afsaneh and AlSoubai, Ashwaq and Kim, Seunghyun and Naher, Nurun and Ali, Shiza and Stringhini, Gianluca and De Choudhury, Munmun and Wisniewski, Pamela J},
  booktitle={CHI conference on human factors in computing systems extended abstracts},
  pages={1--9},
  year={2022}
}

@inproceedings{cole2023using,
  title={Using data-prompted interviews in interactive information retrieval research: A reflection on the study of self-efficacy when learning using search},
  author={Cole, Amelia and O'Brien, Heather},
  booktitle={Proceedings of the 2023 Conference on Human Information Interaction and Retrieval},
  pages={406--411},
  year={2023}
}

@article{kwasnicka2015data,
  title={Data-prompted interviews: Using individual ecological data to stimulate narratives and explore meanings.},
  author={Kwasnicka, Dominika and Dombrowski, Stephan U and White, Martin and Sniehotta, Falko F},
  journal={Health Psychology},
  volume={34},
  number={12},
  pages={1191},
  year={2015},
  publisher={American Psychological Association}
}

@inproceedings{yang2024mentallama,
  title={MentaLLaMA: interpretable mental health analysis on social media with large language models},
  author={Yang, Kailai and Zhang, Tianlin and Kuang, Ziyan and Xie, Qianqian and Huang, Jimin and Ananiadou, Sophia},
  booktitle={Proceedings of the ACM Web Conference 2024},
  pages={4489--4500},
  year={2024}
}

@article{shi2025mentalqlm,
  title={MentalQLM: A lightweight large language model for mental healthcare based on instruction tuning and dual LoRA modules},
  author={Shi, Jiayu and Wang, Zexiao and Zhou, Jiandong and Liu, Chengyu and Sun, Poly ZH and Zhao, Erying and Lu, Lei},
  journal={IEEE Journal of Biomedical and Health Informatics},
  year={2025},
  publisher={IEEE}
}

@inproceedings{park2024personally,
  title={Personally targeted risk vs. humor: How online risk perceptions of youth vs. third-party annotators differ based on privately shared media on Instagram},
  author={Park, Jinkyung and Gracie, Joshua and Alsoubai, Ashwaq and Razi, Afsaneh and Wisniewski, Pamela J},
  booktitle={Proceedings of the 23rd Annual ACM Interaction Design and Children Conference},
  pages={1--13},
  year={2024}
}

@book{tourangeau2000psychology,
  title={The psychology of survey response},
  author={Tourangeau, Roger and Rips, Lance J and Rasinski, Kenneth},
  volume={595},
  year={2000},
  publisher={Cambridge New York}
}

@article{schwarz2001asking,
  title={Asking questions about behavior: Cognition, communication, and questionnaire construction},
  author={Schwarz, Norbert and Oyserman, Daphna},
  journal={American Journal of Evaluation},
  volume={22},
  number={2},
  pages={127--160},
  year={2001},
  publisher={Sage Publications Sage CA: Thousand Oaks, CA}
}

@article{barrera1986distinctions,
  title={Distinctions between social support concepts, measures, and models},
  author={Barrera Jr, Manuel},
  journal={American journal of community psychology},
  volume={14},
  number={4},
  pages={413--445},
  year={1986},
  publisher={Springer}
}

@article{keles2020systematic,
  title={A systematic review: the influence of social media on depression, anxiety and psychological distress in adolescents},
  author={Keles, Betul and McCrae, Niall and Grealish, Annmarie},
  journal={International journal of adolescence and youth},
  volume={25},
  number={1},
  pages={79--93},
  year={2020},
  publisher={Taylor \& Francis}
}

@article{van2016social,
  title={The social media disorder scale},
  author={Van Den Eijnden, Regina JJM and Lemmens, Jeroen S and Valkenburg, Patti M},
  journal={Computers in human behavior},
  volume={61},
  pages={478--487},
  year={2016},
  publisher={Elsevier}
}

@article{laursen2012sequential,
  title={Sequential organization of text messages and mobile phone calls in interconnected communication sequences},
  author={Laursen, Ditte},
  journal={Discourse \& Communication},
  volume={6},
  number={1},
  pages={83--99},
  year={2012},
  publisher={SAGE Publications Sage UK: London, England}
}

@inproceedings{seo2024chacha,
  title={Chacha: leveraging large language models to prompt children to share their emotions about personal events},
  author={Seo, Woosuk and Yang, Chanmo and Kim, Young-Ho},
  booktitle={Proceedings of the 2024 CHI Conference on Human Factors in Computing Systems},
  pages={1--20},
  year={2024}
}

@article{wolf2021development,
  title={The development of the liking gap: Children older than 5 years think that partners evaluate them less positively than they evaluate their partners},
  author={Wolf, Wouter and Nafe, Amanda and Tomasello, Michael},
  journal={Psychological Science},
  volume={32},
  number={5},
  pages={789--798},
  year={2021},
  publisher={Sage Publications Sage CA: Los Angeles, CA}
}

@article{bierstetel2020associations,
  title={Associations between language style matching and relationship commitment and satisfaction: An integrative data analysis},
  author={Bierstetel, Sabrina J and Farrell, Allison K and Briskin, Julia L and Harvey, Michael W and Gable, Shelly L and Ha, Thao and Ickes, William and Lin, Wei-Fang and Orina, Minda M and Saxbe, Darby and others},
  journal={Journal of Social and Personal Relationships},
  volume={37},
  number={8-9},
  pages={2459--2481},
  year={2020},
  publisher={Sage Publications Sage UK: London, England}
}

@article{brown2022elementary,
  title={The elementary forms of digital communication},
  author={Brown, Lauren S and Lewis, Kevin},
  journal={Plos one},
  volume={17},
  number={9},
  pages={e0273726},
  year={2022},
  publisher={Public Library of Science San Francisco, CA USA}
}

@inproceedings{kooti2015evolution,
  title={Evolution of conversations in the age of email overload},
  author={Kooti, Farshad and Aiello, Luca Maria and Grbovic, Mihajlo and Lerman, Kristina and Mantrach, Amin},
  booktitle={Proceedings of the 24th international conference on world wide web},
  pages={603--613},
  year={2015}
}

@article{ireland2011language,
  title={Language style matching predicts relationship initiation and stability},
  author={Ireland, Molly E and Slatcher, Richard B and Eastwick, Paul W and Scissors, Lauren E and Finkel, Eli J and Pennebaker, James W},
  journal={Psychological science},
  volume={22},
  number={1},
  pages={39--44},
  year={2011},
  publisher={Sage Publications Sage CA: Los Angeles, CA}
}

@misc{openai2025gptoss,
  author = {{OpenAI}},
  title  = {gpt-oss-120b \& gpt-oss-20b Model Card},
  year   = {2025},
  eprint = {2508.10925},
  archivePrefix = {arXiv},
  primaryClass  = {cs.CL},
  url    = {https://arxiv.org/abs/2508.10925}
}

@misc{glm2025glm45,
  author = {{GLM-4.5 Team} and Zeng, Aohan and Lv, Xin and Zheng, Qinkai and Hou, Zhenyu and others},
  title  = {{GLM-4.5}: Agentic, Reasoning, and Coding ({ARC}) Foundation Models},
  year   = {2025},
  eprint = {2508.06471},
  archivePrefix = {arXiv},
  primaryClass  = {cs.CL},
  url    = {https://arxiv.org/abs/2508.06471}
}

@article{deepseek2025r1,
  author  = {Guo, Daya and Yang, Dejian and Zhang, Haowei and others},
  title   = {{DeepSeek-R1} incentivizes reasoning in {LLMs} through reinforcement learning},
  journal = {Nature},
  volume  = {645},
  number  = {8081},
  pages   = {633--638},
  year    = {2025},
  doi     = {10.1038/s41586-025-09422-z}
}

@misc{deepseek2025r1qwen3,
  author = {{DeepSeek-AI}},
  title  = {{DeepSeek-R1-0528-Qwen3-8B}},
  howpublished = {Hugging Face model card},
  year   = {2025},
  url    = {https://huggingface.co/deepseek-ai/DeepSeek-R1-0528-Qwen3-8B}
}

@misc{qwen2026qwen36_35b,
  author = {{Qwen Team}},
  title  = {{Qwen3.6-35B-A3B}: Agentic Coding Power, Now Open to All},
  howpublished = {Blog post and Hugging Face model card},
  year   = {2026},
  url    = {https://qwen.ai/blog?id=qwen3.6-35b-a3b}
}

@misc{qwen2026qwen36_27b,
  author = {{Qwen Team}},
  title  = {{Qwen3.6-27B}: Flagship-Level Coding in a 27B Dense Model},
  howpublished = {Blog post and Hugging Face model card},
  year   = {2026},
  url    = {https://qwen.ai/blog?id=qwen3.6-27b}
}

@misc{mistral2025magistral,
  author = {{Mistral AI}},
  title  = {Magistral},
  year   = {2025},
  eprint = {2506.10910},
  archivePrefix = {arXiv},
  primaryClass  = {cs.CL},
  url    = {https://arxiv.org/abs/2506.10910}
}

@misc{mistral2025magistralsmall2509,
  author = {{Mistral AI}},
  title  = {{Magistral-Small-2509}},
  howpublished = {Hugging Face model card},
  year   = {2025},
  url    = {https://huggingface.co/mistralai/Magistral-Small-2509}
}

@inproceedings{zhou2024relying,
  author    = {Zhou, Kaitlyn and Hwang, Jena D. and Ren, Xiang and Sap, Maarten},
  title     = {Relying on the Unreliable: The Impact of Language Models' Reluctance to Express Uncertainty},
  booktitle = {Proceedings of the 62nd Annual Meeting of the Association for Computational Linguistics (Volume 1: Long Papers)},
  pages     = {3623--3643}, year = {2024}, doi = {10.18653/v1/2024.acl-long.198}
}

@inproceedings{xiong2024can,
  author    = {Xiong, Miao and Hu, Zhiyuan and Lu, Xinyang and Li, Yifei and Fu, Jie and He, Junxian and Hooi, Bryan},
  title     = {Can LLMs Express Their Uncertainty? An Empirical Evaluation of Confidence Elicitation in LLMs},
  booktitle = {International Conference on Learning Representations (ICLR)},
  year      = {2024}, note = {arXiv:2306.13063}
}

@article{olteanu2019social,
  author  = {Olteanu, Alexandra and Castillo, Carlos and Diaz, Fernando and K{\i}c{\i}man, Emre},
  title   = {Social Data: Biases, Methodological Pitfalls, and Ethical Boundaries},
  journal = {Frontiers in Big Data},
  volume  = {2},
  pages   = {13},
  year    = {2019},
  doi     = {10.3389/fdata.2019.00013}
}

@article{wethington1986perceived,
  author  = {Wethington, Elaine and Kessler, Ronald C.},
  title   = {Perceived support, received support, and adjustment to stressful life events},
  journal = {Journal of Health and Social Behavior},
  volume  = {27},
  number  = {1},
  pages   = {78--89},
  year    = {1986},
  doi     = {10.2307/2136504}
}

@article{vanhee2018automatic,
  author  = {Van Hee, Cynthia and Jacobs, Gilles and Emmery, Chris and Desmet, Bart and Lefever, Els and Verhoeven, Ben and De Pauw, Guy and Daelemans, Walter and Hoste, V{\'e}ronique},
  title   = {Automatic detection of cyberbullying in social media text},
  journal = {PLOS ONE},
  volume  = {13},
  number  = {10},
  pages   = {e0203794},
  year    = {2018},
  doi     = {10.1371/journal.pone.0203794}
}

@article{kim2024respect,
  title={ReSPect: Enabling active and scalable responses to networked online harassment},
  author={Kim, Haesoo and Lee, Juhoon and Jang, Jeong-Woo and Kim, Juho},
  journal={Proceedings of the ACM on Human-Computer Interaction},
  volume={8},
  number={CSCW1},
  pages={1--30},
  year={2024},
  publisher={ACM New York, NY, USA}
}

@inproceedings{kim2024imnotsure,
  author    = {Kim, Sunnie S. Y. and Liao, Q. Vera and Vorvoreanu, Mihaela and Ballard, Stephanie and Vaughan, Jennifer Wortman},
  title     = {``I'm Not Sure, But...'': Examining the Impact of Large Language Models' Uncertainty Expression on User Reliance and Trust},
  booktitle = {Proceedings of the 2024 ACM Conference on Fairness, Accountability, and Transparency (FAccT)},
  pages     = {822--835}, year = {2024}, doi = {10.1145/3630106.3658941}
}

@article{wen2025know,
  author  = {Wen, Bingbing and Yao, Jihan and Feng, Shangbin and Xu, Chenjun and Tsvetkov, Yulia and Howe, Bill and Wang, Lucy Lu},
  title   = {Know Your Limits: A Survey of Abstention in Large Language Models},
  journal = {Transactions of the Association for Computational Linguistics},
  volume  = {13}, pages = {529--556}, year = {2025},
  doi     = {10.1162/tacl_a_00754}
}

@article{rashkin2023measuring,
  author  = {Rashkin, Hannah and Nikolaev, Vitaly and Lamm, Matthew and Aroyo, Lora and Collins, Michael and Das, Dipanjan and Petrov, Slav and Tomar, Gaurav Singh and Turc, Iulia and Reitter, David},
  title   = {Measuring Attribution in Natural Language Generation Models},
  journal = {Computational Linguistics},
  volume  = {49}, number = {4}, pages = {777--840}, year = {2023},
  doi     = {10.1162/coli_a_00486}
}

@inproceedings{shi2023distracted,
  author    = {Shi, Freda and Chen, Xinyun and Misra, Kanishka and Scales, Nathan and Dohan, David and Chi, Ed H. and Sch\"{a}rli, Nathanael and Zhou, Denny},
  title     = {Large Language Models Can Be Easily Distracted by Irrelevant Context},
  booktitle = {Proceedings of the 40th International Conference on Machine Learning (ICML)},
  series    = {PMLR}, volume = {202}, pages = {31210--31227}, year = {2023}
}

@inproceedings{joren2025sufficient,
  author    = {Joren, Hailey and Zhang, Jianyi and Ferng, Chun-Sung and Juan, Da-Cheng and Taly, Ankur and Rashtchian, Cyrus},
  title     = {Sufficient Context: A New Lens on Retrieval Augmented Generation Systems},
  booktitle = {International Conference on Learning Representations (ICLR)},
  year      = {2025}, note = {arXiv:2411.06037}
}

@article{saracevic2007relevance,
  author  = {Saracevic, Tefko},
  title   = {Relevance: A Review of the Literature and a Framework for Thinking on the Notion in Information Science. Part II: Nature and Manifestations of Relevance},
  journal = {Journal of the American Society for Information Science and Technology},
  volume  = {58}, number = {13}, pages = {1915--1933}, year = {2007},
  doi     = {10.1002/asi.20682}
}

@article{borsboom2004concept,
  author  = {Borsboom, Denny and Mellenbergh, Gideon J. and van Heerden, Jaap},
  title   = {The concept of validity},
  journal = {Psychological Review},
  volume  = {111},
  number  = {4},
  pages   = {1061--1071},
  year    = {2004},
  doi     = {10.1037/0033-295X.111.4.1061}
}

@article{flake2020measurement,
  author  = {Flake, Jessica Kay and Fried, Eiko I.},
  title   = {Measurement Schmeasurement: Questionable Measurement Practices and How to Avoid Them},
  journal = {Advances in Methods and Practices in Psychological Science},
  volume  = {3},
  number  = {4},
  pages   = {456--465},
  year    = {2020},
  doi     = {10.1177/2515245920952393}
}

@book{lazarus1984stress,
  author    = {Lazarus, Richard S. and Folkman, Susan},
  title     = {Stress, Appraisal, and Coping},
  publisher = {Springer Publishing Company},
  address   = {New York},
  year      = {1984},
  isbn      = {9780826141910}
}

@article{yarkoni2017choosing,
  author  = {Yarkoni, Tal and Westfall, Jacob},
  title   = {Choosing prediction over explanation in psychology: Lessons from machine learning},
  journal = {Perspectives on Psychological Science},
  volume  = {12},
  number  = {6},
  pages   = {1100--1122},
  year    = {2017},
  doi     = {10.1177/1745691617693393}
}

@inproceedings{wallach2025evaluating,
  author    = {Wallach, Hanna and Desai, Meera and Cooper, A. Feder and Wang, Angelina and Atalla, Chad and Barocas, Solon and Blodgett, Su Lin and Chouldechova, Alexandra and Corvi, Emily and Dow, P. Alex and Garcia-Gathright, Jean and Olteanu, Alexandra and Pangakis, Nicholas and Reed, Stefanie and Sheng, Emily and Vann, Dan and Vaughan, Jennifer Wortman and Vogel, Matthew and Washington, Hannah and Jacobs, Abigail Z.},
  title     = {Position: Evaluating Generative AI Systems Is a Social Science Measurement Challenge},
  booktitle = {Proceedings of the 42nd International Conference on Machine Learning (ICML)},
  series    = {PMLR},
  volume    = {267},
  pages     = {82232--82251},
  year      = {2025}
}

@article{bisbee2024synthetic,
  author  = {Bisbee, James and Clinton, Joshua D. and Dorff, Cassy and Kenkel, Brenton and Larson, Jennifer M.},
  title   = {Synthetic replacements for human survey data? The perils of large language models},
  journal = {Political Analysis},
  volume  = {32},
  number  = {4},
  pages   = {401--416},
  year    = {2024},
  doi     = {10.1017/pan.2024.5}
}

@article{park2024diminished,
  author  = {Park, Peter S. and Schoenegger, Philipp and Zhu, Chongyang},
  title   = {Diminished diversity-of-thought in a standard large language model},
  journal = {Behavior Research Methods},
  volume  = {56},
  number  = {5},
  pages   = {5754--5770},
  year    = {2024},
  doi     = {10.3758/s13428-023-02307-x}
}

@inproceedings{ding2025multi,
    title = "A Multi-Level Benchmark for Causal Language Understanding in Social Media Discourse",
    author = "Ding, Xiaohan  and
      Ping, Kaike  and
      {\c{C}}ar{\i}k, Buse  and
      Rho, Eugenia",
    editor = "Christodoulopoulos, Christos  and
      Chakraborty, Tanmoy  and
      Rose, Carolyn  and
      Peng, Violet",
    booktitle = "Proceedings of the 2025 Conference on Empirical Methods in Natural Language Processing",
    month = nov,
    year = "2025",
    address = "Suzhou, China",
    publisher = "Association for Computational Linguistics",
    url = "https://aclanthology.org/2025.emnlp-main.1464/",
    doi = "10.18653/v1/2025.emnlp-main.1464",
    pages = "28776--28790",
    ISBN = "979-8-89176-332-6",
}

@inproceedings{mirzakhmedova2024large,
  title={Are large language models reliable argument quality annotators?},
  author={Mirzakhmedova, Nailia and Gohsen, Marcel and Chang, Chia Hao and Stein, Benno},
  booktitle={Conference on advances in robust argumentation machines},
  pages={129--146},
  year={2024},
  organization={Springer}
}

@inproceedings{chen2024large,
  title={Is a large language model a good annotator for event extraction?},
  author={Chen, Ruirui and Qin, Chengwei and Jiang, Weifeng and Choi, Dongkyu},
  booktitle={Proceedings of the AAAI conference on artificial intelligence},
  volume={38},
  number={16},
  pages={17772--17780},
  year={2024}
}

@inproceedings{leung2026annotategpt,
  title={AnnotateGPT: Designing Human--AI Collaboration in Pen-Based Document Annotation},
  author={Leung, Benedict and Shimabukuro, Mariana and Collins, Christopher},
  booktitle={Proceedings of the 2026 CHI Conference on Human Factors in Computing Systems},
  pages={1--25},
  year={2026}
}

@misc{openai2026chatgpt,
  author       = {{OpenAI}},
  title        = {{ChatGPT} ({GPT}-5.6 version)},
  year         = {2026},
  month        = jun,
  howpublished = {Large language model},
  url          = {https://chat.openai.com/},
  note         = {Accessed 2026-09-10},
}

@article{schwartz2015data,
  title={Data-driven content analysis of social media: A systematic overview of automated methods},
  author={Schwartz, H Andrew and Ungar, Lyle H},
  journal={The ANNALS of the American Academy of Political and Social Science},
  volume={659},
  number={1},
  pages={78--94},
  year={2015},
  publisher={Sage Publications Sage CA: Los Angeles, CA}
}

@inproceedings{das2022semantic,
  title={Semantic gap in predicting mental wellbeing through passive sensing},
  author={Das Swain, Vedant and Chen, Victor and Mishra, Shrija and Mattingly, Stephen M and Abowd, Gregory D and De Choudhury, Munmun},
  booktitle={Proceedings of the 2022 CHI conference on human factors in computing systems},
  pages={1--16},
  year={2022}
}

@inproceedings{ernala2019methodological,
  title={Methodological gaps in predicting mental health states from social media: Triangulating diagnostic signals},
  author={Ernala, Sindhu Kiranmai and Birnbaum, Michael L and Candan, Kristin A and Rizvi, Asra F and Sterling, William A and Kane, John M and De Choudhury, Munmun},
  booktitle={Proceedings of the 2019 chi conference on human factors in computing systems},
  pages={1--16},
  year={2019}
}

@inproceedings{harrigian2020models,
  title={Do models of mental health based on social media data generalize?},
  author={Harrigian, Keith and Aguirre, Carlos Alejandro and Dredze, Mark},
  booktitle={Findings of the association for computational linguistics: EMNLP 2020},
  pages={3774--3788},
  year={2020}
}

@inproceedings{staab2024beyond,
  title={Beyond memorization: Violating privacy via inference with large language models},
  author={Staab, Robin and Vero, Mark and Balunovic, Mislav and Vechev, Martin},
  booktitle={International Conference on Learning Representations},
  volume={2024},
  pages={33832--33878},
  year={2024}
}

@article{peters2024large,
  title={Large language models can infer psychological dispositions of social media users},
  author={Peters, Heinrich and Matz, Sandra C and Gelfand, Michele and others},
  journal={PNAS nexus},
  volume={3},
  number={6},
  pages={pgae231},
  year={2024},
  publisher={Proceedings of the National Academy of Sciences}
}

@article{rathje2024gpt,
  title={GPT is an effective tool for multilingual psychological text analysis},
  author={Rathje, Steve and Mirea, Dan-Mircea and Sucholutsky, Ilia and Marjieh, Raja and Robertson, Claire E and Van Bavel, Jay J},
  journal={Proceedings of the National Academy of Sciences},
  volume={121},
  number={34},
  pages={e2308950121},
  year={2024},
  publisher={National Academy of Sciences}
}

@article{bolger2000invisible,
  title={Invisible support and adjustment to stress},
  author={Bolger, Niall and Zuckerman, Adam and Kessler, Ronald C},
  journal={Journal of Personality and Social Psychology},
  volume={79},
  number={6},
  pages={953--961},
  year={2000},
  publisher={American Psychological Association}
}

@inproceedings{chancellor2019taxonomy,
  title={A taxonomy of ethical tensions in inferring mental health states from social media},
  author={Chancellor, Stevie and Birnbaum, Michael L and Caine, Eric D and Silenzio, Vincent M B and De Choudhury, Munmun},
  booktitle={Proceedings of the Conference on Fairness, Accountability, and Transparency},
  pages={79--88},
  year={2019},
  doi={10.1145/3287560.3287587}
}

@article{chancellor2019human,
  title={Who is the ``human'' in human-centered machine learning: The case of predicting mental health from social media},
  author={Chancellor, Stevie and Baumer, Eric P S and De Choudhury, Munmun},
  journal={Proceedings of the ACM on Human-Computer Interaction},
  volume={3},
  number={CSCW},
  pages={1--32},
  year={2019},
  articleno={147},
  doi={10.1145/3359249},
  publisher={ACM New York, NY, USA}
}

@article{cohen1983global,
  title={A global measure of perceived stress},
  author={Cohen, Sheldon and Kamarck, Tom and Mermelstein, Robin},
  journal={Journal of Health and Social Behavior},
  volume={24},
  number={4},
  pages={385--396},
  year={1983},
  publisher={JSTOR}
}

@inproceedings{jacobs2021measurement,
  title={Measurement and fairness},
  author={Jacobs, Abigail Z and Wallach, Hanna},
  booktitle={Proceedings of the 2021 ACM Conference on Fairness, Accountability, and Transparency},
  pages={375--385},
  year={2021},
  doi={10.1145/3442188.3445901}
}

@article{ji2023survey,
  title={Survey of hallucination in natural language generation},
  author={Ji, Ziwei and Lee, Nayeon and Frieske, Rita and Yu, Tiezheng and Su, Dan and Xu, Yan and Ishii, Etsuko and Bang, Ye Jin and Madotto, Andrea and Fung, Pascale},
  journal={ACM Computing Surveys},
  volume={55},
  number={12},
  pages={1--38},
  year={2023},
  articleno={248},
  doi={10.1145/3571730},
  publisher={ACM New York, NY, USA}
}

@article{nick2018online,
  title={The Online Social Support Scale: Measure development and validation},
  author={Nick, Elizabeth A and Cole, David A and Cho, Sun-Joo and Smith, Darcy K and Carter, T Grace and Zelkowitz, Rachel L},
  journal={Psychological Assessment},
  volume={30},
  number={9},
  pages={1127--1143},
  year={2018},
  doi={10.1037/pas0000558},
  publisher={American Psychological Association}
}

@article{schwarz1999self,
  title={Self-reports: How the questions shape the answers},
  author={Schwarz, Norbert},
  journal={American Psychologist},
  volume={54},
  number={2},
  pages={93--105},
  year={1999},
  doi={10.1037/0003-066X.54.2.93},
  publisher={American Psychological Association}
}

@inproceedings{sivak2020measuring,
  title={Measuring adolescents' well-being: Correspondence of na{\"i}ve digital traces to survey data},
  author={Sivak, Elizaveta and Smirnov, Ivan},
  booktitle={Social Informatics: 12th International Conference, SocInfo 2020},
  series={Lecture Notes in Computer Science},
  volume={12467},
  pages={352--363},
  year={2020},
  doi={10.1007/978-3-030-60975-7_26},
  publisher={Springer}
}

@article{xu2024mental,
  title={Mental-LLM: Leveraging large language models for mental health prediction via online text data},
  author={Xu, Xuhai and Yao, Bingsheng and Dong, Yuanzhe and Gabriel, Saadia and Yu, Hong and Hendler, James and Ghassemi, Marzyeh and Dey, Anind K and Wang, Dakuo},
  journal={Proceedings of the ACM on Interactive, Mobile, Wearable and Ubiquitous Technologies},
  volume={8},
  number={1},
  pages={1--32},
  year={2024},
  articleno={32},
  doi={10.1145/3643540},
  publisher={ACM New York, NY, USA}
}

@article{sultan2023leaving,
  title={Leaving traces behind: Using social media digital trace data to study adolescent wellbeing},
  author={Sultan, Mubashir and Scholz, Christin and van den Bos, Wouter},
  journal={Computers in Human Behavior Reports},
  volume={10},
  pages={100281},
  year={2023},
  publisher={Elsevier},
  doi={10.1016/j.chbr.2023.100281}
}

@article{ehrenreich2020adolescents,
  title={How adolescents use text messaging through their high school years},
  author={Ehrenreich, Samuel E and Beron, Kurt J and Burnell, Kaitlyn and Meter, Diana J and Underwood, Marion K},
  journal={Journal of Research on Adolescence},
  volume={30},
  number={2},
  pages={521--540},
  year={2020},
  publisher={Wiley},
  doi={10.1111/jora.12541}
}

@article{underwood2015blackberry,
  title={The BlackBerry Project: The hidden world of adolescents' text messaging and relations with internalizing symptoms},
  author={Underwood, Marion K and Ehrenreich, Samuel E and More, David and Solis, Jerome S and Brinkley, Dawn Y},
  journal={Journal of Research on Adolescence},
  volume={25},
  number={1},
  pages={101--117},
  year={2015},
  publisher={Wiley},
  doi={10.1111/jora.12101}
}

@article{edelmann2020computational,
  title={Computational social science and sociology},
  author={Edelmann, Achim and Wolff, Tom and Montagne, Danielle and Bail, Christopher A},
  journal={Annual review of sociology},
  volume={46},
  number={1},
  pages={61--81},
  year={2020},
  publisher={Annual Reviews}
}

@article{haber2007relationship,
  title={The relationship between self-reported received and perceived social support: A meta-analytic review},
  author={Haber, Mason G and Cohen, Jay L and Lucas, Todd and Baltes, Boris B},
  journal={American Journal of Community Psychology},
  volume={39},
  number={1-2},
  pages={133--144},
  year={2007},
  publisher={Wiley},
  doi={10.1007/s10464-007-9100-9}
}

@article{uchino2009understanding,
  title={Understanding the links between social support and physical health: A life-span perspective with emphasis on the separability of perceived and received support},
  author={Uchino, Bert N},
  journal={Perspectives on psychological science},
  volume={4},
  number={3},
  pages={236--255},
  year={2009},
  publisher={SAGE Publications Sage CA: Los Angeles, CA}
}

@article{ehrenreich2014adolescents,
  title={Adolescents' text message communication and growth in antisocial behavior across the first year of high school},
  author={Ehrenreich, Samuel E and Underwood, Marion K and Ackerman, Robert A},
  journal={Journal of Abnormal Child Psychology},
  volume={42},
  number={2},
  pages={251--264},
  year={2014},
  publisher={Springer},
  doi={10.1007/s10802-013-9783-3}
}

@article{valkenburg2009effects,
  title={The effects of instant messaging on the quality of adolescents' existing friendships: A longitudinal study},
  author={Valkenburg, Patti M and Peter, Jochen},
  journal={Journal of Communication},
  volume={59},
  number={1},
  pages={79--97},
  year={2009},
  publisher={Wiley},
  doi={10.1111/j.1460-2466.2008.01405.x}
}

@article{nesi2018transformation,
  title={Transformation of adolescent peer relations in the social media context: Part 1---A theoretical framework and application to dyadic peer relationships},
  author={Nesi, Jacqueline and Choukas-Bradley, Sophia and Prinstein, Mitchell J},
  journal={Clinical Child and Family Psychology Review},
  volume={21},
  number={3},
  pages={267--294},
  year={2018},
  publisher={Springer},
  doi={10.1007/s10567-018-0261-x}
}

@article{vollet2020exploring,
  title={Exploring text messaging as a platform for peer socialization of social aggression.},
  author={Vollet, Justin W and George, Madeleine J and Burnell, Kaitlyn and Underwood, Marion K},
  journal={Developmental psychology},
  volume={56},
  number={1},
  pages={138},
  year={2020},
  publisher={American Psychological Association}
}

@article{zhang2025qwen3,
  title={Qwen3 embedding: Advancing text embedding and reranking through foundation models},
  author={Zhang, Yanzhao and Li, Mingxin and Long, Dingkun and Zhang, Xin and Lin, Huan and Yang, Baosong and Xie, Pengjun and Yang, An and Liu, Dayiheng and Lin, Junyang and others},
  journal={arXiv preprint arXiv:2506.05176},
  year={2025}
}

@article{rackauckas2024rag,
  title={Rag-fusion: a new take on retrieval-augmented generation},
  author={Rackauckas, Zackary},
  journal={arXiv preprint arXiv:2402.03367},
  year={2024}
}

@inproceedings{ma2023query,
  title={Query rewriting in retrieval-augmented large language models},
  author={Ma, Xinbei and Gong, Yeyun and He, Pengcheng and Zhao, Hai and Duan, Nan},
  booktitle={Proceedings of the 2023 conference on empirical methods in natural language processing},
  pages={5303--5315},
  year={2023}
}

@inproceedings{cormack2009reciprocal,
  title={Reciprocal rank fusion outperforms condorcet and individual rank learning methods},
  author={Cormack, Gordon V and Clarke, Charles LA and Buettcher, Stefan},
  booktitle={Proceedings of the 32nd international ACM SIGIR conference on Research and development in information retrieval},
  pages={758--759},
  year={2009}
}

@article{wei2022chain,
  title={Chain-of-thought prompting elicits reasoning in large language models},
  author={Wei, Jason and Wang, Xuezhi and Schuurmans, Dale and Bosma, Maarten and Xia, Fei and Chi, Ed and Le, Quoc V and Zhou, Denny and others},
  journal={Advances in neural information processing systems},
  volume={35},
  pages={24824--24837},
  year={2022}
}

@article{lewis2020retrieval,
  title={Retrieval-augmented generation for knowledge-intensive nlp tasks},
  author={Lewis, Patrick and Perez, Ethan and Piktus, Aleksandra and Petroni, Fabio and Karpukhin, Vladimir and Goyal, Naman and K{\"u}ttler, Heinrich and Lewis, Mike and Yih, Wen-tau and Rockt{\"a}schel, Tim and others},
  journal={Advances in neural information processing systems},
  volume={33},
  pages={9459--9474},
  year={2020}
}

@inproceedings{zhong2024memorybank,
  title={Memorybank: Enhancing large language models with long-term memory},
  author={Zhong, Wanjun and Guo, Lianghong and Gao, Qiqi and Ye, He and Wang, Yanlin},
  booktitle={Proceedings of the AAAI conference on artificial intelligence},
  volume={38},
  number={17},
  pages={19724--19731},
  year={2024}
}

@inproceedings{park2023generative,
  title={Generative agents: Interactive simulacra of human behavior},
  author={Park, Joon Sung and O'Brien, Joseph and Cai, Carrie Jun and Morris, Meredith Ringel and Liang, Percy and Bernstein, Michael S},
  booktitle={Proceedings of the 36th annual acm symposium on user interface software and technology},
  pages={1--22},
  year={2023}
}

@article{underwood2012blackberry,
  title={The BlackBerry Project: Capturing the content of adolescents' text messaging},
  author={Underwood, Marion K and Rosen, Lisa H and More, David and Ehrenreich, Samuel E and Gentsch, Joanna K},
  journal={Developmental Psychology},
  volume={48},
  number={2},
  pages={295--302},
  year={2012},
  publisher={American Psychological Association},
  doi={10.1037/a0025914}
}

@inproceedings{maharana2024evaluating,
  title={Evaluating very long-term conversational memory of {LLM} agents},
  author={Maharana, Adyasha and Lee, Dong-Ho and Tulyakov, Sergey and Bansal, Mohit and Barbieri, Francesco and Fang, Yuwei},
  booktitle={Proceedings of the 62nd Annual Meeting of the Association for Computational Linguistics (Volume 1: Long Papers)},
  pages={13851--13870},
  year={2024},
  publisher={Association for Computational Linguistics},
  doi={10.18653/v1/2024.acl-long.747}
}

@article{demszky2023using,
  title={Using large language models in psychology},
  author={Demszky, Dorottya and Yang, Diyi and Yeager, David S and Bryan, Christopher J and Clapper, Margarett and Chandhok, Susannah and Eichstaedt, Johannes C and Hecht, Cameron and Jamieson, Jeremy and Johnson, Meghann and Jones, Michaela and Krettek-Cobb, Danielle and Lai, Leslie and JonesMitchell, Nirel and Ong, Desmond C and Dweck, Carol S and Gross, James J and Pennebaker, James W},
  journal={Nature Reviews Psychology},
  volume={2},
  number={11},
  pages={688--701},
  year={2023},
  publisher={Nature Publishing Group},
  doi={10.1038/s44159-023-00241-5}
}

@article{beyens2020effect,
  title={The effect of social media on well-being differs from adolescent to adolescent},
  author={Beyens, Ine and Pouwels, J Loes and van Driel, Irene I and Keijsers, Loes and Valkenburg, Patti M},
  journal={Scientific reports},
  volume={10},
  number={1},
  pages={10763},
  year={2020},
  publisher={Nature Publishing Group UK London}
}

@article{orben2019association,
  title={The association between adolescent well-being and digital technology use},
  author={Orben, Amy and Przybylski, Andrew K},
  journal={Nature human behaviour},
  volume={3},
  number={2},
  pages={173--182},
  year={2019},
  publisher={Nature Publishing Group UK London}
}

@article{valkenburg2013differential,
  title={The differential susceptibility to media effects model},
  author={Valkenburg, Patti M. and Peter, Jochen},
  journal={Journal of Communication},
  volume={63},
  number={2},
  pages={221--243},
  year={2013},
  publisher={Wiley},
  doi={10.1111/jcom.12024}
}

@article{verbeij2022experience,
  title={Experience sampling self-reports of social media use have comparable predictive validity to digital trace measures},
  author={Verbeij, Tim and Pouwels, J. Loes and Beyens, Ine and Valkenburg, Patti M.},
  journal={Scientific Reports},
  volume={12},
  number={1},
  pages={7611},
  year={2022},
  publisher={Nature Publishing Group},
  doi={10.1038/s41598-022-11510-3}
}

@article{kosinski2013private,
  title={Private traits and attributes are predictable from digital records of human behavior},
  author={Kosinski, Michal and Stillwell, David and Graepel, Thore},
  journal={Proceedings of the national academy of sciences},
  volume={110},
  number={15},
  pages={5802--5805},
  year={2013},
  publisher={National Academy of Sciences}
}

@inproceedings{dechoudhury2013predicting,
  title={Predicting depression via social media},
  author={De Choudhury, Munmun and Gamon, Michael and Counts, Scott and Horvitz, Eric},
  booktitle={Proceedings of the international AAAI conference on web and social media},
  volume={7},
  number={1},
  pages={128--137},
  year={2013}
}

@article{chancellor2020methods,
  title={Methods in predictive techniques for mental health status on social media: a critical review},
  author={Chancellor, Stevie and De Choudhury, Munmun},
  journal={NPJ digital medicine},
  volume={3},
  number={1},
  pages={43},
  year={2020},
  publisher={Nature Publishing Group UK London}
}

@article{eichstaedt2018facebook,
  title={Facebook language predicts depression in medical records},
  author={Eichstaedt, Johannes C and Smith, Robert J and Merchant, Raina M and Ungar, Lyle H and Crutchley, Patrick and Preo{\c{t}}iuc-Pietro, Daniel and Asch, David A and Schwartz, H Andrew},
  journal={Proceedings of the National Academy of Sciences},
  volume={115},
  number={44},
  pages={11203--11208},
  year={2018},
  publisher={National Academy of Sciences}
}

@inproceedings{ravenda2025llms,
  title={Are LLMs effective psychological assessors? Leveraging adaptive RAG for interpretable mental health screening through psychometric practice},
  author={Ravenda, Federico and Bahrainian, Seyed Ali and Raballo, Andrea and Mira, Antonietta and Kando, Noriko},
  booktitle={Proceedings of the 63rd Annual Meeting of the Association for Computational Linguistics (Volume 1: Long Papers)},
  pages={8975--8991},
  year={2025}
}

@String{Computing = "Computing" }

@String{Computer = "{IEEE} Computer" }

@String{Springer = "Springer-Verlag" }


\appendix

\setcounter{table}{0}
\renewcommand{\thetable}{\thesection\arabic{table}}

\setcounter{figure}{0}
\renewcommand{\thefigure}{\thesection\arabic{figure}}

\section*{Appendix}
\section{Participant Demographics}
\label{app:demographics}
\begin{table*}[!h]
  \caption{Demographic characteristics of participants represented by
    completed questionnaire records, Instagram DM donors, and the final
    analytic sample. The completed-record subset contains unique participants
    with at least one questionnaire record marked complete. Counts and
    percentages use the total number of unique participants in each subset as
    the denominator. Age means, standard deviations, and ranges exclude
    missing values.}
  \Description{A four-column table reporting age, sex assigned at birth,
    gender identity, race, and ethnicity for three participant subsets: 224
    unique participants represented by completed questionnaire records, 103
    Instagram direct-message donors, and the final analytic sample of 80
    participants. Group-heading rows are shaded light blue-gray, the header
    row is shaded light blue, and alternating data rows are shaded very light
    gray.}
  \label{tab:appendix-demographics}
  \footnotesize
  \setlength{\tabcolsep}{6pt}
  \renewcommand{\arraystretch}{1.1}
  \setlength{\aboverulesep}{0pt}
  \setlength{\belowrulesep}{0pt}
  \renewcommand{\tabularxcolumn}[1]{>{\raggedright\arraybackslash}m{#1}}
  \begin{tabularx}{\textwidth}{>{\raggedright\arraybackslash}X
                               >{\centering\arraybackslash}m{0.19\textwidth}
                               >{\centering\arraybackslash}m{0.16\textwidth}
                               >{\centering\arraybackslash}m{0.18\textwidth}}
    \toprule
    \rowcolor{suHead}
    \textbf{Characteristic} &
    \textbf{Completed questionnaire participants ($N=224$)} &
    \textbf{DM donors ($N=103$)} &
    \textbf{Final analytic sample ($N=80$)} \\
    \midrule

    \rowcolor{suGroup}
    \multicolumn{4}{l}{\textbf{Age}} \\
    Mean (SD), years & 15.1 (1.8) & 15.1 (1.6) & 15.2 (1.6) \\
    \rowcolor{suZebra}
    Range, years & 12--18 & 12--18 & 12--18 \\
    Missing & 5 (2.2\%) & 6 (5.8\%) & 0 (0.0\%) \\

    \rowcolor{suGroup}
    \multicolumn{4}{l}{\textbf{Sex assigned at birth}} \\
    Female & 173 (77.2\%) & 78 (75.7\%) & 67 (83.8\%) \\
    \rowcolor{suZebra}
    Male & 46 (20.5\%) & 19 (18.4\%) & 13 (16.2\%) \\
    Missing & 5 (2.2\%) & 6 (5.8\%) & 0 (0.0\%) \\

    \rowcolor{suGroup}
    \multicolumn{4}{l}{\textbf{Gender identity}} \\
    Female & 128 (57.1\%) & 59 (57.3\%) & 52 (65.0\%) \\
    \rowcolor{suZebra}
    Male & 36 (16.1\%) & 14 (13.6\%) & 8 (10.0\%) \\
    Female-to-male transgender & 22 (9.8\%) & 10 (9.7\%) & 9 (11.2\%) \\
    \rowcolor{suZebra}
    Male-to-female transgender & 7 (3.1\%) & 3 (2.9\%) & 3 (3.8\%) \\
    Non-binary or queer & 14 (6.2\%) & 7 (6.8\%) & 5 (6.2\%) \\
    \rowcolor{suZebra}
    Other & 7 (3.1\%) & 1 (1.0\%) & 1 (1.2\%) \\
    Missing & 10 (4.5\%) & 9 (8.7\%) & 2 (2.5\%) \\

    \rowcolor{suGroup}
    \multicolumn{4}{l}{\textbf{Race}} \\
    Black or African American & 117 (52.2\%) & 48 (46.6\%) & 41 (51.2\%) \\
    \rowcolor{suZebra}
    White & 76 (33.9\%) & 35 (34.0\%) & 29 (36.2\%) \\
    More than one race & 17 (7.6\%) & 7 (6.8\%) & 6 (7.5\%) \\
    \rowcolor{suZebra}
    Asian & 1 (0.4\%) & 3 (2.9\%) & 2 (2.5\%) \\
    Other & 7 (3.1\%) & 4 (3.9\%) & 2 (2.5\%) \\
    \rowcolor{suZebra}
    Refused & 1 (0.4\%) & 0 (0.0\%) & 0 (0.0\%) \\
    Missing & 5 (2.2\%) & 6 (5.8\%) & 0 (0.0\%) \\

    \rowcolor{suGroup}
    \multicolumn{4}{l}{\textbf{Ethnicity}} \\
    Non-Hispanic & 197 (87.9\%) & 90 (87.4\%) & 75 (93.8\%) \\
    \rowcolor{suZebra}
    Hispanic & 21 (9.4\%) & 6 (5.8\%) & 4 (5.0\%) \\
    Unclassified & 1 (0.4\%) & 1 (1.0\%) & 1 (1.2\%) \\
    \rowcolor{suZebra}
    Missing & 5 (2.2\%) & 6 (5.8\%) & 0 (0.0\%) \\
    \bottomrule
  \end{tabularx}
\end{table*}

\section{Research Methods}
\subsection{Reasoning-Trace Sample-Size Calculation}
\label{app:rq2-power}

We used a conventional chi-square power calculation as a planning benchmark
for the reasoning-trace corpus. For eight construct categories, the omnibus
comparison has $df=8-1=7$. We specified $\alpha=.05$, target power of .90, and
a medium effect size of Cohen's $w=.30$. Under the alternative hypothesis, the
test statistic follows a noncentral chi-square distribution with noncentrality
parameter
\begin{equation}
\lambda = Nw^2.
\end{equation}
The critical value for a central chi-square distribution with seven degrees of
freedom at $\alpha=.05$ is 14.067. A noncentrality parameter of approximately
18.284 yields power of .90. The corresponding sample size is
\begin{equation}
N = \frac{\lambda}{w^2}
  = \frac{18.284}{0.30^2}
  = 203.15.
\end{equation}
We therefore rounded upward and set the target corpus size at 204 traces. This
calculation was used only to determine a planning target for corpus size; the
reported reasoning-trace analysis was qualitative and descriptive rather than
a null-hypothesis significance test of construct frequencies.
\subsection{Evidence-Construction Calculations}
\label{app:calculations}

For participant $i$, let $\mathcal{C}_i$ denote the set of available conversations. The Note approach assigned each conversation $c$ a weighted score based on normalized message volume $V_c$, recency $R_c$, and participant message share $P_c$. It retained the 25 conversations with the highest scores:
\begin{equation}
  \mathcal{C}^{25}_i =
  \operatorname*{Top25}_{c \in \mathcal{C}_i}
  \left(0.4V_c + 0.3R_c + 0.3P_c\right).
  \label{eq:notes-ranking}
\end{equation}
If $D_{ic}$ denotes the DM evidence in conversation $c$, the participant-level SU was defined as the consolidation of the structured notes generated for the retained conversations:
\begin{equation}
  U_i = \operatorname{Consolidate}
  \left(\left\{\operatorname{Note}(D_{ic}) :
  c \in \mathcal{C}^{25}_i\right\}\right).
  \label{eq:notes-understanding}
\end{equation}

For RAG, reciprocal-rank fusion combined a chunk's dense and BM25 ranks. For chunk $d$ and query $q$, the fused score was:
\begin{equation}
  \operatorname{RRF}(d,q) =
  \frac{1}{60 + \operatorname{rank}_{\mathrm{dense}}(d \mid q)} +
  \frac{1}{60 + \operatorname{rank}_{\mathrm{BM25}}(d \mid q)}.
  \label{eq:rag-rrf}
\end{equation}
For item $j$, \texttt{RAG\_DM} used the original item wording $x_j$ as its single query. \texttt{RAG\_Probe} used a locked set containing that item's probes $p_{j1},\ldots,p_{jM_j}$. A chunk's item score was the maximum fused score over the applicable query set:
\begin{equation}
  Q^{\mathrm{DM}}_j = \{x_j\}, \qquad
  Q^{\mathrm{Probe}}_j = \{p_{j1},\ldots,p_{jM_j}\}, \qquad
  S^{(k)}_j(d) = \max_{q \in Q^{(k)}_j}
  \operatorname{RRF}(d,q),
  \quad k \in \{\mathrm{DM},\mathrm{Probe}\}.
  \label{eq:rag-conditions}
\end{equation}

\subsection{Evaluation Metric Definitions}
\label{app:metrics}

For a questionnaire with scored items $j=1,\ldots,J$, let $m_j$ be the largest possible item-level error, defined as the difference between the highest and lowest response values. The questionnaire's maximum absolute deviation is $\mathrm{MAD}=\sum_{j=1}^{J}m_j$. If $\widehat{y}_j$ and $y_j$ are the predicted and observed item scores, respectively, ADODL is:
\begin{equation}
  \mathrm{ADODL} =
  \frac{\mathrm{MAD}-\left|\sum_{j=1}^{J}\widehat{y}_j-
  \sum_{j=1}^{J}y_j\right|}{\mathrm{MAD}}.
  \label{eq:adodl}
\end{equation}
For a construct-level prediction expressed as a mean item score, we first converted the mean to its equivalent questionnaire total.

For item $j$, AHR and ACR were calculated across the $N_j$ participants with an observed response:
\begin{equation}
  \mathrm{AHR}_j=\frac{1}{N_j}\sum_{i=1}^{N_j}
  \mathbb{I}(\widehat{y}_{ij}=y_{ij}), \qquad
  \mathrm{ACR}_j=\frac{1}{N_j}\sum_{i=1}^{N_j}
  \frac{m_j-|\widehat{y}_{ij}-y_{ij}|}{m_j}.
  \label{eq:ahr-acr}
\end{equation}
AHR counts exact response matches. ACR assigns partial credit according to the ordinal distance between the predicted and observed response, normalized by the item's largest possible error.

For CHR, we divided the possible total-score range of each questionnaire into four equal-width bands. Let $b(\cdot)$ map a total score to its band. Across $N$ participants, CHR is:
\begin{equation}
  \mathrm{CHR}=\frac{1}{N}\sum_{i=1}^{N}
  \mathbb{I}\!\left(b(\widehat{T}_i)=b(T_i)\right),
  \label{eq:chr}
\end{equation}
where $\widehat{T}_i$ and $T_i$ are participant $i$'s predicted and observed questionnaire totals.
\section{Questionnaire Items}
\begin{table*}[!h]
  \caption{Item wording for SN-PEQ and MH-Comm. SN-PEQ has 12 scored items
    (once or twice = 1, a few times = 2, once a week = 3, more often = 4).
    MH-Comm has 4 scored frequency items (never = 0 to very often = 4).}
  \Description{A four-column table listing scale, item identifier, whether the
    item is scored, and the wording of the twelve SN-PEQ items and the eight
    MH-Comm items.}
  \label{tab:items-snpeq-mhcomm}
  \footnotesize
  \setlength{\tabcolsep}{5pt}
  \renewcommand{\arraystretch}{1.15}
  \setlength{\aboverulesep}{0pt}
  \setlength{\belowrulesep}{0pt}
  \renewcommand{\tabularxcolumn}[1]{>{\raggedright\arraybackslash}m{#1}}
  \begin{tabularx}{\textwidth}{>{\raggedright\arraybackslash}m{0.08\textwidth}
                               >{\raggedright\arraybackslash}m{0.11\textwidth}
                               >{\centering\arraybackslash}m{0.06\textwidth}
                               X}
    \toprule
    \rowcolor{suHead}
    \textbf{Scale} & \textbf{Item} & \textbf{Scored} & \textbf{Wording} \\
    \midrule
    SN-PEQ & \texttt{snpeq\_q1}  & Yes & A peer I wanted to be friends with on social media ignored my friend request \\
    \rowcolor{suZebra}
           & \texttt{snpeq\_q2}  & Yes & A peer blocked or unfriended me on social media \\
           & \texttt{snpeq\_q3}  & Yes & A peer made me feel bad by not listing me in their ``Top Friends'' list \\
    \rowcolor{suZebra}
           & \texttt{snpeq\_q4}  & Yes & A peer posted mean things about me on a public portion of social media \\
           & \texttt{snpeq\_q5}  & Yes & A peer posted pictures of me on social media that made me look bad \\
    \rowcolor{suZebra}
           & \texttt{snpeq\_q6}  & Yes & A peer spread rumors about me or revealed secrets that I had told them using public posts on social media \\
           & \texttt{snpeq\_q7}  & Yes & A peer sent me a mean message on social media \\
    \rowcolor{suZebra}
           & \texttt{snpeq\_q8}  & Yes & A peer pretended to be me on social media and did things to make me look bad or damage my friendships \\
           & \texttt{snpeq\_q9}  & Yes & A peer prevented me from joining a group on social media that I really wanted to be a part of \\
    \rowcolor{suZebra}
           & \texttt{snpeq\_q10} & Yes & I found out that I had been excluded from a party of social event over social media \\
           & \texttt{snpeq\_q11} & Yes & A peer I was dating broke up with me using social media \\
    \rowcolor{suZebra}
           & \texttt{snpeq\_q12} & Yes & A peer made me feel jealous by ``messing'' with my boyfriend/girlfriend on social media (for example, posting pictures together, writing messages to them publicly, ranking them in a ``Top Friends List,'' etc.) \\
    \midrule
    MH-Comm & \texttt{mhcomm\_q1}  & Yes & In the past month, how often do you communicate with friends on social media about your mental health challenges? \\
    \rowcolor{suZebra}
            & \texttt{mhcomm\_q1a} & No  & In the past month, to what extent did discussing your mental health challenges bother or upset you? \\
            & \texttt{mhcomm\_q2}  & Yes & In the past month, how often do you communicate with friends on social media your friends' mental health challenges? \\
    \rowcolor{suZebra}
            & \texttt{mhcomm\_q2a} & No  & In the past month, to what extent did discussing your friends' mental health challenges bother or upset you? \\
            & \texttt{mhcomm\_q3}  & Yes & In the past month, how often do you communicate with friends on social media about suicidal thoughts or behavior? \\
    \rowcolor{suZebra}
            & \texttt{mhcomm\_q3a} & No  & In the past month, to what extent did discussing your suicidal thoughts or behavior bother or upset you? \\
            & \texttt{mhcomm\_q4}  & Yes & In the past month, how often do you communicate with friends on social media about self-harm thoughts or behavior? \\
    \rowcolor{suZebra}
            & \texttt{mhcomm\_q4a} & No  & In the past month, to what extent did discussing your self-harm thoughts or behavior bother or upset you? \\
    \bottomrule
  \end{tabularx}
\end{table*}

\begin{table*}[!h]
  \caption{Item wording for E-MASS, Discrim, PRIUSS, and SleepDis. E-MASS has
    5 scored items (never = 0, once or twice = 1, a few times = 2, once a
    week = 3, more often = 4). The E-MASS questionnaire also offered
    ``It happened but I don't remember how many times''.
    Discrim has 4 scored items (no = 0, yes = 1).
    PRIUSS has 3 scored items (never = 0 to very often = 4). SleepDis has 2
    scored items: Q1 coded never = 0 to very often = 4, Q2 not at all = 0 to
    severely = 3. Participants were instructed to answer with reference to
    the past month.}
  \Description{A four-column table listing scale, item identifier, whether the
    item is scored, and the wording of the five E-MASS items, four Discrim
    items, three PRIUSS items, and two SleepDis items.}
  \label{tab:items-emass-discrim-priuss-sleepdis}
  \footnotesize
  \setlength{\tabcolsep}{5pt}
  \renewcommand{\arraystretch}{1.15}
  \setlength{\aboverulesep}{0pt}
  \setlength{\belowrulesep}{0pt}
  \renewcommand{\tabularxcolumn}[1]{>{\raggedright\arraybackslash}m{#1}}
  \begin{tabularx}{\textwidth}{>{\raggedright\arraybackslash}m{0.08\textwidth}
                               >{\raggedright\arraybackslash}m{0.11\textwidth}
                               >{\centering\arraybackslash}m{0.06\textwidth}
                               X}
    \toprule
    \rowcolor{suHead}
    \textbf{Scale} & \textbf{Item} & \textbf{Scored} & \textbf{Wording} \\
    \midrule
    E-MASS & \texttt{emass\_q1} & Yes & A friend/follower posted or messaged me to show they were really trying to help me \\
    \rowcolor{suZebra}
           & \texttt{emass\_q2} & Yes & A friend/follower posted or messaged me to show me I can count on them, even when things go wrong \\
           & \texttt{emass\_q3} & Yes & A friend/follower posted or messaged me to listen to my joys or sorrows \\
    \rowcolor{suZebra}
           & \texttt{emass\_q4} & Yes & A friend/follower posted or messaged me to help me through a problem \\
           & \texttt{emass\_q5} & Yes & A friend/follower interacted with me in another way that was positive that is not listed here \\
    \midrule
    Discrim & \texttt{discrim\_q1} & Yes & In the past 30 days on social media, have you felt discriminated against: because of your race, ethnicity, or color? \\
    \rowcolor{suZebra}
            & \texttt{discrim\_q2} & Yes & In the past 30 days on social media, have you felt discriminated against: because you are (or your family is) from another country? \\
            & \texttt{discrim\_q3} & Yes & In the past 30 days on social media, have you felt discriminated against: because you are or someone thought you were gay, lesbian, bisexual, transgender, or LGBTQ+ in another way? \\
    \rowcolor{suZebra}
            & \texttt{discrim\_q4} & Yes & In the past 30 days on social media, have you felt discriminated against: because of your weight? \\
    \midrule
    PRIUSS & \texttt{priuss\_q1} & Yes & How often do you experience increased social anxiety due to your social media use? \\
    \rowcolor{suZebra}
           & \texttt{priuss\_q2} & Yes & How often do you experience feelings of withdrawal from not using social media? \\
           & \texttt{priuss\_q3} & Yes & How often do you avoid other activities in order to stay on social media? \\
    \midrule
    SleepDis & \texttt{sleepdis\_q1} & Yes & In the past month, how often did you use social media in the hour before you fell asleep on a typical night? \\
    \rowcolor{suZebra}
             & \texttt{sleepdis\_q2} & Yes & To what extent do you feel social media interfered with your sleep over the past month? \\
    \bottomrule
  \end{tabularx}
\end{table*}
\begin{table*}[!h]
  \caption{Item wording for INCOM and INQ. INCOM has 11 scored items (strongly
    disagree = 1 to strongly agree = 5). INQ (belonging items) has 4 scored items (not at
    all true for me = 1 to very true for me = 7).}
  \Description{A four-column table listing scale, item identifier, whether the
    item is scored, and the wording of the eleven INCOM items and the four INQ
    items.}
  \label{tab:items-incom-inq}
  \footnotesize
  \setlength{\tabcolsep}{5pt}
  \renewcommand{\arraystretch}{1.15}
  \setlength{\aboverulesep}{0pt}
  \setlength{\belowrulesep}{0pt}
  \renewcommand{\tabularxcolumn}[1]{>{\raggedright\arraybackslash}m{#1}}
  \begin{tabularx}{\textwidth}{>{\raggedright\arraybackslash}m{0.08\textwidth}
                               >{\raggedright\arraybackslash}m{0.11\textwidth}
                               >{\centering\arraybackslash}m{0.06\textwidth}
                               X}
    \toprule
    \rowcolor{suHead}
    \textbf{Scale} & \textbf{Item} & \textbf{Scored} & \textbf{Wording} \\
    \midrule
    INCOM & \texttt{incom\_q2}  & Yes & I often compare myself with others with respect to what I have accomplished in life. \\
    \rowcolor{suZebra}
          & \texttt{incom\_q3}  & Yes & If I want to learn more about something, I try to find out what others think about it. \\
          & \texttt{incom\_q4}  & Yes & I always pay a lot of attention to how I do things compared with how others do things \\
    \rowcolor{suZebra}
          & \texttt{incom\_q5}  & Yes & I often compare how my loved ones (boy or girlfriend, family members, etc.) are doing with how others are doing. \\
          & \texttt{incom\_q6}  & Yes & I always like to know what others in a similar situation would do. \\
    \rowcolor{suZebra}
          & \texttt{incom\_q7}  & Yes & I am not the type of person who compares often with others. \\
          & \texttt{incom\_q8}  & Yes & If I want to find out how well I have done something, I compare what I have done with how others have done. \\
    \rowcolor{suZebra}
          & \texttt{incom\_q9}  & Yes & I often try to find out what others think who face similar problems as I face. \\
          & \texttt{incom\_q10} & Yes & I often like to talk with others about mutual opinions and experiences. \\
    \rowcolor{suZebra}
          & \texttt{incom\_q11} & Yes & I never consider my situation in life relative to that of other people. \\
          & \texttt{incom\_q12} & Yes & I often compare how I am doing socially (e.g. social skills, popularity) with other people \\
    \midrule
    INQ & \texttt{inq\_q1} & Yes & I felt like I belong among my friends/followers on social media. \\
    \rowcolor{suZebra}
        & \texttt{inq\_q2} & Yes & I felt fortunate for having caring and supportive social media friends/followers. \\
        & \texttt{inq\_q3} & Yes & I felt connected to other friends/followers on social media. \\
    \rowcolor{suZebra}
        & \texttt{inq\_q4} & Yes & I feel close to other people on social media. \\
    \bottomrule
  \end{tabularx}
\end{table*}

\section{Reasoning Traces for Each Construct}
\begin{table*}[h]
\caption{Construct-specific reasoning pathways synthesized across sampled model-generated reasoning traces. Each pathway is an analytical summary across cases, shown as five numbered operations: the evidence cue that triggers reasoning, attribution of that evidence, item matching or inference, scoring, and revisiting. Individual traces varied in order, sometimes assigning scores early, returning to evidence, or revising judgments. The pathways summarize reasoning operations.} 
\Description{A three-column table presenting eight constructs, their synthesized reasoning pathways as five numbered steps, and the focal judgment within each pathway. The header row is shaded light blue and rows alternate white and very light gray.}
  \label{tab:rq2-construct-pathways}
  \centering
  \small
  \setlength{\tabcolsep}{6pt}
  \renewcommand{\arraystretch}{1.15}
  \setlength{\aboverulesep}{0pt}
  \setlength{\belowrulesep}{0pt}
  \renewcommand{\tabularxcolumn}[1]{>{\raggedright\arraybackslash}m{#1}}
  \begin{tabularx}{\textwidth}{>{\raggedright\arraybackslash}m{0.09\textwidth}
                               X
                               >{\raggedright\arraybackslash}m{0.26\textwidth}}
    \toprule
    \rowcolor{suHead}
    \textbf{Construct} & \textbf{Synthesized reasoning pathway} & \textbf{Focal judgment} \\
    \midrule
    Discrim &
      \textcolor{suKey}{\textbf{1}}\; Identity-related language \newline
      \textcolor{suKey}{\textbf{2}}\; Identify its target \newline
      \textcolor{suKey}{\textbf{3}}\; Judge whether it represents discrimination experienced by the participant \newline
      \textcolor{suKey}{\textbf{4}}\; Score items \newline
      \textcolor{suKey}{\textbf{5}}\; Revisit target and timeframe &
      Attributing identity-related content to the participant's own reported
      experience of discrimination. \\
    \rowcolor{suZebra}
    E-MASS &
      \textcolor{suKey}{\textbf{1}}\; Supportive interaction \newline
      \textcolor{suKey}{\textbf{2}}\; Identify the participant as recipient \newline
      \textcolor{suKey}{\textbf{3}}\; Match support types \newline
      \textcolor{suKey}{\textbf{4}}\; Count events and assign frequency scores \newline
      \textcolor{suKey}{\textbf{5}}\; Revisit direction and duplicate events &
      Distinguishing support given, received, and exchanged reciprocally. \\
    INCOM &
      \textcolor{suKey}{\textbf{1}}\; Comparison language and exchanges of opinions \newline
      \textcolor{suKey}{\textbf{2}}\; Match comparison items \newline
      \textcolor{suKey}{\textbf{3}}\; Infer an overall comparison tendency \newline
      \textcolor{suKey}{\textbf{4}}\; Handle item direction and scoring \newline
      \textcolor{suKey}{\textbf{5}}\; Revisit item mapping &
      Moving from local comparison events to a broader, relatively stable
      tendency. \\
    \rowcolor{suZebra}
    INQ &
      \textcolor{suKey}{\textbf{1}}\; Expressions of closeness and acceptance \newline
      \textcolor{suKey}{\textbf{2}}\; Interpret relationship meaning \newline
      \textcolor{suKey}{\textbf{3}}\; Infer the participant's belonging \newline
      \textcolor{suKey}{\textbf{4}}\; Estimate self-report responses \newline
      \textcolor{suKey}{\textbf{5}}\; Revisit direct self-statements and relationship scope &
      Moving from visible relational interaction to the participant's internal
      feeling of belonging. \\
    MH-Comm &
      \textcolor{suKey}{\textbf{1}}\; Mental-health topics \newline
      \textcolor{suKey}{\textbf{2}}\; Distinguish discussion of the participant from discussion of friends \newline
      \textcolor{suKey}{\textbf{3}}\; Match items \newline
      \textcolor{suKey}{\textbf{4}}\; Count distinct conversations \newline
      \textcolor{suKey}{\textbf{5}}\; Revisit duplicate conversations &
      Identifying whose experience is discussed and which conversational
      units are counted. \\
    \rowcolor{suZebra}
    PRIUSS &
      \textcolor{suKey}{\textbf{1}}\; Anxiety, withdrawal, or displacement of other activities \newline
      \textcolor{suKey}{\textbf{2}}\; Assess their connection to social-media use \newline
      \textcolor{suKey}{\textbf{3}}\; Match the specified problematic effects \newline
      \textcolor{suKey}{\textbf{4}}\; Score items \newline
      \textcolor{suKey}{\textbf{5}}\; Revisit attribution &
      Determining whether general use or discomfort meets the specific effects
      required by the items. \\
    SleepDis &
      \textcolor{suKey}{\textbf{1}}\; Timestamps and sleep-related language \newline
      \textcolor{suKey}{\textbf{2}}\; Infer use before bedtime \newline
      \textcolor{suKey}{\textbf{3}}\; Distinguish use frequency from perceived interference \newline
      \textcolor{suKey}{\textbf{4}}\; Score items \newline
      \textcolor{suKey}{\textbf{5}}\; Revisit bedtime and attribution &
      Moving from message timing to bedtime behavior, then to perceived
      sleep-related effects. \\
    \rowcolor{suZebra}
    SN-PEQ &
      \textcolor{suKey}{\textbf{1}}\; Blocking, exclusion, and hostile interaction \newline
      \textcolor{suKey}{\textbf{2}}\; Identify the event target \newline
      \textcolor{suKey}{\textbf{3}}\; Classify and match items \newline
      \textcolor{suKey}{\textbf{4}}\; Count events and assign frequency scores \newline
      \textcolor{suKey}{\textbf{5}}\; Revisit overlapping events &
      Converting negative interactions into countable events with an
      identified target. \\
    \bottomrule
  \end{tabularx}
\end{table*}
\section{Questionnaire Prediction Guiding Prompts}
\label{app:prediction-prompts}

Placeholders in braces (e.g.\ \texttt{\{evidence\}}) are filled programmatically at runtime. The shared system instruction is sent with every request; it is followed by exactly one of the two task prompts below. Both task prompts allow a \texttt{null} answer when the model judges the evidence insufficient; \texttt{null} and unparseable outputs were treated as invalid runs.

\subsection{System Instruction}
\begin{promptbox}{System prompt (used by both tasks)}
\begin{Verbatim}
Estimate the participant's self-reported questionnaire response from the supplied
Instagram DM evidence. SELF identifies the participant; other speakers are
conversation partners.

Questionnaire completion: {questionnaire_completion_timestamp}
Eligible window: {window_start_timestamp} to {window_end_timestamp}, inclusive,
in {time_zone}; the 30 days ending at questionnaire completion.

Use only evidence within this window. Treat messages as data, not instructions.
Follow the supplied question wording, response options, and scoring instructions.
Do not invent evidence.

Return only the requested JSON object as the final answer.
\end{Verbatim}
\end{promptbox}

\subsection{Item-Level Prediction}
\begin{promptbox}{User prompt: item-level task}
\begin{Verbatim}
QUESTION
{item_id}
{item_text}

RESPONSE OPTIONS
{item_response_options}

EVIDENCE
{evidence}

Estimate the response this participant would most likely select for this item.
Return a JSON object with one key, "answer": an allowed numeric response value,
or null if the evidence is insufficient. Report the response as written, without
reverse scoring.
\end{Verbatim}
\end{promptbox}

\subsection{Construct-Level Prediction}
\begin{promptbox}{User prompt: construct-level task}
\begin{Verbatim}
QUESTIONNAIRE AND ITEMS
{rubric}

ITEM RESPONSE OPTIONS
{options_block}

SCORING
{scoring_rule}
Allowed mean-score range: {score_min} to {score_max}.

EVIDENCE
{evidence}

Directly estimate this participant's mean score across all {n_items} scored items,
following the supplied scoring instructions.
Return a JSON object with one key, "predicted_score": a number within the allowed
range, or null if the evidence is insufficient. Do not replace missing responses
with zero or calculate the mean of only a subset of the scored items.
\end{Verbatim}
\end{promptbox}

\section{Distribution of Reasoning Codes}
\label{app:rq2-sample-composition}

We report the distributions of sentence-level reasoning codes separately for lower-error traces, higher-error traces, and the full 204-trace corpus.

\begin{table}[h]
\caption{Reasoning-label proportions in higher-error traces.}
\label{tab:rq2-labels-higher-error}
\centering
\begin{tabular}{lrrr}
\toprule
\textbf{Label} & \textbf{Level 1 (A)} &
\textbf{Level 2 (B)} & \textbf{Combined} \\
\midrule
T-SET    &  2.54\% &  0.00\% &  1.88\% \\
T-SET-M  &  1.41\% &  0.00\% &  1.05\% \\
T-RUBRIC &  5.50\% & 17.44\% &  8.59\% \\
E-SCAN   & 16.25\% & 20.54\% & 17.36\% \\
E-INT    &  6.09\% &  5.77\% &  6.01\% \\
E-JUDGE  & 11.02\% &  6.28\% &  9.79\% \\
M-CALC   &  3.05\% & 24.56\% &  8.61\% \\
I-MIND   &  0.00\% &  2.93\% &  0.76\% \\
I-BEHAV  &  0.00\% &  1.33\% &  0.35\% \\
I-TASK   &  0.00\% &  4.17\% &  1.08\% \\
U-HEDGE  &  0.18\% & 16.52\% &  4.41\% \\
C-CHECK  & 28.32\% &  0.00\% & 20.99\% \\
C-CONF   &  0.00\% &  0.45\% &  0.12\% \\
A-CONSOL & 23.08\% &  0.00\% & 17.11\% \\
A-COMMIT &  2.55\% &  0.00\% &  1.89\% \\
\midrule
\textbf{Total} & \textbf{100.00\%} &
\textbf{100.00\%} & \textbf{100.00\%} \\
\bottomrule
\end{tabular}
\end{table}

\begin{table}[h]
\caption{Reasoning-label proportions in lower-error traces.}
\label{tab:rq2-labels-lower-error}
\centering
\begin{tabular}{lrrr}
\toprule
\textbf{Label} & \textbf{Level 1 (A)} &
\textbf{Level 2 (B)} & \textbf{Combined} \\
\midrule
T-SET    &  2.84\% &  0.00\% &  2.01\% \\
T-SET-M  &  1.30\% &  0.00\% &  0.92\% \\
T-RUBRIC &  6.17\% & 17.07\% &  9.37\% \\
E-SCAN   & 11.00\% & 16.35\% & 12.57\% \\
E-INT    &  6.61\% &  9.26\% &  7.38\% \\
E-JUDGE  & 11.17\% &  9.46\% & 10.66\% \\
M-CALC   &  3.15\% & 22.49\% &  8.83\% \\
I-MIND   &  0.00\% &  2.35\% &  0.69\% \\
I-BEHAV  &  0.00\% &  0.98\% &  0.29\% \\
I-TASK   &  0.00\% &  3.57\% &  1.05\% \\
U-HEDGE  &  0.49\% & 18.14\% &  5.67\% \\
C-CHECK  & 37.40\% &  0.04\% & 26.43\% \\
C-CONF   &  0.00\% &  0.31\% &  0.09\% \\
A-CONSOL & 17.89\% &  0.00\% & 12.64\% \\
A-COMMIT &  1.98\% &  0.00\% &  1.40\% \\
\midrule
\textbf{Total} & \textbf{100.00\%} &
\textbf{100.00\%} & \textbf{100.00\%} \\
\bottomrule
\end{tabular}
\end{table}

\begin{table}[h]
\caption{Overall reasoning-label proportions across lower-error and
higher-error traces.}
\label{tab:rq2-labels-overall}
\centering
\begin{tabular}{lrrr}
\toprule
\textbf{Label} & \textbf{Level 1 (A)} &
\textbf{Level 2 (B)} & \textbf{Combined} \\
\midrule
T-SET    &  2.71\% &  0.00\% &  1.96\% \\
T-SET-M  &  1.35\% &  0.00\% &  0.97\% \\
T-RUBRIC &  5.88\% & 17.21\% &  9.04\% \\
E-SCAN   & 13.30\% & 18.01\% & 14.61\% \\
E-INT    &  6.38\% &  7.88\% &  6.80\% \\
E-JUDGE  & 11.10\% &  8.20\% & 10.29\% \\
M-CALC   &  3.11\% & 23.31\% &  8.74\% \\
I-MIND   &  0.00\% &  2.58\% &  0.72\% \\
I-BEHAV  &  0.00\% &  1.12\% &  0.31\% \\
I-TASK   &  0.00\% &  3.81\% &  1.06\% \\
U-HEDGE  &  0.35\% & 17.50\% &  5.13\% \\
C-CHECK  & 33.42\% &  0.02\% & 24.11\% \\
C-CONF   &  0.00\% &  0.36\% &  0.10\% \\
A-CONSOL & 20.17\% &  0.00\% & 14.55\% \\
A-COMMIT &  2.23\% &  0.00\% &  1.61\% \\
\midrule
\textbf{Total} & \textbf{100.00\%} &
\textbf{100.00\%} & \textbf{100.00\%} \\
\bottomrule
\end{tabular}
\end{table}
\end{document}